\documentclass[useAMS,usenatbib]{mnras}
 \usepackage{graphicx}
 \usepackage{epstopdf}
 \usepackage[colorinlistoftodos]{todonotes}
 \usepackage{natbib}
 \usepackage{morefloats}
 \usepackage{amsmath}
 \usepackage{dirtytalk}
 \usepackage{subfigure}
 \usepackage{verbatim}
 \usepackage{multirow}
\newcommand{\lsim}{\mathrel{\rlap{\lower4pt\hbox{$\sim$}}\raise1pt\hbox{$<$}}}

\newcommand{\change}[1]{#1}

\title[21CMNest]{Bayesian Model Selection with Future 21cm Observations of The Epoch of Reionisation}
\author[T. Binnie and J.R.Pritchard]{T. Binnie$^{1}$\thanks{E-mail:
t.binnie16@imperial.ac.uk} and J.~R.~Pritchard$^{1}$\thanks{E-mail:
j.pritchard@imperial.ac.uk}\\
$^{1}$Blackett Laboratory, Imperial College, London, SW7 2AZ, UK}

\begin{document}


\maketitle

\label{firstpage}

\begin{abstract}
We apply Bayesian statistics to perform model selection on different reionisation scenarios via the Multinest algorithm. 
Initially, we recover the results shown by 21CMMC for the parameter estimation of 21cmFAST models. 
We proceed to test several toy models of the Epoch of Reionisation (EoR) defined in contrasting morphology and scale. 
We find that LOFAR observations are unlikely to allow model selection even with long integration times.
HERA would require 61 dipoles to perform the same analysis in 1080 hours, and becomes comparable to the SKA with 217 dipoles. 
We find the SKA requires only 324 hours of observation to conclusively distinguish between our models.
Once model selection is achievable, an analysis of observational priors is performed finding that neutral fraction checks at specific redshifts add little to no inference.
We show the difficulties in model selection at the level of distinguishing fiducial parameters within a model or distinguishing galaxies with a constant versus power law mass-to-light ratio.
Finally, we explore the use of the Savage-Dickey density ratio to show the redundancy of the parameter $R_{\rm mfp}$ within 21cmFAST. 
\end{abstract}


\section{Introduction}\label{sec: intro}

The Epoch of Reionisation (EoR) is the most recent phase change of the Universe.
During this process, the hydrogen (HI) in the intergalactic medium (IGM) that has not formed stars and galaxies becomes heated by these very objects and eventually becomes ionised. 
We expect to uncover astrophysics involving radiative transfer, star formation, X-ray heating and the percolation of ionised hydrogen (HII) bubbles. 
Due to a lack of observation however, we remain in the dark.

An absorption signal in the Cosmic Microwave Background (CMB) radiation at 78 MHz was claimed to be observed for the first time by the EDGES team \citep{2bc8363417fb44159194d3775d6e51bd} in February 2018. Around the same time an upper bound of $(79.6~ \rm mk)^2$ on the EoR power spectrum has been achieved by the LOFAR team with a 13 hour observation of the NCP field within the redshift range $z=9.6-10.6$ at $k=0.053~h ~\rm c \rm Mpc^{-1}$ \citep{2017ApJ...838...65P}.

Although these limits on the power spectrum are still several orders of magnitude larger than the signal, this paves the way for the first observations of the EoR with a new generation of radio interferometer. 
The Square Kilometre Array (SKA) will be capable of tomographically mapping primordial hydrogen from $z=6$ to $27$ \citep{2013ExA....36..235M, 2015aska.confE...1K}.
The aim of this work is to explore the application of Bayesian model selection in the context of EoR models to explore the possibility for ruling out specific reionisation scenarios. 

The redshifted 21cm line is set to be a powerful tool for analysing 
the high redshift universe \citep{2006NewAR..50..157F}. 
This atomic line originates from the spin-flip transition of the coupled magnetic moments between the proton and the electron in a hydrogen atom, releasing a 21cm (lab frame) wavelength photon which is then redshifted by the Hubble flow. 
The CMB acts as the initial backlight for the transition.  
By observing the signal at different frequencies, a tomographical image of how the primordial hydrogen evolves can be constructed. 
This methodology allows us to trace the formation of the first stars and galaxies. 
For a full discussion of how physically informative the 21cm signal could be, see e.g. \citet{2012RPPh...75h6901P}.

The current cosmological picture is bound by two sources of observational data: The CMB, at $z \sim 1100$ and high-redshift quasars $z \lsim 7$. 
The Big Bang model suggests that protons and electrons form separately.
By $z \sim 1090$ the expanding universe has cooled enabling $99.9999\%$ of the universe's baryonic matter to form hydrogen atoms \citep{1968ApJ...153....1P,1969Ap&SS...4..301Z}. 
Currently this is thought to remain the case until stars start to form $\sim 150 \rm Myr$ after the Big Bang.
The latter source of data includes measuring the Gunn-Peterson trough \citep{1965ApJ...142.1633G} within quasar spectra. 
This shows that by $z \sim 6$\change{,} $99.96 \pm 0.03 \%$ of the hydrogen in the IGM is ionised \citep{2006AJ....132..117F}.
In between these two milestones the universe \textit{re}-ionises the hydrogen left over from recombination.
A defining feature of EoR toy models is the morphology of reionisation \change{\citep{FHZ, MHR}}. 
In this work we consider two possibilities:
\begin{itemize}
\item \textit{Inside-Out} - Where low density gas is ionised last, due to star formation being most dominant in dense regions of gas. The density field and radiation field correlate with position. 
\item \textit{Outside-In} - Where high density gas is ionised last, due to the recombination rate of dense gas dominating over radiation. The density field and radiation field anti-correlate with position (in the IGM).
\end{itemize}
The other is the scale on which these behaviours are associated with:
\begin{itemize}
\item \textit{Local} - Implementation of the ionisation criteria is performed pixel by pixel in the simulation. i.e. with no dependency on the surrounding area. 
\item \textit{Global} - A scale dependency is included when implementing the ionisation criteria. Typically the surrounding region is influenced up to the mean free path of a photon.
\end{itemize}
These descriptions are approximate. 
In reality reionisation will progress with a variety of these behaviours at different times and scales, as is observed in detailed simulations \citep{technicolor}.
For a detailed description of the astrophysics involved in the EoR, please see e.g. \citet{FGU} \change{or \citet{2018PhR...780....1D}}.

Data is now becoming available from instruments such as LOFAR \citep{2017ApJ...838...65P}, MWA \citep{2015PhRvD..91l3011D}, PAPER \citep{2015ApJ...809...61A}, and HERA \citep{2017PASP..129d5001D}; all of which hope to detect the 21cm power spectrum \citep{2012MNRAS.423.2518C}. 
In the meantime mock data is created by simulations. 
The convention for finding the best fit parameters for these models is done using Markov Chain Mont\'e Carlo (MCMC) algorithms \citep{2017MNRAS.468..122H, 21CMMC}.
Our sampler of choice is called Multinest \citep{Multinest}, a nested sampling algorithm.
This subset of the MCMC family focuses on calculating the Bayesian Evidence, a quantity which measures how well a model fits a data set \change{- see e.g. \citet{sivia2006data} for a comprehensive introduction to MCMC statistical analyses}. 
Unfortunately full radiative transfer simulations are too computationally expensive for the number of samples required for an MCMC analysis\footnote{Typically \protect$\sim 10^4$ samples guarantees convergence in this context.} \citep{zahn}. 
Instead semi-numerical simulations, such as 21cmFAST \citep{DEXM, 21cmFAST} are used. 
These use approximate techniques, often combined with observations, to save on computation time - whilst producing consistent results to the full radiative transfer codes \citep{2012MNRAS.424.1335F, 2010MNRAS.406.2421S}. 
Interestingly the application of machine learning algorithms, such as Convolutional Neural Networks, are beginning to show promise in reducing the computation times of these simulations even further \citep{2018arXiv180502699G, 2017MNRAS.468.3869S, 2018arXiv181008211L, 2017ApJ...848...23K, 2019MNRAS4832524H}. 
With the first light on the SKA due around 2024, next generation radio interferometers are sure to reveal many secrets from the unseen universe. 
We are entering into an age of precision astrophysics and cosmology, particularly if synergies between telescopes such as SPICA\footnote{http://www.spica-mission.org/} and JWST\footnote{https://www.jwst.nasa.gov/} can be achieved. Sophisticated analysis techniques, such as Bayesian Inference, are becoming a necessity. 

This paper is structured as follows: Section \ref{sec: 21cmmc} introduces the base of this work, 21CMMC, including a summary of the recipes used for simulating the 21cm data; Section \ref{sec: 21cmnest} highlights 21cmNest - the additions made to 21CMMC including the toy models; Section \ref{sec: BayInf} is an introduction to the Bayesian Evidence and how it can be used for model selection. 
The results are split \change{into} seven subsections: Initially, we show agreement between the 21CMMC and Multinest in parameter estimation (Section \ref{sec: agreement}), including cross checks of the methodology; then we perform model selection with LOFAR, HERA and SKA; 
in the context of HERA we successfully distinguish between EoR scenarios (Section \ref{sec: HERAmain}) and show the redundancy of average neutral fraction checks as a way of ruling out toy models (Section \ref{sec: 21cmvsQSOs}); 
Finally, we attempt to distinguish two parameterisations of 21cmFAST (Section \ref{sec: 21cmfastmodels}) - including the use of nested models in establishing redundant parameters with the Savage-Dickey density ratio (Section \ref{sec: nested parameters}).

The main aims of this work look at the feasibility of ruling out toy models with the three selected observatories - These can be neatly summarised as the following three questions:
\begin{itemize}
\item How long would observations need to be to effectively distinguish between the toy models with LOFAR? (Section \ref{sec: LOFAR})
\item Given a fixed observing timescale (1080 hours) how many dipoles are necessary to effectively distinguish between the toy models with HERA? (Section \ref{sec: heradipoles})
\item What is the minimum observation time needed to effectively distinguish between the toy models with SKA? (Section \ref{sec: SKA}).
\end{itemize}

We assume a $\Lambda$CDM cosmology throughout this work with the following parameters: ($\Omega_M$, $\Omega_\Lambda$, $\Omega_b$, $n_S$, $\sigma_8$, $H_0$) $=$ (0.27, 0.73, 0.046, 0.96, 0.82, 70 km s$^{-1}$ Mpc$^{-1}$) \citep{2016A&A...594A..13P}.


\section{The State of the Art: 21CMMC} 
\label{sec: 21cmmc}

21CMMC is a Bayesian parameter estimation tool for the EoR \citep{21CMMC}. It combines Cosmohammer \citep{2013A&C.....2...27A} - a framework for using the affine invariant MCMC algorithm Emcee \citep{emcee}; with 21cmFAST \citep{21cmFAST} - a numerical implementation of the analytic model described by Furlanetto, Zaldarraiga, and Hernquist \citep{FHZ}, referred to as FZH.

A streamlined version of 21cmFAST is used, where evolved matter density ($\delta$) and velocity ($dv_{ \rm r}/dr$) fields are pre-calculated for the IGM. 
The 21cm signal intensity is directly proportional to the differential brightness temperature $\delta T_{\rm b}$  which can be approximated  \citep{furlanetto06b} as
\begin{equation}
\begin{aligned}
	\delta T_{\rm b} & \approx 27 x_{\rm HI} \left( 1+\delta \right) \left( \frac{H}{ \frac{dv_{ \rm r} }{dr}+H}\right) \left( 1-\frac{T_{\rm CMB}}{T_{\rm s}}\right)  \\ 
    & \times \left(\frac{1+z}{10} \frac{0.15}{\Omega_{\rm M}h^2} \right)^{\frac{1}{2}} \left( \frac{\Omega_{\rm b}h^2}{0.023} \right) \rm mK  .
\end{aligned}    \label{Tbright}
\end{equation}
where $x_{\rm HI}$ is the hydrogen neutral fraction; $T_{\rm CMB}$ is the CMB temperature; $T_{\rm s}$ is the spin temperature; and all other parameters follow the convention of a $\Lambda$CDM cosmology. 
In this original version of 21CMMC the simplifying assumption of $T_s \gg T_{\rm CMB}$ is used to ease computation (see \citet{21CMMC_EoH} for the inclusion of $T_{\rm s}$ and X-ray heating - which is also omitted here). 
This is valid during most of reionisation $z \sim [6,10]$ \citep{furlanettoTb,  baek, ChenMiralde-Ecude} and is known as the \textit{post-heating} regime.

This model uses the excursion set formalism in order to identify HII regions. A smoothing scale ($R$) is iterated down from the photon mean free path, $R_{\rm mfp}$ to the size of the cell.
This parameter is called the photon \textit{mean free path} for historical reasons. 
It is more accurately a photon's mean horizon since only the photons whose path are stopped by ionising HI are important in the EoR - a simple way to account for recombinations in the ionised IGM\footnote{The newest 21CMMC version \citep{Luminosity21cmmc} computes inhomogeneous recombinations explicitly.}.
What's important is the number of ionising photons verses the number of HI per smoothing scale.
For a more complete discussion please see \citet{21CMMC, 21CMMC_EoH}.
The defined criteria for ionisation is
\begin{equation}
	\hspace{2cm} \zeta f_{\rm coll}(\textbf{x}, z, R, \bar{M}_{\rm min}) \geq 1  \label{fcollzeta} .
\end{equation}
with \textbf{x} being position and $ \bar{M}_{\rm min}$ as the minimum virial halo mass defined by the atomic cooling threshold.
The collapse fraction is dependent on the virial temperature through the virial mass. 
We assume the relationship derived in \citet{barkanaloeb2001} where $M_{\rm vir} \sim T_{\rm vir}^{\frac{2}{3}} $. 
$f_{\rm coll}$ is the fraction of matter collapsed into dark matter halos \citep{P-S,excursionset,S-T} and is calculated by the following integral of the Press-Schechter halo mass function (a numerical correction factor to return on average the Sheth-Tormen result for ellipsoidal dark matter halo collapse is applied)
\begin{equation}\label{eq: integralfcoll}
f_{\rm coll} = \frac{1}{\rho_M} \int^{\infty}_{M_{\rm vir}} m~\frac{dn}{dm}~dm,
\end{equation}

where $\rho_M$ is the matter density of the chosen cosmology.
$\zeta$ is the ionising efficiency of galaxies at high redshift, performed via a step function 

\begin{align}
     \hspace{1.5cm}\zeta = & \begin{cases}
   	    		\zeta_{0},& \text{if }T_{\rm vir}\geq T_{\rm vir}^{\rm feed}\\
    			0,              & \text{otherwise} \\     
			\end{cases} \label{eq: zetaA}     
\end{align}

where $T_{\rm vir}$ is the virial temperature and 

\begin{equation}
    \zeta_{0} = 30\left(\frac{f_{\rm esc}}{0.3}\right) \left(\frac{f_{\star}}{0.05}\right) \left(\frac{N_{\gamma}}{4000}\right)\left(\frac{2}{1+n_{\rm rec}}\right) \label{ioneff}
\end{equation}
to encompass general astrophysical properties of galaxies: 
A`turn-on' mass is defined by the aforementioned atomic cooling threshold, $T_{\rm vir}^{\rm feed}= 10^4K$ \citep{barkanaloeb2001}; 
$N_\gamma$ is the number of ionising photon per baryon within a star, for $\rm popII$ stars only - $N_{\gamma} \approx 4000$ which is currently assumed to be the case during the EoR \citep{barkanaloeb2005}; 
$n_{\rm rec}$ is the typical number of recombinations per hydrogen atom, assumed to be $n_{\rm rec} \approx 1$ \citep{sobacchimesinger2014} for a \textit{photon-starved} end point for the EoR; 
$f_{\star}$ is the fraction of galactic gas in stars, which is taken to be $0.05$ \citep{LoebFerrara2013}; and $f_{\rm esc}$ is the fraction of ionising photons escaping into the IGM. Since both $f_{\star}$ and $f_{\rm esc}$ are observationally uncertain \citep{Gnedin, Wise_Cen, LoebFerrara2013}. We will return to these in Section \ref{sec: Pconsiderations}, when we discuss the parameter priors.

21CMMC features the option to relax the single galaxy population assumed so far i.e. to depart from a constant $\zeta$. This is done by the addition of a 4th parameter, $\alpha$, enabling a power law relationship between $\zeta$ and $T_{\rm vir}$ as,
	\begin{equation}\label{eq: zetaB}
		\zeta =  \begin{cases}
   	    		\zeta_0  \left( \frac{T_{\rm vir}}{ T_{\rm vir}^{\rm feed} }\right)^ \alpha,	& \text{if }T_{\rm vir}\geq T_{\rm vir}^{\rm feed}\\
    			0,              & \text{otherwise}     .
				\end{cases}   
	\end{equation}

During the MCMC analysis the likelihood is defined to be a $\chi^2$ statistic between a mock 21cm observed power spectra and a theoretically calculated one, which is combined with observational priors (\ref{sec: obscheck}). 
The mock power spectrum uses a fiducial choice of parameters while the theoretical models' best fit parameters are found via the MCMC analysis. 
\change{This is done for $8$ k bins from a foreground corruption limit of $k = 0.15 ~ \rm Mpc^{-1}$ to a shot noise limit of $k = 1.0~ \rm Mpc^{-1}$ - this is maintained in this work unless specified otherwise.} 

The spherically averaged power spectrum, $\Delta _{21}^2 $, is chosen to be the aforementioned likelihood statistic and is defined via the brightness temperature as 
	\begin{align}
		\Delta _{21}(k)^2 & \equiv  \frac{k^3}{2\pi^2V} \overline{\delta T_b}^2(z)\langle | \delta_{21}(\mathbf{k},z)|^2 \rangle  \rm mK^2
		\label{spherical_PS} \\ \vspace{0.2cm}
		& \delta_{21}(\mathbf{x}, z)  \equiv \frac{\delta T_b(\mathbf{x},z)}{\overline{\delta T_b}(z)} -1   \nonumber .
	\end{align}

The bar denotes a spatial average, while the angular brackets denote a k-space average. 
This likelihood computation chain is summarised in Figure \ref{lhoodchain}.

\subsection{Observational data priors}\label{sec: obscheck}

Depending on the choice of redshift range 21CMMC implements up to three observational checks:
\begin{itemize}
\item The Thomson optical depth of the IGM was observed by Planck to be: $\tau =  0.058 \pm 0.012$, assuming an instantaneous EoR \citep{Planck}. During each MCMC call, 21cmFAST can produce estimates of $\tau$ by interpolating the neutral hydrogen fraction across the desired redshifts. This allows testing between the observed and simulated values of $\tau$. 
\vspace{0.1cm}
\item The location of the Gunn-Peterson trough in QSO Ly-$\alpha$ forest data: \citet{McGreer} show that the EoR must be $> 90\%$ complete by $z=5.9$. Therefore for all \change{$z \leq 5.9$, $x_{\rm HI} = 0$, and for $z>5.9$} the estimated neutral HI fraction is tested against an half-Gaussian with mean $\bar{x}_{\rm HI} = 0.06$ and variance $\sigma^2 = 0.05$.
\vspace{0.1cm}
\item The Red Ly-$\alpha$ damping wing allows the estimation of the $x_{\rm HI}$ surrounding QSO ULASJ1120+0641 \citep{Grieg}. Provided $z = 7.08$ is included in the redshift interpolation range, the neutral fraction at this redshift is checked against \change{the observational measurement $\bar{x}_{\rm HI} = 0.4 ^{+ 0.41}_{- 0.32}$ at $2\sigma$}.
\end{itemize}
All tests are performed with a $\chi^2$ between $x_{\rm HI}$ (and $\tau$) which are then combined into the likelihood linearly as: $ln \mathcal{L} = \chi^2_{21 \rm cm} + \chi^2_{\rm Planck} + \chi^2_{\rm McGreer} + \chi^2_{\rm Greig}$. 

\subsection{Telescope sensitivity with 21cmSense}\label{sec:noise}

The Python code 21cmSense\footnote{\protect\textcolor{blue}{https://github.com/jpober/21cmSense}, 21cmSense uses heavily the Astrophysical Interferometry Python code AIPY \protect\textcolor{blue}{https://github.com/AaronParsons/aipy}.} \citep{21cmsense} is used throughout this work to calculate the telescope noise profiles. 
In each U-V bin the thermal noise is modelled as: 
\begin{equation}
\Delta^2_{\rm N}(\mathbf{k}) \approx X^2Y \frac{\mathbf{k}^3}{2 \pi ^2} \frac{\Omega_{\rm Eff}}{2t_{\rm int}} T_{\rm sys}^2 .
\end{equation}
Where $X^2Y$ converts the solid angle of observing bandwidth to a comoving distance, $\Omega_{\rm Eff}$ represents the effective beam solid angle as derived in \citet{2014ApJ...788..106P}. $t_{\rm int}$ is the total integration time (per observed frequency).   
$T_{\rm sys}$ is the sum of the sky temperature \citep{2017isra.book.....T} \change{and the} receiver temperature $T_{\rm R}$.
Cosmic variance is included assuming it is a Gaussian error - we can express the total uncertainty by spherically averaging an inversely weighted summation across all the k modes.
Hence in Sections \ref{sec: LOFAR} and \ref{sec: SKA} (when $t_{\rm int}$ is altered) we expect the telescope noise to scale roughly as $t_{\rm int}^{-0.5}$.
The specifications of LOFAR are taken from \citep{2013A&A...556A...2V}. We use the 48 core high-band antenna stations only.
HERA's dipoles are structured \change{in a filled hexagon \citep{redundentantenna, 2015ApJ...800..128B}}.
For the SKA we follow the specifications of LOW Phase 1 with 512 stations\footnote{https://www.skatelescope.org/key-documents/}. 
\change{The middle 224 stations are randomly distributed in a circular core ($\sim$ 400m in radius) with the remaining 288 stations split in three spiral arms extending outwards.}
For all telescopes, unless stated otherwise, we assume 6 hours of observing per day for 180 days for $t_{\rm int}$. The cosmological bandwidth is assumed to be $8 ~\rm MHz$ throughout.
The telescope details are summarised in Table \ref{Tscopetable}.
Fundamental concerns with detecting the 21cm signal are due to foregrounds eg. \citet{2015MNRAS.447.1705P}. We adopt the `\textit{Moderate}' foreground setting within 21cmSense\footnote{The other options are: \textit{Pessimistic} - \change{Where} baselines are added incoherently and no k modes are included in the horizon or the buffer zone; or \textit{Optimistic} - All k modes in the primary field of view are used.}. This means that all baselines are added coherently but no k modes from the horizon or a buffer zone (chosen to be the default $0.1  h \rm Mpc^{-1}$) are used. 
See e.g. \citet{2013ApJ...768L..36P, 2014ApJ...782...66P, 2014ApJ...788...96P} for more detail.

\begin{table}
 \begin{tabular}{ | l | c | c | c |}
    \hline 
    \small \textbf{Parameter} & \small LOFAR  &  \small HERA & \small SKA-Central \\ \hline \hline
    Number of dipole & 48 & 331 & 296 \\ Stations \\ \hline
    Station Diameter [m] & 31 & 14 & 35 \\ \hline
    Collecting Area [$\rm m^2$] & 35,762 & 50,953 &  492,602 \\ \hline
    $T_{\rm R}$[K] & 140 & 100 &  $40+\frac{T_{\rm sky}}{10}$ \\ \hline
    Observing Range & [110,240] & [50,250] & [50,350] \\ ~[MHz] \\ \hline
    Observation Time  & 1080 & 1080 & 1080  \\ ~[hrs] \\ \hline 
    Scan Type & 1-hr Track & Drift & 1-hr Track \\ \hline 
  \end{tabular}
 \vspace{0.2cm}
    \caption{\label{Tscopetable} A summary of the different telescope specifications used in this paper. Note that of the SKA's 512 stations we only simulate using the `central' 296 stations, where sensitivity to the EoR frequencies is dominant. } 
\end{table}


\section{21CMNest} 
\label{sec: 21cmnest}

The primary objective of this work\footnote{\textcolor{blue}{https://binnietom@bitbucket.org/binnietom/21cmnest\_1.0.git}} is to apply Bayesian model comparison to EoR scenarios. 
This has been achieved with two alterations to the 21CMMC framework. 
Firstly, we add various toy models (detailed in Section \ref{sec: further}); and secondly we replace the Emcee sampler \citep{emcee} (from Cosmohammer) with pyMultinest (Section \ref{sec: BayInf}). 


\begin{figure}
\centering
\includegraphics[scale=0.5]{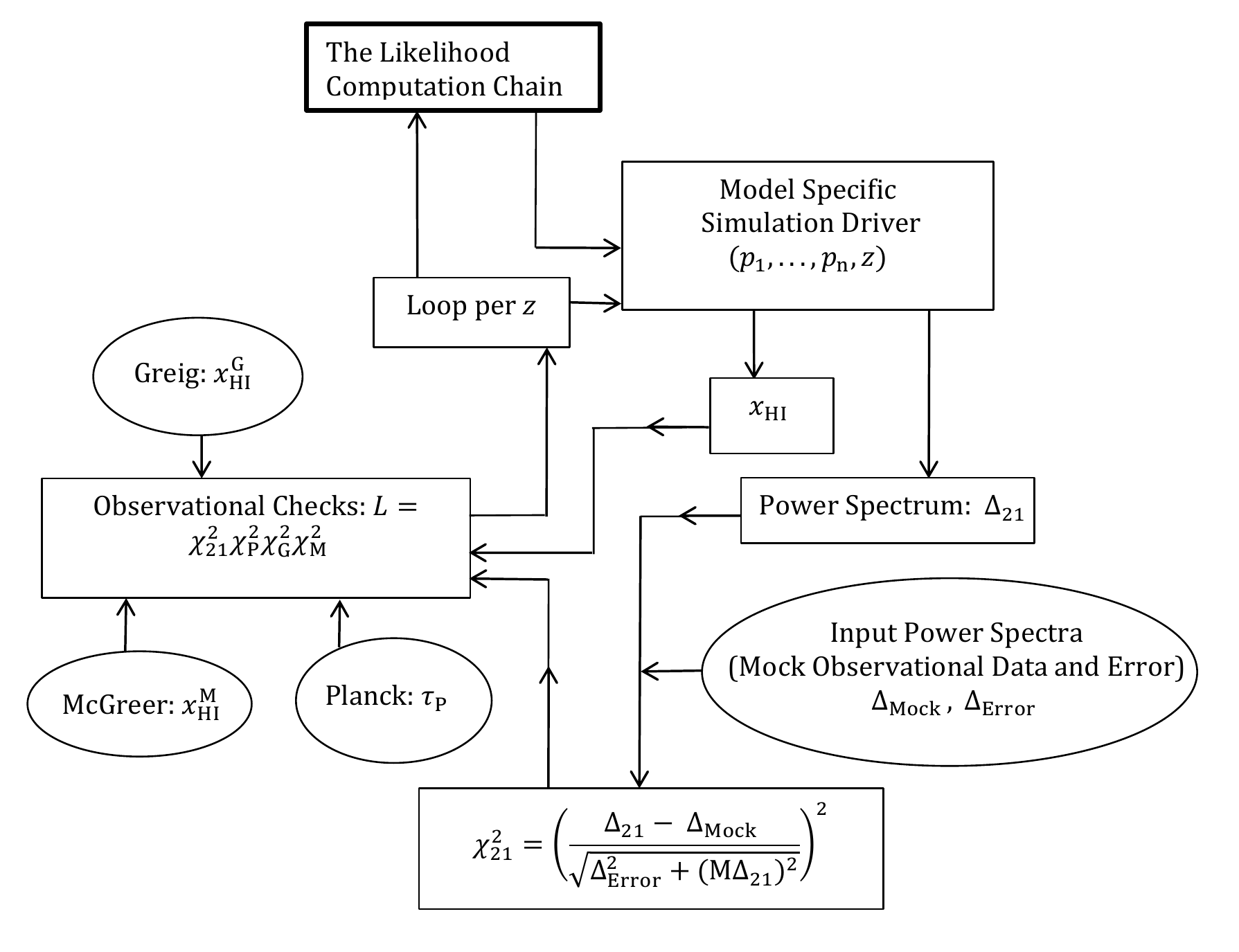}
\caption{\label{lhoodchain} The likelihood computation chain as performed by 21CMMC and 21cmNest. $\rm M$ is a user defined model uncertainty. All other quantities are defined throughout the text.}
\end{figure}

\subsection{Flexibility and toy EoR models} \label{sec: further}

\begin{figure*}
    \centering      
    \subfigure[\label{Tb_FZH}]{\includegraphics[scale=0.35]{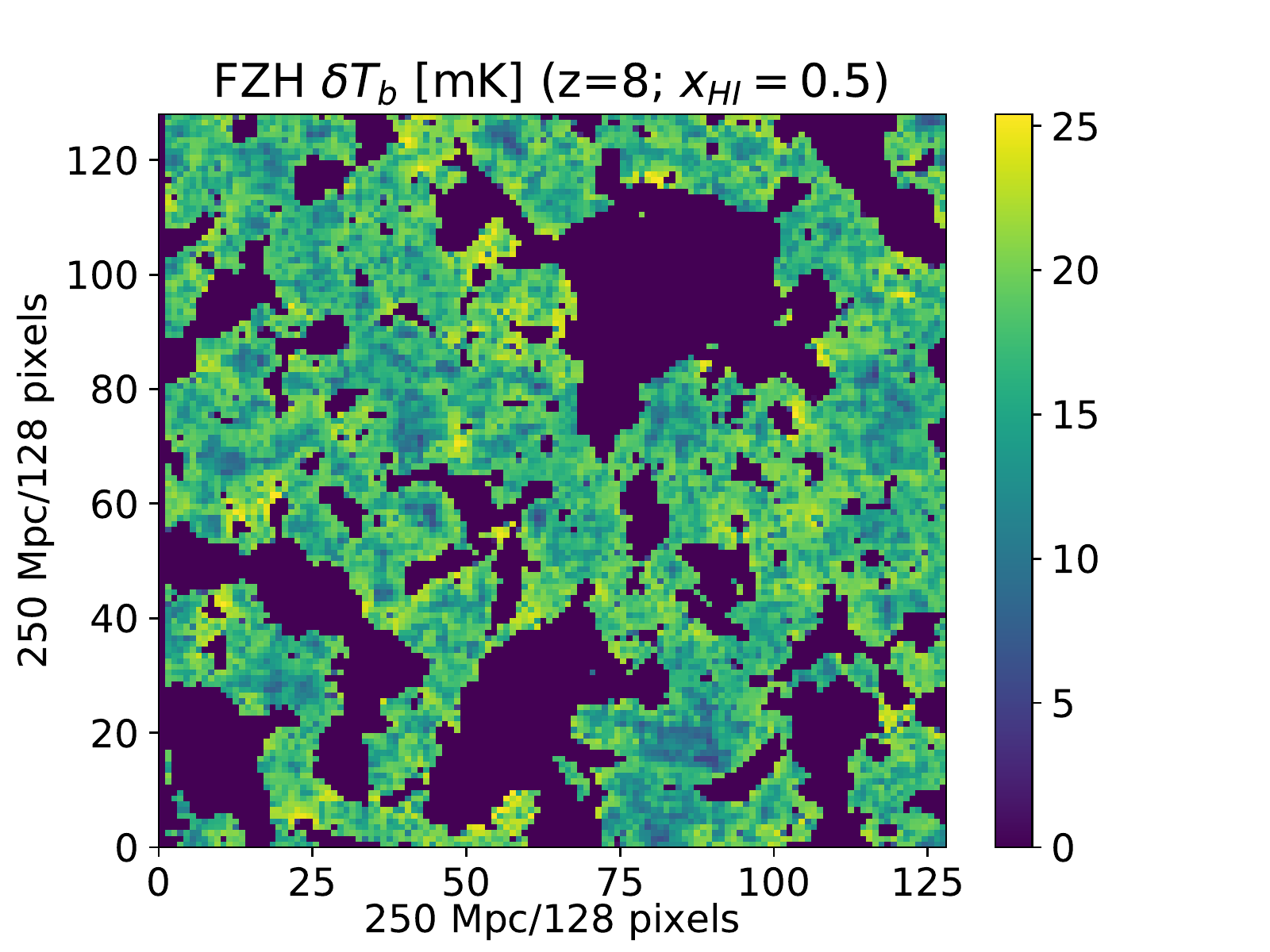}\hfill}
    \subfigure[\label{Tb_FZHinv}]{\includegraphics[scale=0.35]{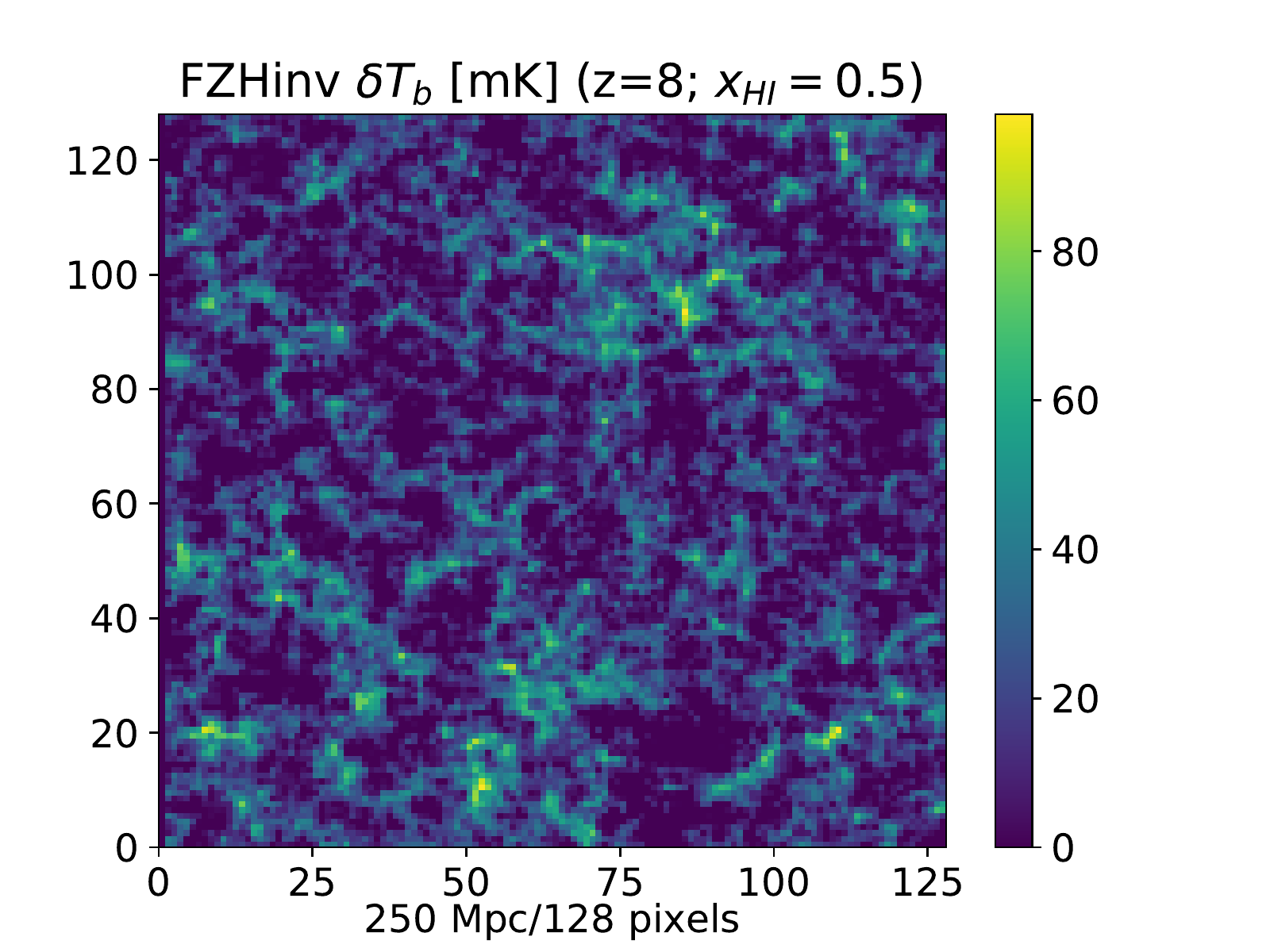}\hfill}
    \subfigure[\label{Tb_MHR}]{\includegraphics[scale=0.35]{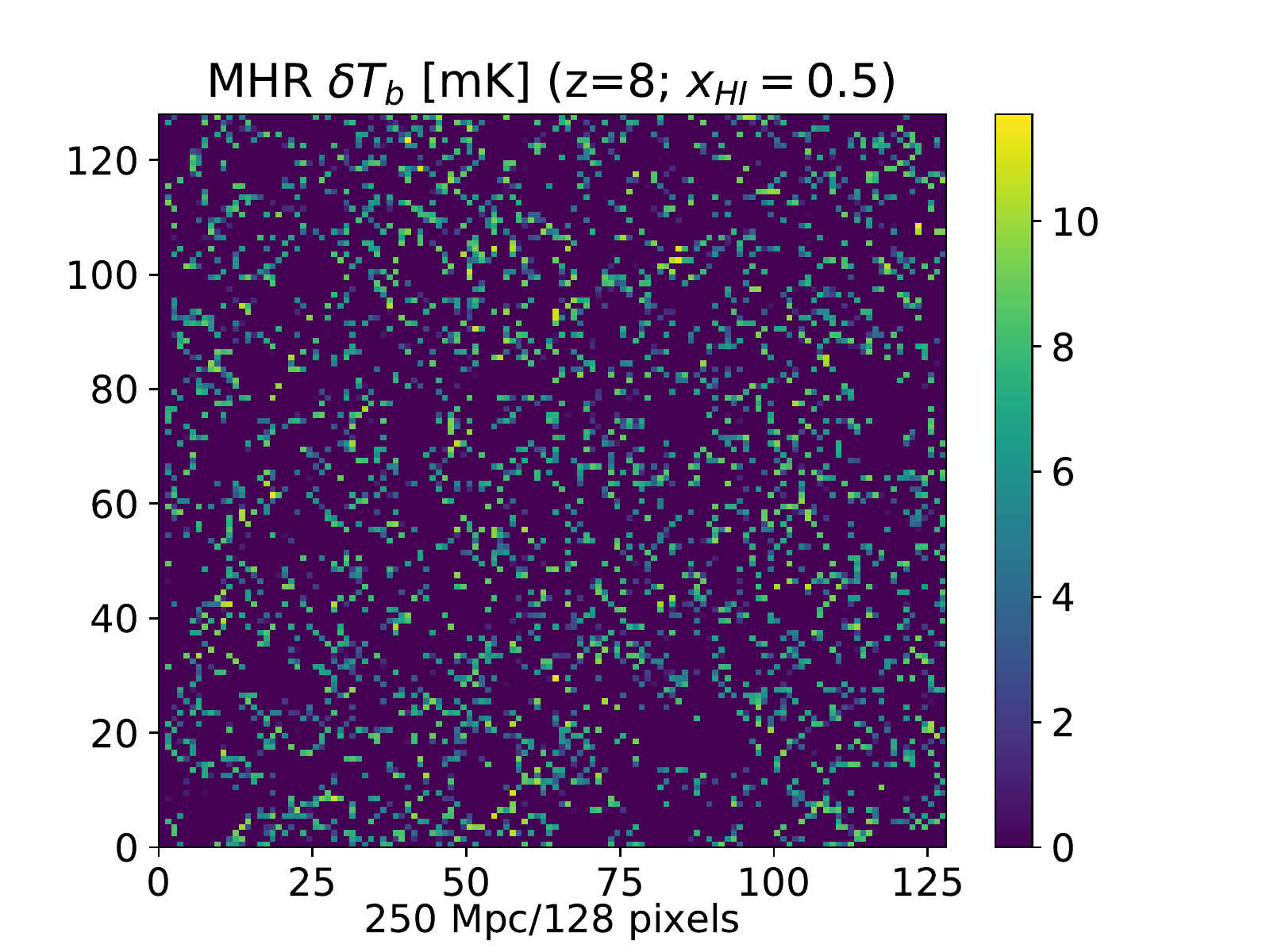}\hfill}
    \subfigure[\label{Tb_F_MHRinv}]{\includegraphics[scale=0.35]{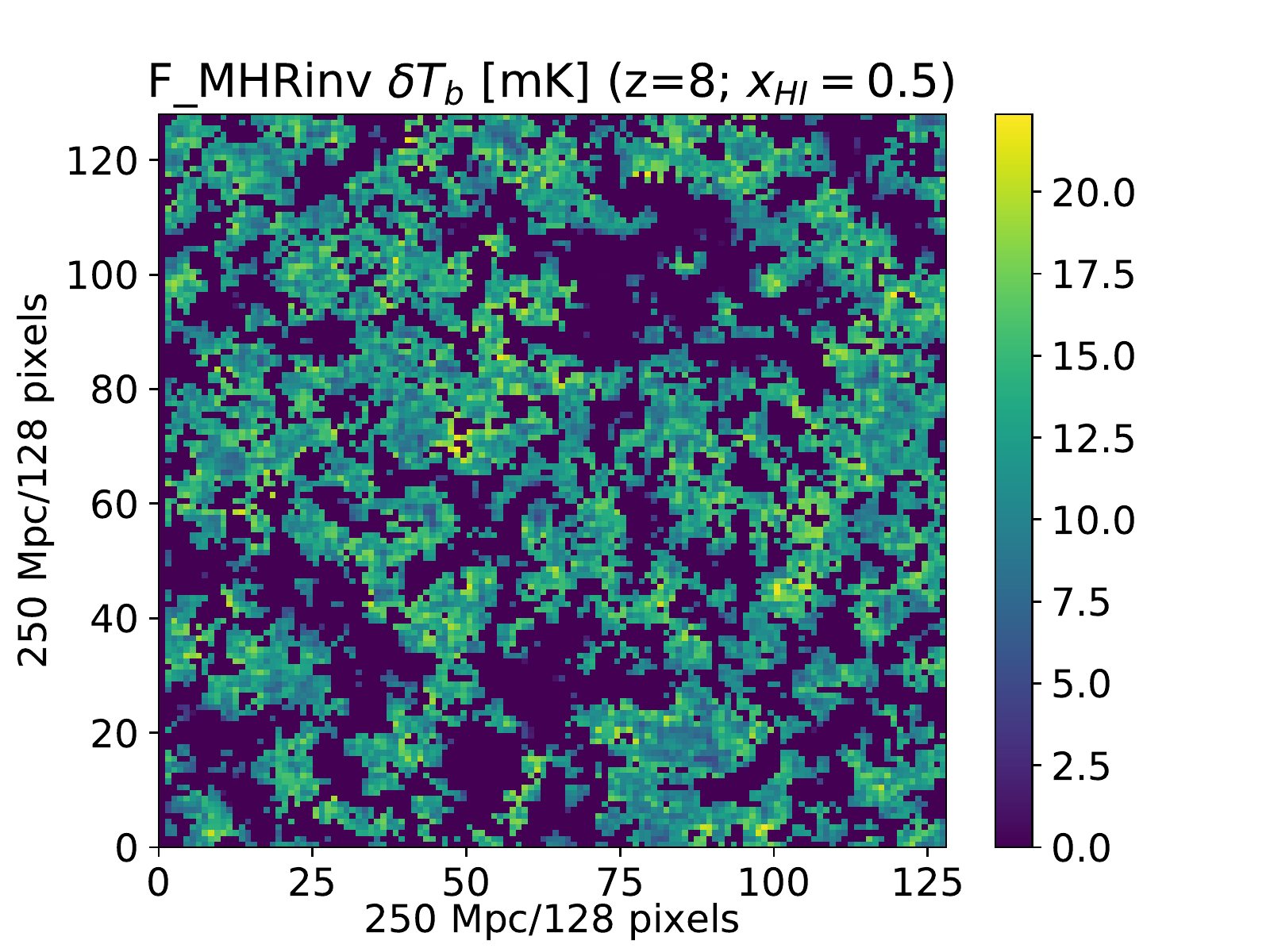}\hfill}
    \subfigure[\label{Tb_F_MHR}]{\includegraphics[scale=0.35]{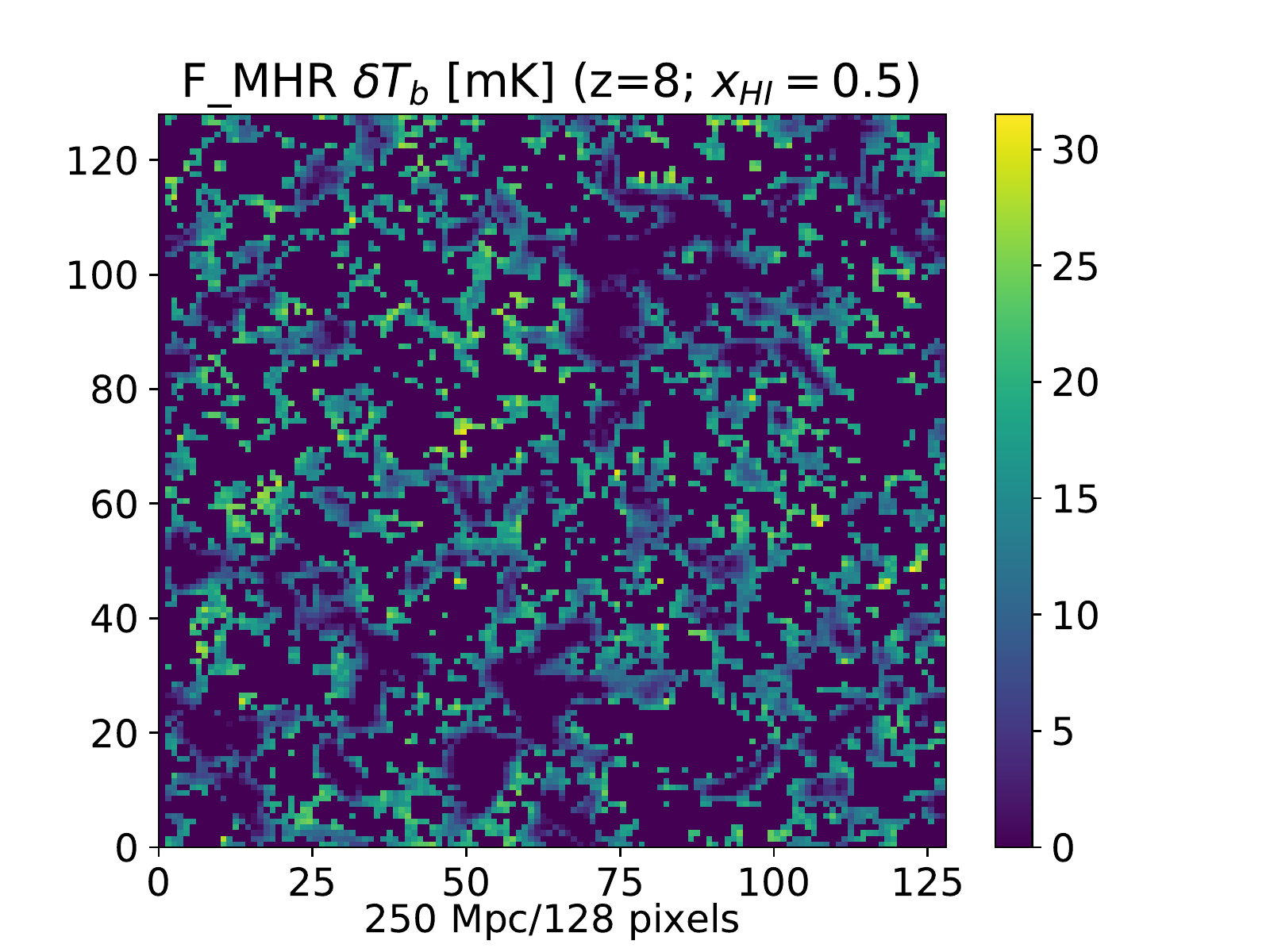}\hfill}
    \subfigure[\label{Tb_MHRinv}]{\includegraphics[scale=0.35]{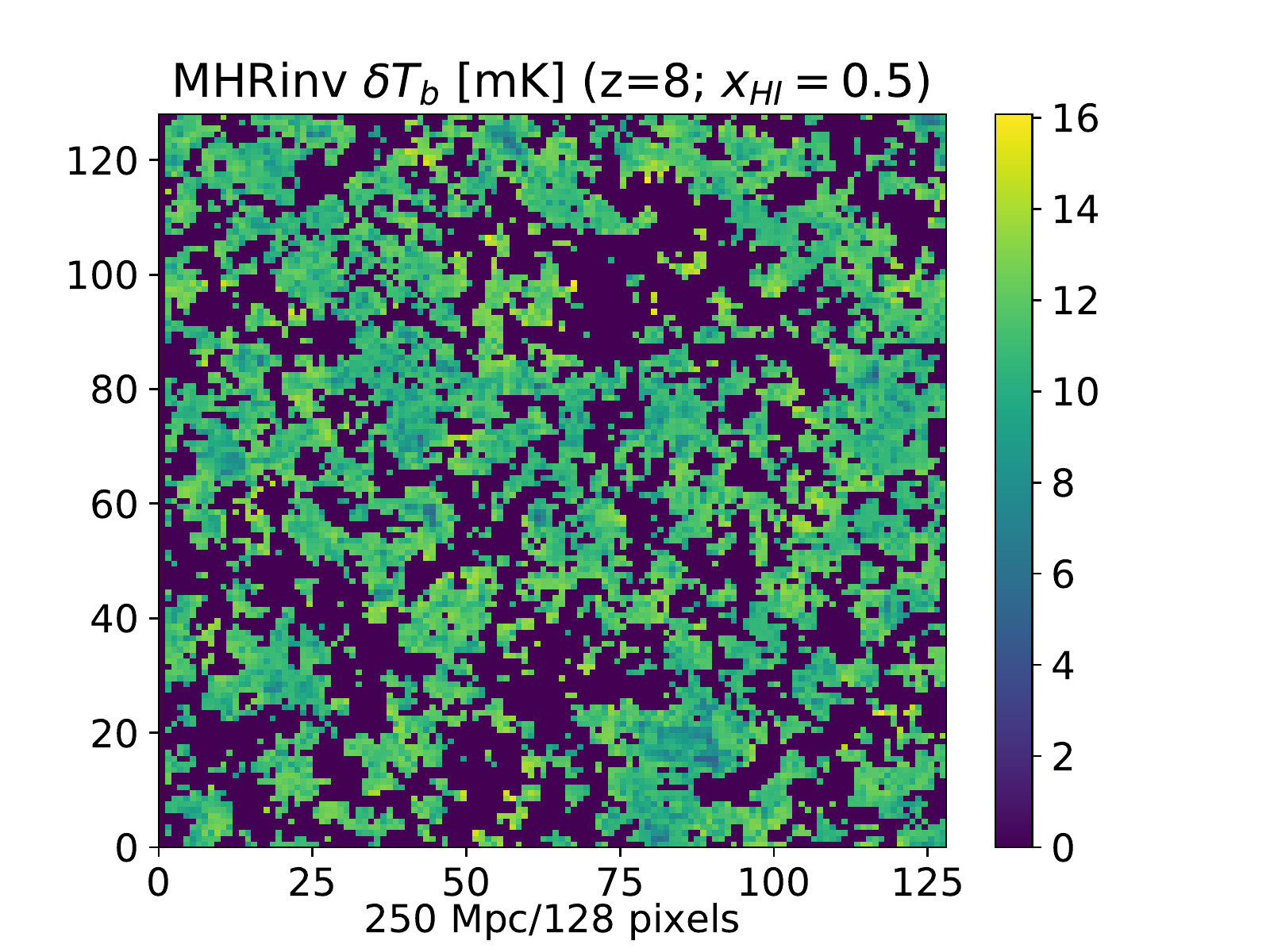}\hfill}
    \caption{Plots \change{\ref{Tb_FZH}, \ref{Tb_FZHinv}, \ref{Tb_MHR}, \ref{Tb_F_MHRinv}, \ref{Tb_F_MHR}, and \ref{Tb_MHRinv}} are brightness temperature slices for the toy models referred to as FZH, Inv FZH, MHR, F Inv MHR, F MHR, and Inv MHR. 
    A summary of these models is given in Table \ref{models},  Section \ref{sec: further} contains the full detail.}
    \label{slices}
\end{figure*}

\begin{figure}
    \centering
    \subfigure[\label{Eor_hist}]{\includegraphics[scale=0.5]{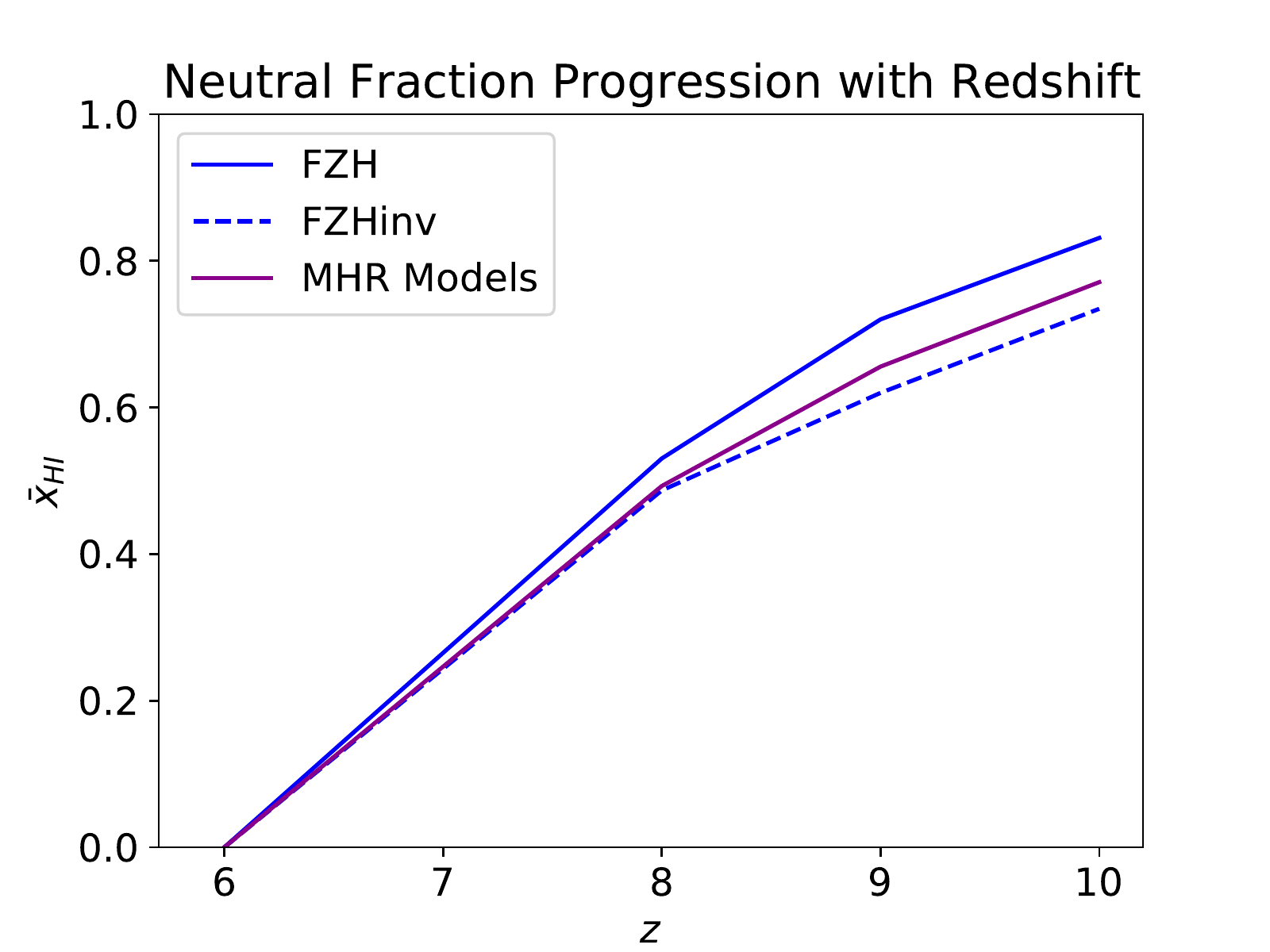}\hfill}
    \subfigure[\label{PS_HALF}]{\includegraphics[scale=0.5]{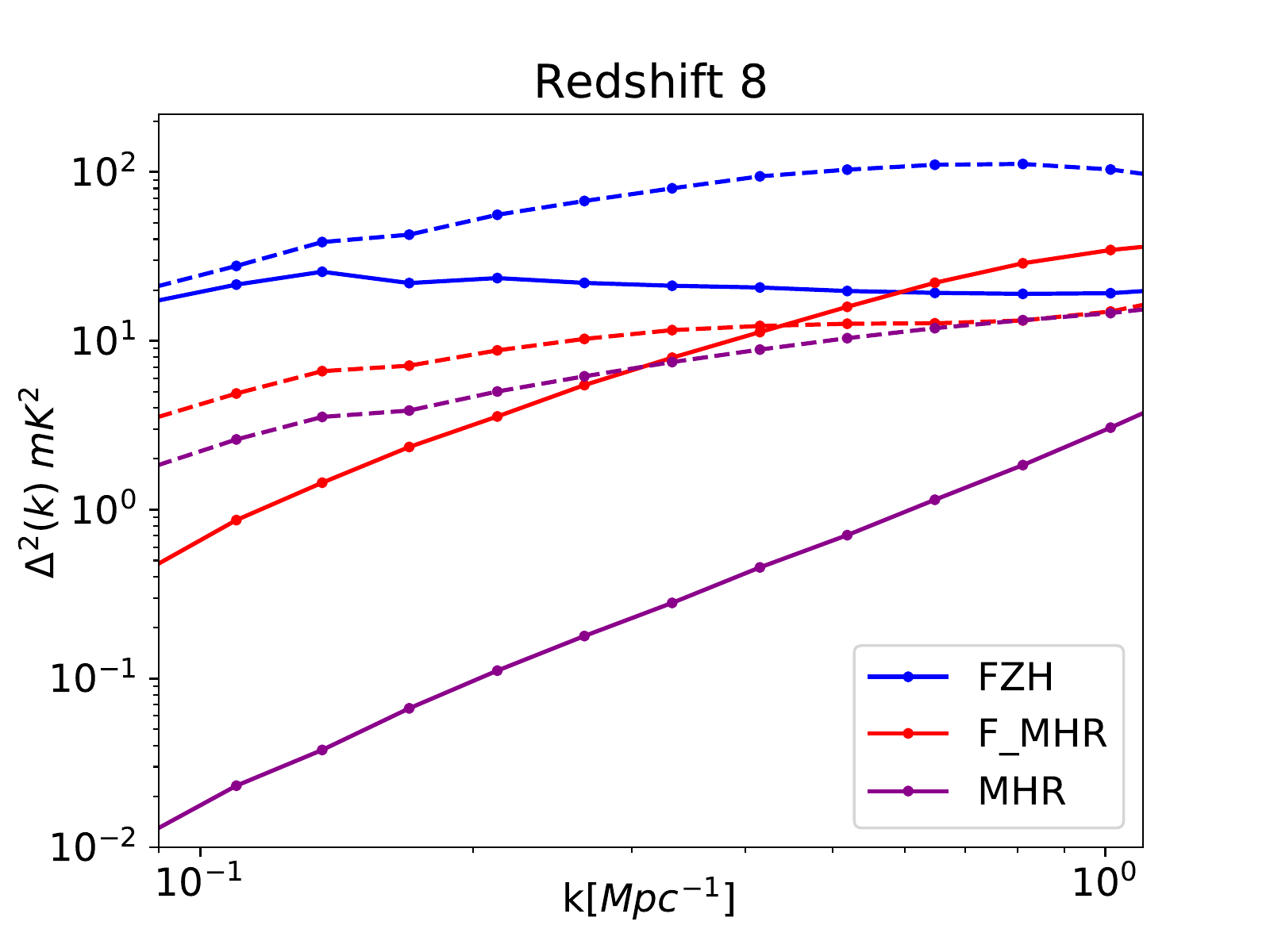}\hfill}
    \caption{\change{\ref{Eor_hist} and \ref{PS_HALF} are the 21cm power spectra and EoR histories for the toy models respectively. Blue, purple and red represent the 21cmFAST model (FZH), MHR and F MHR models respectively. 
    Dotted lines indicate the inversion of that model's ionisation criteria.
    These are performed at the values used to produce $\bar{x}_{\rm HI} \approx 0.5$ at at $z=8$.
    Note that, given the same $\zeta$ and $T_{\rm vir}$, all MHR based models have the same EoR history by construction (hence only one solid purple line). 
    The models are summarised in Table \ref{models}, see Section \ref{sec: further} for more detail.}}
    \label{referee}
\end{figure}

Since there is still uncertainty in the detailed physics of the EoR we explore several toy models. 
These toy models are explained below. 
Initially we use the same four models as \citet{Catherine}, summarised in Table \ref{models}. 
See Figure \ref{slices} for a slice of brightness temperature from each models' box at $z=8$ for $\bar{x}_{\rm HI} = 0.5$. 
Figure \ref{referee} shows the corresponding 21cm power spectra and EoR ionisation histories for these slices.
Although Table \ref{models} contains physical motivations for all the models, we stress here that only the FZH and MHR models are physically liable - the others are included to test the proof-of-concept for Bayesian model selection. 
Reionisation may begin with the \textit{local outside-in} scenario but \textit{global inside-out} reionisation will eventually dominate \change{assuming UV radiation drives reionisation \citep{2019MNRAS.483.5301G}}.  
Given we understand the fluctuations of the IGM density on linear (Mpc) scales, this is easy to show by comparing the recombination and ionisation rates with e.g. 21cmFAST.
Table \ref{j_ob_params} shows the maximum a posteriori (MAP) parameters for the toy models fit to the Greig, McGreer and Planck priors detailed in Section \ref{sec: obscheck}. 
Before 21cm data is included the allowed EoR parameter space is ample \citep{2017MNRAS.465.4838G}.

\begin{table*} \centering
\begin{tabular}{ | p{1.5cm} | p{6cm} | p{5cm} | p{4cm} |}
    \hline  
     \textbf{Model} & \textit{Physical motivation} &  \textit{How is ionisation defined?} &  \textit{Resulting type of reionisation} \\
      \hline \hline
   	 FZH & Over-dense regions collapse to form stars, UV radiation is therefore the driving force behind the ionised bubbles. The size of these bubbles dictates the neutral fraction and therefore the brightness temperature. & Over-density increases above the critical barrier ($\delta_{\rm crit}$). & \textit{Global Inside-Out} \\ \hline
   	 Inverted (Inv) FZH &  Over-dense regions remain neutral due to a high recombination rate, hence reionisation begins in underdense regions. These then grow and dominate as hard radiation becomes the dominant radiation background and the IGM becomes ionised.  &  Over-density decreases  below the critical barrier.  &  \textit{Global Outside-In} \\ \hline
     MHR & Stars are formed as above, ionising the circum-galactic gas. Ionisation then proceeds a long underdense regions because the atoms recombine faster than they are ionised in the dense regions. Eventually the background radiation is dense enough to dominate. & A pixel-by-pixel implementation of an under density threshold. &  \textit{Local Outside-In}  \\  \hline
     Inverted (Inv) MHR & Stars are formed as above, ionising the circum-galactic gas. Over-dense regions become ionised. It is easy to implement rather than being physically viable since even dense regions that do not have a line of sight to a radiation source will become ionised. &  A pixel-by-pixel implementation of an over density threshold. & \textit{Local Inside-Out} \\   \hline
\end{tabular}
\vspace{0.2cm}
\centering \caption{\label{models} This table summarises the 4 main EoR models considered in this work, see \citet{Catherine} for more details.
It is worth noting here that \change{filtering} (F) the MHR models changes the scale of reionisation from \textit{local} to \textit{global}.
}\end{table*}

\begin{table}
\centering \begin{tabular}{| l | c | c | r |}
    \hline 
    \small  & \small $\zeta$ & $R_{\rm mfp}$ &  $\rm log_{10}[T_{\rm vir}]$ \\ \hline \hline
    FZH & 95.4 $\pm$ 61.9 & 12.3 $\pm$ 4.3 & 5.17 $\pm$ 0.29 \\ \hline
    Inv FZH & 71.8 $\pm$ 68.5 & 14.9 $\pm$ 5.4 & 4.72 $\pm$ 0.34 \\ \hline
    F MHR & 413 $\pm$ 289 & 9.3 $\pm$ 4.3 & 5.80 $\pm$ 0.31 \\ \hline
    F Inv MHR & 268 $\pm$ 290 & 18.3 $\pm$ 4.4 & 5.61 $\pm$ 0.34 \\ \hline
    MHR & 404 $\pm$ 293 & - & 5.72 $\pm$ 0.33 \\ \hline
    Inv MHR & 799 $\pm$ 291 & - & 5.88 $\pm$ 0.33 \\ \hline
\end{tabular}  
\vspace{0.2cm}
    \caption{\change{The MAP parameter values to $1\sigma$ (standard deviation) for the toy models fit to all three observational prior checks (without 21cm data).
    Note that the large errors (compared to each respective parameter) hint that this data will not be constraining in the context of distinguishing toy EoR scenario (discussed further in Section \ref{sec: 21cmvsQSOs}}).
    \label{j_ob_params}  
    }
\end{table}

\subsubsection{FZH}\label{FZH}
The 21cmFAST simulation is a numerical implementation of the model developed by Furlanetto, Hernquist and Zaldarriagan (FZH), as summarised in Section \ref{sec: 21cmmc}. This assumes\footnote{The simplest scenario \citep{barkanaloeb2001}.} that the mass of the collapsed object is relatable to the mass of the ionised region by the ionising efficiency ($\zeta$) as follows, 

\begin{equation}
\label{eq: m_ion}
m_{\rm ion} = \zeta m_{\rm gal} . 
\end{equation}

Using the Press-Schechter collapse fraction, $f_{\rm coll}$ (Equation \ref{eq: integralfcoll}) can be written as

\begin{equation}\label{eq: fcoll}
    f_{\rm coll}   =  \text{erfc} \left\{ \frac{ \delta_{\rm crit}  -  \delta }  {  \sqrt[]{   2[  \sigma^2 (m_{\rm min})-\sigma^2  (m) ]  }  }    \right\} .
 \end{equation}
This implies that the barrier for ionisation can be rewritten as a density constraint instead:

\begin{equation}
\label{eq: barriera}
\delta \geq \delta_{\rm crit} - \sqrt[]{2[  \sigma_{\rm min}^2 (m_{\rm min})-\sigma^2  (m)]} ~\text{erfc}^{-1}\left(1-\frac{1}{\zeta}\right) . 
\end{equation}
Note that the factor of 2 in Equations \ref{eq: fcoll} and \ref{eq: barriera} comes from a symmetry argument. For every excursion that reaches the critical barrier, only half of these will collapse to form virialised objects due to the nature of \textit{random} walks. 

Numerically, FZH is applied by smoothing around each pixel from large to small scale. If Equation \ref{eq: barriera} is achieved before the filter scale reaches the pixel size is met the central pixel is flagged as ionised. Since this is a \textit{global} prescription, not all regions are purely ionised or neutral. After the filtering scale has reached the pixel size 21cmFAST sets the neutral fraction of $x_{\rm HI} = 1- \zeta f_{\rm coll-per-pixel}$ to each pixel that is yet to be fully ionised\footnote{This includes rounding to $0~if~x_{\rm HI}<0$ and $1~if~x_{\rm HI}>1$.}.

\subsubsection{Inverted (Inv) FZH}\label{sec: invFZH}

Inverting the ionisation threshold (Equation \ref{eq: barriera}) allows us to produce an \textit{outside-in} model sensitive to \textit{global} scales. 
Within this model underdense regions will be ionised first to form bubbles, the edge of which will remain in equilibrium\footnote{Similar to a Str\"omgren sphere but on galactic scales.} until the background ionising radiation increases. Eventually this will dominate over the recombination rate in overdense regions anticipating the completion of reionisation.
Firstly we redefine $\zeta$ as $\zeta'$ in order to keep the parameter framework within 21CMMC unchanged. $\zeta'$ is now the \textit{background} ionising radiation efficiency. Next we must turn to the critical barrier definition, which we choose to be $\delta _{\rm crit} ' = -\delta_{\rm crit} $. 
Assuming Gaussian perturbations for our regions of over-density, as is done for the FZH models (see the form of Equation \ref{eq: integralfcoll} becomes \ref{eq: fcoll}), all we need to do is calculate the fraction of matter that satisfies $\delta < \delta _{\rm crit} '$. In other words a new collapse fraction $f_{\rm coll}'$, below which HI structures are sparse enough to be ionised. This can be done by changing the limits of the original $f_{\rm coll}$ integral as is shown below 
\begin{eqnarray}
\label{limit_fcoll} 
f_{\rm coll}' =& \frac{1}{\rho_{\rm M}} \int^{M_{\rm vir}}_{0} m~\frac{dn}{dm}~dm  \\ \label{newbarrier}
 =& \text{erfc} \left[ \frac{ \delta_{\rm crit}  +  \delta }  {  \sqrt[]{   2(  \sigma^2 (m_{\rm min})-\sigma^2  (m) )  }  }    \right] .
\end{eqnarray}
Now we have an analogous ionisation threshold, $f_{\rm coll}' \zeta ' \geq 1.$, that can be implemented easily within the 21CMMC framework. As before, any partially ionised pixels left after using the excursion set formalism are assigned $x_{\rm HI} = 1 - \zeta'f_{\rm coll}'$. 

\subsubsection{MHR}

For an alternate method of implementing ionisation we turn to the work of Miralda-Escud\'e, Haenelt and Rees \change{referred to as MHR \citep{MHR}}. 
This allows us to a produce \textit{local outside-in} model of reionisation - underdense regions of the IGM are ionised first due to a recombination rate that is dependent on the density of the gas. The idea is that the denser a region of gas, the faster it can recombine and therefore the dense regions of the IGM are ionised last. Circumgalactic gas, within the halo, is ionised first allowing ionising photons to seep into the IGM. HII regions then expand in the directions of low gas density and the EoR is defined to end when these regions overlap.
Typically this is achieved with a background of X-ray photons, due to their large mean free paths. 
Firstly, to define the aforementioned ionisation threshold we prescribe a neutral fraction $\bar{x}_{\rm HI} = 1 - f_{\rm coll}\zeta$. Secondly, we order the pixels by HI density. We then ensure consistency between the $x_{\rm HI}$ by finding the $i$'th pixel (of total $N_{\rm p}$) that satisfies 
\begin{equation}
\label{eq: MHRcondition}
\frac{i}{N_{\rm p}} = 1 - \bar{x}_{\rm HI} .
\end{equation}
Once the \textit{i}'th pixel has been identified, we can set the ionisation definition as a function of density to be $\delta < \delta_i$. In summary the only deviation from 21cmFAST is the definition of ionisation via a new density threshold. 
Prescribing the $\bar{x}_{\rm HI}$ directly is referred to as the 1-parameter (1p) MHR model. 

\subsubsection{Inverted (Inv) MHR}\label{sec: invMHR}

Here we produce a \textit{local inside-out} model of reionisation. The procedure follows as above, but with the greater than sign reversed to produce an over-density threshold: $\delta > \delta_j$. The choice of the \textit{j}'th pixel using: $j = N_{\rm p} - i$, ensures (again) that the neutral fraction remains consistent between simulations. 

\subsubsection{Filtered (F) MHR}

In order to test \textit{21cmNest} more thoroughly any skewing effect from extra parameters (see Section \ref{sec: jeff}) in the Bayesian Evidence, a third parameter is added to the basic MHR model. 
This takes the form of a Gaussian k-space filter performed across the density field. 
The integrated volume of this Gaussian equates to that of a real space top-hat with radius $R_{\rm Filter}$. 
This smoothing has a tendency to reduce the range of values within the density field, which can have a large effect on the number and density of ionised structures. 
We expect this added flexibility to relax the \textit{local} scenario, allowing the filtered model to fit the scale of the fiducial model. 
This means producing a \textit{global} model when fitting for an FZH fiducial power spectrum, but being a redundant parameter when fitting for an MHR fiducial power spectrum (Section \ref{sec: f3}). 
\change{The} F Inv MHR proves to be the most challenging toy model to test, since it shares morphology and scale with FZH. In the F MHR model the background ionising efficiency, $\zeta$, is pushed below that of the MHR model. 
With the opposite being true for the F Inv MHR. 
As for the filter scale itself, this tends to the mean sizes of HII bubbles.


\section{Bayesian Inference}\label{sec: BayInf}

Bayes' theorem can be written as 

\begin{equation}\label{PEbayes}
P(\mathbf{\Theta} | D, M)=\frac{P(D|\mathbf{\Theta},M)P(\mathbf{\Theta}|M)}{P(D|M)} ,
\end{equation}
\change{where} $\mathbf{\Theta}$ are the parameters of a given model $M$, and $D$ is the data set to be fit against given an hypothesis. In this context the hypothesis is implied via the choice of model under consideration. 
$P(\mathbf{\Theta} \mid D, M)$ is the posterior parameter distribution, $\mathcal{P}$; $P(D \mid \mathbf{\Theta},M)$ is the likelihood $\mathcal{L}$; the prior knowledge $\Pi$ is $P(\mathbf{\Theta} \mid M)$; and the Bayesian Evidence\footnote{Alternatively known as the model likelihood or marginal likelihood.}, $\mathcal{Z}$ is defined to be $P(D \mid M)$. 
Bayesian inference for parameter estimation is centred around the likelihood, as is the case for Cosmohammer within 21CMMC.
In most MCMC algorithms (including \change{21CMMC}) the Evidence is omitted because normalising the posterior is a redundant overhead in parameter estimation and the integral in Equation \ref{Z_2} is typically difficult to evaluate.

We now turn to the Bayesian Evidence, $\mathcal{Z}$, as this takes centre stage for model selection. 
By evaluating a $\mathcal{Z}_{i}$ for the i'th model, $M_{i}$, we can define what is called the Bayes factor as follows:

\begin{equation}
\mathcal{B}_{12} = \frac{P(D|M_1)}{P(D|M_2)} = \frac{{\mathcal{Z}}_{1}}{{\mathcal{Z}}_{2}} \label{Bayes}  .
\end{equation}

Bayes factors can be interpreted as the odds on which model is thought to be correct, given the data.
We can expand $\mathcal{Z}_i$ as

\begin{eqnarray}\label{Z} 
\mathcal{Z}_i =\int P(D \mid \Theta_{1},...,{\Theta_{N}}, M_i)P(\Theta_{1},...,{\Theta_{N}}|M_i)d^{N}\Theta  \\  \label{Z_2}
 = \int \mathcal{L}(\mathbf{\Theta})\Pi( \mathbf{\Theta})d\mathbf{\Theta} \\
 = \int_0^1\mathcal{L}(X)dX     \label{X}.
\end{eqnarray}

Nested sampling shortcuts this difficult calculation of $\mathcal{Z}_i$ by transforming the integral to one-dimension (Equation \ref{X}), the fraction of prior volume ($X$) which reduces per sampling iteration. 
For more specific information on nested sampling see \citep{skilling}. 
We discuss the interpretation of the Bayes factor in Section \ref{sec: jeff}. 
We choose this machinery over simpler options, such as the \change{Akaike Information Criterion \citep{1100705} or Bayesian Information Criterion \citep{schwarz1978}}, due to their lack of consistency \change{- e.g. on WMAP data \citep{2007MNRAS.377L..39M, 2007MNRAS.377L..74L}}.

The use of Multinest over other nested sampling algorithms is due to the low number of dimensions in our selected models ($\leqslant 4$). 
As dimensionality increases the acceptance fraction geometrically approaches $0$ until the ellipsoidal rejection sampling within Multinest becomes inefficient.
At higher dimension ($\sim 10$), the competing algorithm Polychord \citep{polychord} might be a better choice. 

In this work, 2000 live points\footnote{Live points are used to calculate iso-likelihood contours within nested sampling algorithms.} were found to be enough to produce consistent results within the statistical error bars produced by Multinest. 
The Multinest-obtained posterior distributions plotted throughout this text have been checked for convergence using NestCheck \citep{nestcheck}\footnote{\textcolor{blue}{https://github.com/ejhigson/nestcheck} see the documentation for more details.}.
Dynamic nested sampling algorithms such as Dynesty \citep{Dynamic_NS} result in a maximum live point number of 1500 (to two s.f.) when run in this context. We use this as confirmation that nothing has been missed by our choice of 2000 live points within Multinest.
The errors plotted in the Bayes factor figures in this work are those produced by Multinest, arisen purely from the numerical uncertainties within the algorithm (e.g. using the trapezium rule\change{) - see \citet{skilling} for more details}. It is worth emphasising that these errors do not reflect the sensitivity to Prior choice or the observational measurements. 

\subsection{The Savage-Dickey density ratio}\label{S-D}

One convenient approach to analysing the degeneracy of parameters is via the Savage-Dickey density ratio \citep{10.2307/2958475, doi:10.1080/01621459.1995.10476554}. Since this can only be applied to models that are nested within each other we hold one parameter, $\Theta_{*}$, constant and calculate the Bayesian Evidence. In this notation $d^{N-1}\Theta$ refers to the differential $d\Theta^N$ without the $d\Theta_{*}$ term (e.g. $d\Theta_1 ... d\Theta_N=d^{N-1}\Theta d\Theta_{*}$). Comparing the Bayesian Evidence factor of each of these $\mathcal{Z}(\Theta_{*})$ to $\mathcal{Z}$ (as in Equation \ref{Z}) gives us a parameter dependent Bayes factor

\begin{eqnarray}\label{eq: sddr}
\mathcal{B}(\Theta_*) \equiv \frac{\mathcal{Z}(\Theta_{*})}{\mathcal{Z}} \textcolor{white}{. . . . . . . . . . . . . . }\\ \vspace{0.2cm}
 = \frac{\int \mathcal{L}(\mathbf{\Theta})\Pi( \mathbf{\Theta})d\Theta^{N-1}}{\mathcal{Z}} = \int \mathcal{P}(\mathbf{\Theta}) d\Theta^{N-1} ,
\end{eqnarray}
\change{which}, as shown above (and in Section \ref{sec: nested parameters}), reflects the shape of the 1-dimensional posterior for the parameter $\Theta_{*}$ doubling as a cross check for calculating the Evidence integral.

\subsection{The Jeffreys' scale and potential pitfalls of the Bayes factor }
\label{sec: jeff}

The largest counter argument to using Bayesian statistics is the possibility of skewing results from a naive choice of prior distribution\footnote{See \change{\citet{jaynes03}} for a very well illustrated example in which results exactly reflect choices of prior distribution.}. 
One of the many attractive properties Bayesian model selection has to offer is its inherent inclusion of Occam's Razor - more parameters should not be used than are necessary. Because of this if any parameter prior region is chosen naively the added hyper-volume can skew the result via Equation \ref{Z_2}. We therefore try to keep to observationally constrained information, particularly when considering the prior ranges. 
By doing so we can approach this as a way of rewarding good parameters' rather than penalising redundant ones \citep{2013JCAP...08..036N}.

The Jeffreys' scale is conventionally used as a rough guide when treating this dilemma - since large Evidences must be obtained before conclusions are drawn. 
We use the Jeffreys' scale throughout this work to ease interpretation, keeping in mind that the analytic odds are the real conclusions.
We adopt the following version \citep{al2009bayesian} with three levels of significance:
\begin{itemize}
    \item \textit{Strong} - $\mathcal{B}_{12} > 150$ then  model 1 has outperformed model 2 and is objectively better at describing the data in hand. 
    \item \textit{Moderate} - $10 < \mathcal{B}_{12} < 150 $ then the two models are likely to be distinguishable by this method, but care must be taken in assuring no skew has been introduced. 
    \item \textit{Weak} - $\mathcal{B}_{12} < 10 $ then the two models are likely to be indistinguishable by this method. 
\end{itemize}

\section{Considerations of sensitivity to the parameter priors }\label{sec: Pconsiderations}

\begin{table}
 \begin{tabular}{ | l | c | c | c | r |}
    \hline 
    \small \textbf{Model} & \small$\zeta$ &  \small $R$ \small & \small $\rm Log[T_{\rm vir}]$  & \small $\alpha$ \\ \hline \hline
    3pFZH & [5.,250.] & [5.,20.] & [4.0,5.3] & - \\ \hline
    4pFZH & [5.,250.] & [5.,20.] & [4.0,5.3] & [-3.,3.] \\ \hline
    InvFZH & [5.,5000.] & [0.1,10.]  & [4.,7.] & - \\ \hline
    MHR & [5.,4000.] & - & [4.,7.]  & - \\ \hline
    Inv MHR  & [5.,1200.] & - & [4.,6.]  & - \\ \hline
    F MHR &  [5.,1000.] & [0.1,10.] & [4.,6.] & - \\ \hline
    F Inv MHR & [5.,1200.] & [0.1,10.] & [4.,6.] & - \\ \hline
  \end{tabular}
 \vspace{0.2cm}
    \caption{\label{priortable} A summary of the uniform prior ranges used for the 21cmFAST and toy models' Evidence values. For a detailed description of these models and their parameters see Section \ref{sec: further}.}
\end{table}

Throughout this work we apply only uniform (uninformative) prior distributions across all parameter ranges. Log priors have been tested (not shown) to produce the same values for $\mathcal{Z}$ (within error bars) for the 3pFZHf1, 3pFZHf2 and 3pFZH(with $\alpha=0.4$)f2 models with the same ranges of prior. 
We follow similar assumptions of prior boundaries taken from the original 21CMMC. 
The only differences from the initial paper are at the higher end of the $\zeta$ parameter, which we increase from $[5.,100.]$ to $[5.,250.]$ for the 21cmFast models. 
The upper end of this parameter is under debate due to the lack of observational data surrounding $f_{\star}$ and $f_{\rm esc}$ (see Section \ref{sec: 21cmmc}\change{)}. 
The original choice of $f_{\rm esc}$ is in the range $[0.05, 1]$ implying that $f_{\star}=0.05$. 
This is motivation by a theoretical modelling performed by \change{\citet{2014MNRAS.445.2545D}} and is confirmed to be adequate by the observational constraints performed in \citet{2017arXiv171004152G}. 
The latter of these two papers sums up the problem nicely: `the escape fraction is related to too many astrophysical parameters to allow us to use a complete and fully satisfactory model'. 
An updated parameterisation of $\zeta$ has been established \citep{Luminosity21cmmc}. 
We defer use of this to future work where we will analyse (similarly to Section \ref{sec: 21cmvsQSOs}) likelihood contributions via the galaxy luminosity function.
The atomic cooling threshold mentioned in Section \ref{sec: 21cmmc} is used as the lower boundary for $T_{\rm vir}$. 
The upper limit is set to be consistent with the observations of Lyman break galaxies and the cooling thresholds for ionised gas. 
These are limited due to feedback mechanisms as galaxies above this threshold are too small to maintain star formation  \citep{complete21cmhistory}. 
This results in the prior on $T_{\rm vir}[K]$ being $[10^4, 2 \times 10^5]$. 
When relaxing this upper bound to find the posterior peaks of our toy models, the exponential drop off of the halo-mass function means there are insignificant changes to the Evidence. 
$R_{\rm mfp}$ is motivated from a simulation stand point. 
The mean free path of photons should depend only on the instantaneous recombination rate of hydrogen in the IGM; hence this scale must be captured. 
Typically the bubbles in a simulation such as 21cmFast will be of the order $10~h^{-1}~\rm Mpc$ and so we would expect this to be roughly where this value should lie (see Figure \ref{R_Bfactors} in Section \ref{sec: nested parameters} for more discussion). 
To allow some flexibility in the scales in which the physics can proceed this is set to $[5.,20.]~h^{-1}\rm ~Mpc$ (as in 21CMMC). 
We note that nothing is gained by extending the prior beyond this. 

To test the impact of these choices we explore the skewing of the results with widened prior ranges and with $\delta$ function priors (blue and red points respectively, Figure \ref{fig: bfactors+bias} - Section \ref{sec: 21cmfastmodels} contains the full discussion). 
The widened prior values are changed as follows: $\zeta \sim [0.,1000.]$; $R_{\rm mfp} \sim [0.,30.]$; $\rm Log[T_{\rm vir}] \sim [3.5,6.0]$; and $\alpha \sim [-3.0,3.0]$. 

In the context of the toy models, we have expanded the priors in the plots to find the peaks of the likelihood distributions - this is to maximise their Evidence values for the sake of argument. 
Values obtained by integrating the full likelihood distributions are overwhelming compared to the penalties of using these large priors (large compared to FZH). 
The conclusions obtained from any Bayes factors are not challenged by any change in prior hyper-volume that is performed in this work. 
Any false Jeffreys' scale conclusions can be disputed by physical motivations.
The Evidences shown in this paper use the priors given in Table \ref{priortable} - in general as long as the width of the prior captures the peak of the likelihood, Bayes factor conclusions are repeatable.

Finally we have included two reparameterisations of the MHR models (1p MHR and F MHR, both detailed in Section \ref{sec: further} with the results discussed in Section \ref{sec: f3prior}) as well as nested versions of the 3pFZH model (Section \ref{sec: nested parameters}) to quantitatively asses the impact of Occam's razor implicit in calculating the Evidence.


\section{Results} 
\label{sec: results2}

Unless specified otherwise all results are tested against an FZH power spectrum with one of two fiducial sets of parameters as the simulated data, namely: f1 $[\zeta=20.,~R_{\rm mfp}=15.~\rm Mpc ,~ T_{\rm vir}=30000~\rm K]$; and f2 $[\zeta=15.,~R_{\rm mfp}=15., ~T_{\rm vir}=50000,~ \alpha=0.4]$\change{, as is done in the original 21CMMC.} 

In this work, the 21cmFast models is the collective name we use for FZH simulations, with 3p referring to those with $\alpha=0$ (therefore implying Equation \ref{eq: zetaA}) and 4p to any with $\alpha \neq 0$  (Equation \ref{eq: zetaB}). 
The toy models are FZH, Inv FZH, MHR, Inv MHR, F MHR and F Inv MHR collectively.
For example, the notation 3pFZHf1 refers to a 3 parameter FZH model fitting for the f1 fiducial power spectrum. 

\begin{figure}
    \centering
    \includegraphics[scale=0.5]{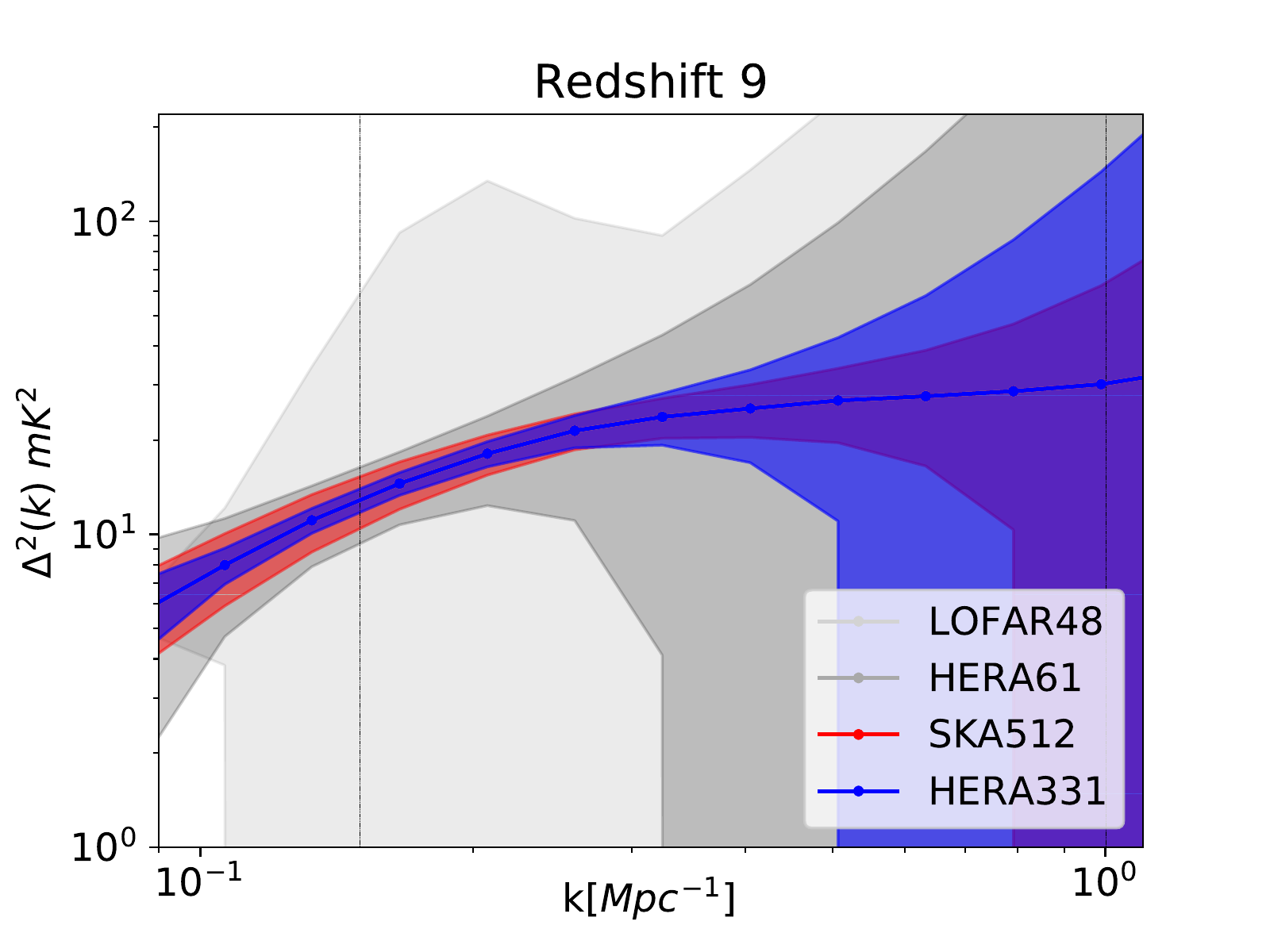}
    \caption{Uncertainties on the f1 power spectra (defined in Section \ref{sec: agreement}) as produced by 21cmSense for LOFAR, HERA-61, SKA and HERA-331. 
    Only observations between the black lines are used, these limits are defined by foregrounds (lower) and shot noise (upper).}
    \label{fig: errors}
\end{figure}

The noise profile of HERA-331 with 1080 hours of observation is used throughout this work, unless specified otherwise. Sections \ref{sec: LOFAR}, \ref{sec: heradipoles}, and \ref{sec: SKA} respectively summarise the results that follow when simulating with LOFAR-48, multiple HERA configurations, and the SKA-512. The error on the f1 power spectrum for LOFAR, HERA-61, HERA-331 and the SKA at $z=9$ are shown in Figure \ref{fig: errors}. 
\change{The SKA has longer baselines than HERA and is more sensitive at small scales. 
Due to HERA-331 having many more small baselines it has greater large scale sensitivity making it comparable to the SKA at model selection via the 21cm power spectrum.}

\subsection{Direct comparisons of 21cmNest vs 21cmMC}\label{sec: agreement}

\begin{figure}
\centering
\subfigure[\label{fig:5ai} 3pFZHf1]{\includegraphics[width=0.36\textwidth]{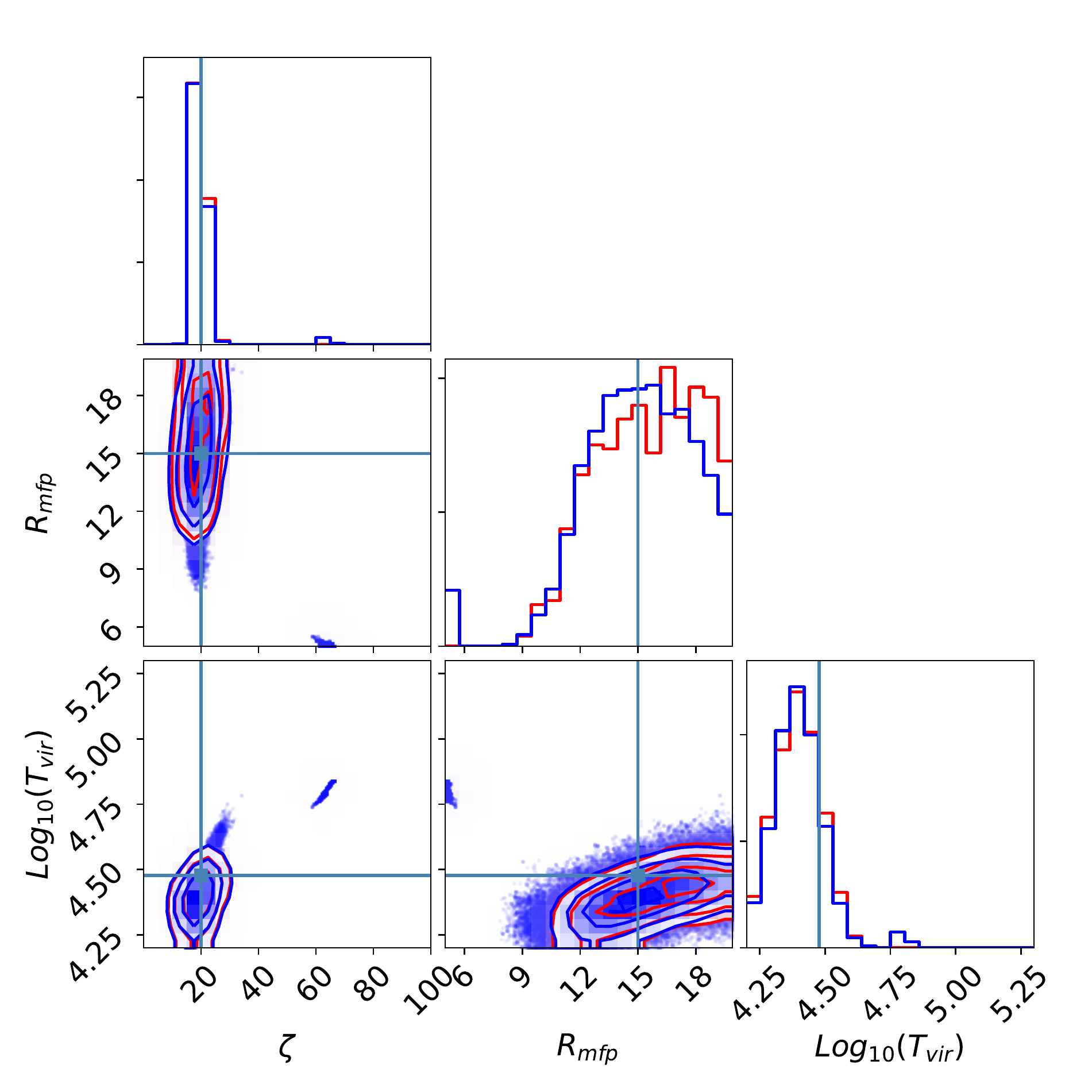}}
\subfigure[\label{fig:5b} 3pFZHf2]{\includegraphics[width=0.36\textwidth]{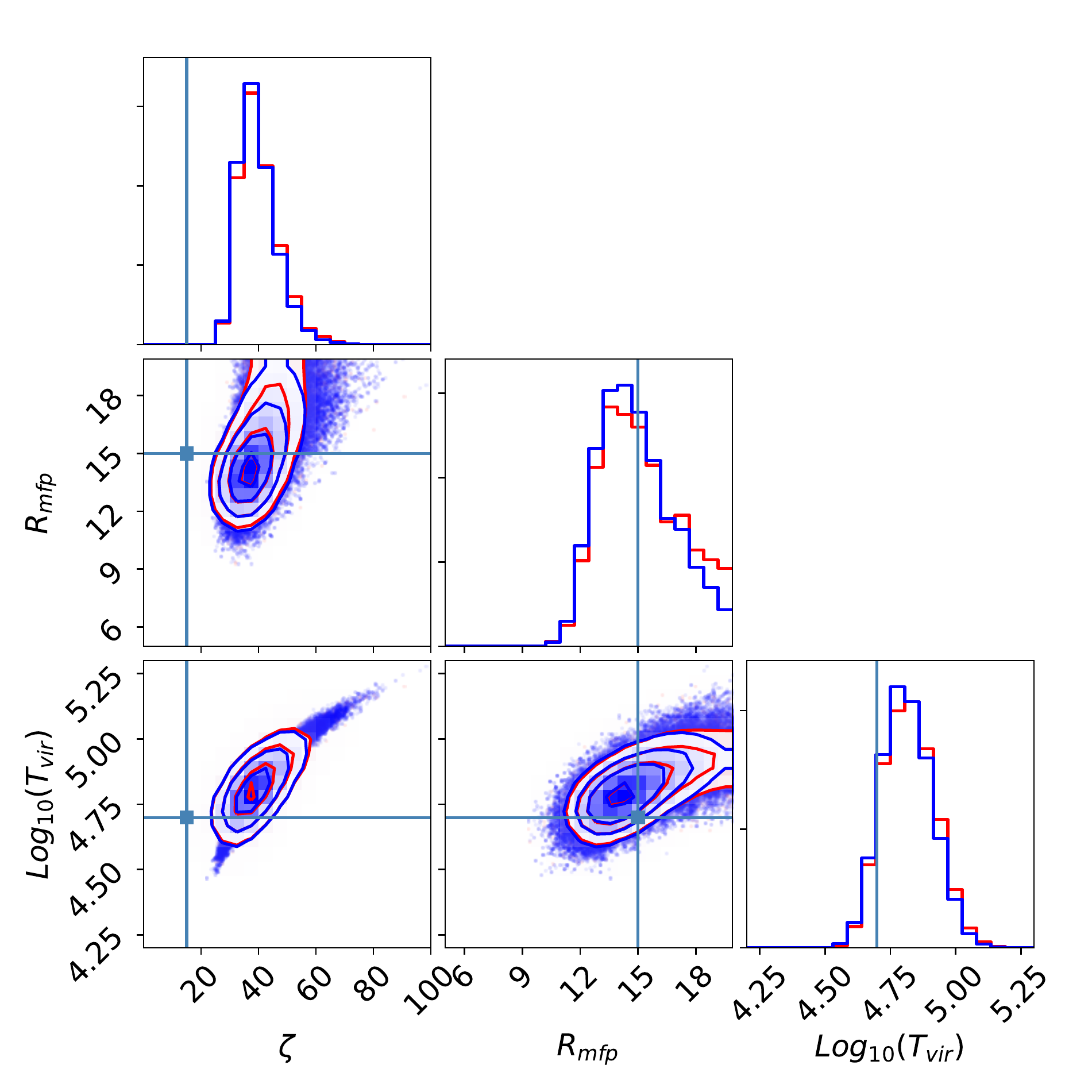}}
\caption{\label{mydata} Here we show that both the 21cmNest (red) and 21CMMC (blue) parameter posteriors for the 3 parameter 21cmFAST models are in agreement. \ref{fig:5ai} This simple model can re-obtain the parameters from its own fiducial simulation (shown by the pale blue lines). \ref{fig:5b} The 3p model requires $\zeta$ to compensate when recovering the f2 values which can be seen by an offset which is quantified in Section \ref{sec: Z_eff}. When $\alpha$ is fixed to this fiducial value the 3p model can easily recover the 4p fiducial parameters (not plotted, see Table \ref{datapoints}). The `islands' of points away from the modes (only in blue) are \change{a pitfall of the Emcee algorithm as walkers can get stuck in regions of low likelihood (see the Emcee API for more detail)}. Typically they contain $\sim 2\%$ of points evaluated and are removed from 21CMMC with a likelihood cut.}
\end{figure}

We first demonstrate our ability to reproduce the results of 21CMMC.
These are done in comparison to two fiducial power spectra across three redshifts ($z~=~8,9,10$). 
As is done in \citet{21CMMC}, we attempt to recover the f1 and f2 fiducial parameter sets (specified in the previous section) with the 21cmFast models. Figure \ref{mydata} shows the overlaid posterior distributions from 21CMMC and 21cmNest for the 3pFZHf1 and 3pFZHf2 results. 
A full summary of all runs shown in Table \ref{datapoints}. 
\change{The} 4pFZHf1, 4pFZHf2 and 3pFZH(with $\alpha = 0.4$)f2 exhibit similarly good agreement to Figure \ref{fig:5ai} however these are not shown.

Multinest is a more efficient algorithm for model selection purposes than Emcee is. 
Less likelihood evaluations are necessary due to the weighting of points calculated by the change in prior volume at each step of the nested sampling algorithm. 
In comparison, correlation in the MCMC chain must be removed for algorithms, such as MCEvidence \citep{ALAN}, that approximate the Evidence using the points in the chain - this requires an order of magnitude more points than nested sampling algorithms do.
At low dimensionality (up to 4 is considered in this work) Multinest performs as well as Emcee for parameter estimation purposes also, as can be seen by the matching posteriors in Figure \ref{mydata}. 
None of the EoR models we consider consist of likelihood distributions with sharp edges that would typically cause trouble for the ellipsoidal rejection sampling used in Multinest \citep{gambit}. 
The MCMC chains produced by Emcee are run until the passing of a Gelmin-Rubin \citep{1992StaSc...7..457G} convergence test. 
\change{We used these settings for 21CMMC and Multinest: walkersRatio=16, burninIterations=250, sampleIterations=3000, threadCount=20, reuseBurnin=False; n\_live\_points=2000, max\_iter=0, multimodal=True, evidence\_tolerance=0.5, sampling\_efficiency=0.3 respectively.
On a 20 core machine, Emcee obtained 39000 likelihood evaluations in 32 hours while Multinest produced 50031 samples in 15 hours. 
For the problem at hand Multinest is faster, but the scaling to higher dimensions needs exploring within the EoR context.}

\subsubsection{Cross check: $\zeta_{\rm eff}$ the effective ionising efficiency}
\label{sec: Z_eff}

Figure \ref{mydata} shows agreement between the two statistical algorithms well in Figure \ref{fig:5ai}. 
In plot \ref{fig:5b} however both exhibit a posterior peaked at a value that is different from the fiducial (shown by the pale blue lines). 
This is because the FZH model prescribes an effective $\zeta_{\rm eff}$ that is calculated as 
\begin{equation}\label{eq: Z_eff}
\zeta_{\rm eff} = \frac{\int^{\infty}_{M_{\rm vir}}~ dm \frac{dn}{dm} m ~\zeta(m)}{\int^{\infty}_{M_{\rm vir}}~ dm \frac{dn}{dm} m}
\end{equation}
(note the denominator can be rewritten $f_{\rm coll}~\rho_{\rm M}$ from Equation \ref{eq: fcoll}). Since the f2 set was created with $\alpha \neq 0$ it is impossible (by definition) for the 3p model to retrieve the exact fiducial values that satisfy this mock data.
This is evident in \ref{fig:5b}. 
The consistency between the $\zeta$ of 3pf1 and the $\zeta_{\rm eff}$ of 4pf1 in Table \ref{datapoints} shows how the added parameter allows a compensation between $\zeta$ and $\rm log_{10}[T_{\rm vir}]$. 
Therefore the standard deviations of $\zeta$ ($\sigma_{\zeta}$) are much larger when $\alpha$ is allowed to vary (the 4p models in Table \ref{datapoints}). In particular note that the $\zeta_{\rm eff}$ agrees for all of the f2 simulations and all of the f1 simulations within $\sigma_{\zeta}$ (rather than for the inputted $\zeta$).

\begin{table}
\centering \begin{tabular}{| l | c | c | c | r |}
    \hline 
    \small  & \small $\zeta$ & $\rm log_{10}[T_{\rm vir}]$ & $\alpha$ &  $\zeta_{\rm eff} $ \\ \hline \hline
    3pFZHf1 & 18.7 $\pm$ 2.0 & 4.366 $\pm$ 0.09 & (0.0) & 18.7 \\ \hline
    4pFZHf1 & 5.8 $\pm$ 27.6 & 4.148 $\pm$ 0.16 & 0.36 $\pm$ 0.52 & 6.6 \\ \hline
    f1 & 20. $\pm$ 5. & 4.447 $\pm$ 1.1 & (0.0)  & 20.  \\ \hline \hline
    3pFZHf2 & 34.9 $\pm$ 3.2 & 4.747 $\pm$ 0.05 & (0.0) & 34.9 \\  \hline 
    3pFZHf2 & 12.2 $\pm$ 4.2 & 4.576 $\pm$ 0.13 & (0.4)  & 66.1 \\  \hline 
    4pFZHf2 & 5.3 $\pm$ 27.3 & 4.473 $\pm$ 0.15 & 0.624 $\pm$ 0.40 & 7.8 \\ \hline 
    f2 & 15. $\pm$ 3.8 & 4.699 $\pm$ 1.2 & 0.40 $\pm$ 0.10 & 39.6  \\ \hline
\end{tabular}  
\vspace{0.2cm}
    \caption{\label{datapoints} The MAP parameter values to $1\sigma$ for the 5 simulations performed in Section \ref{sec: agreement} (two of which are shown in Figure \ref{mydata}, with fiducial results marked on). 
    $\zeta_{\rm eff}$ is the actual ionisation efficiency that the galaxies produce, via the implementation of the $\zeta$ parameter (via Equation \ref{eq: Z_eff}). 
    Note that in the 3pFZH cases ($\alpha=0$) use the relationship of Equation \ref{eq: zetaA}, while for 4pFZH we use Equation \ref{eq: zetaB} ($\alpha \neq 0$). 
    For the fiducial values, f1 and f2, $\sigma_{\zeta}$ has been approximated by using 25\% of each value as suggested in the original 21cmFAST paper \protect\citep{21cmFAST}; this is motivated by the discrepancies between semi-numerical and full radiative transfer simulations. 
    Deviations from recovering the exact fiducial values are due to the observational priors inputted in the likelihood.
    }
\end{table}

\subsubsection{Cross check: recovering an Inv MHR fiducial power spectrum with the toy models}
\label{sec: f3}

To further test our methodology, a fiducial power spectrum (mock data) has been generated by the Inv MHR model with $[\zeta=30, \rm Log[T_{\rm vir}]=4.5]$. 
This showed that different input models can be recovered - proof of concept for Bayesian model selection. 


\subsubsection{Cross Check: Do alternate parameterisations of MHR skew the Bayes factor?}
\label{sec: f3prior}

Testing for the Inv MHR fiducial as in the previous section, we assess the choice of prior width with the reparameterisations of MHR.
We can recover the Inv MHR mock data power spectrum using the F Inv MHR and 1p Inv MHR models detailed in Section \ref{models}.

The 1p Inv MHR is a simplified version of Inv MHR, and in particular has a vastly reduced prior space (the parameters' volume has been reduced from $\sim 2000$ to $1$). 
But this should significantly reduce the predictive power of the Inv MHR model for two reasons: without the use of $f_{\rm coll}$ there is no intrinsic redshift dependence via the cosmological parameters required to calculate Equation \ref{eq: integralfcoll}; and mainly, we expect the neutral fraction to decrease with redshift as reionisation progresses. 

On the other hand with F MHR by smoothing the density field via the addition of a third parameter $R_{\rm Filter}$, the filtered models can adapt to the scaling of the ionised bubbles. 
When fitting for f1 this complexity makes the F MHR and F Inv MHR harder to distinguish than their unfiltered counterparts.
In this context however (when fitting mock data from Inv MHR), increasing the flexibility of this model increases the prior volume without any improvement to the fit (because the fiducial power spectrum was made without filtering). 
The 1p Inv MHR and F Inv MHR models achieve Bayes factors (odds) of $10^{-25}$ and $10^{-29}$ respectively. 
This illustrates that the added parameter is penalised because it is redundant in this context (for comparison 3pFZH obtains $10^{-28}$; Inv MHR has $\mathcal{B} = 1$, by definition). 

\subsection{Can LOFAR-48 perform model selection?}\label{sec: LOFAR}

\begin{figure}
    \centering
    \subfigure[\label{fig: lofarPSs}]{\includegraphics[scale=0.4]{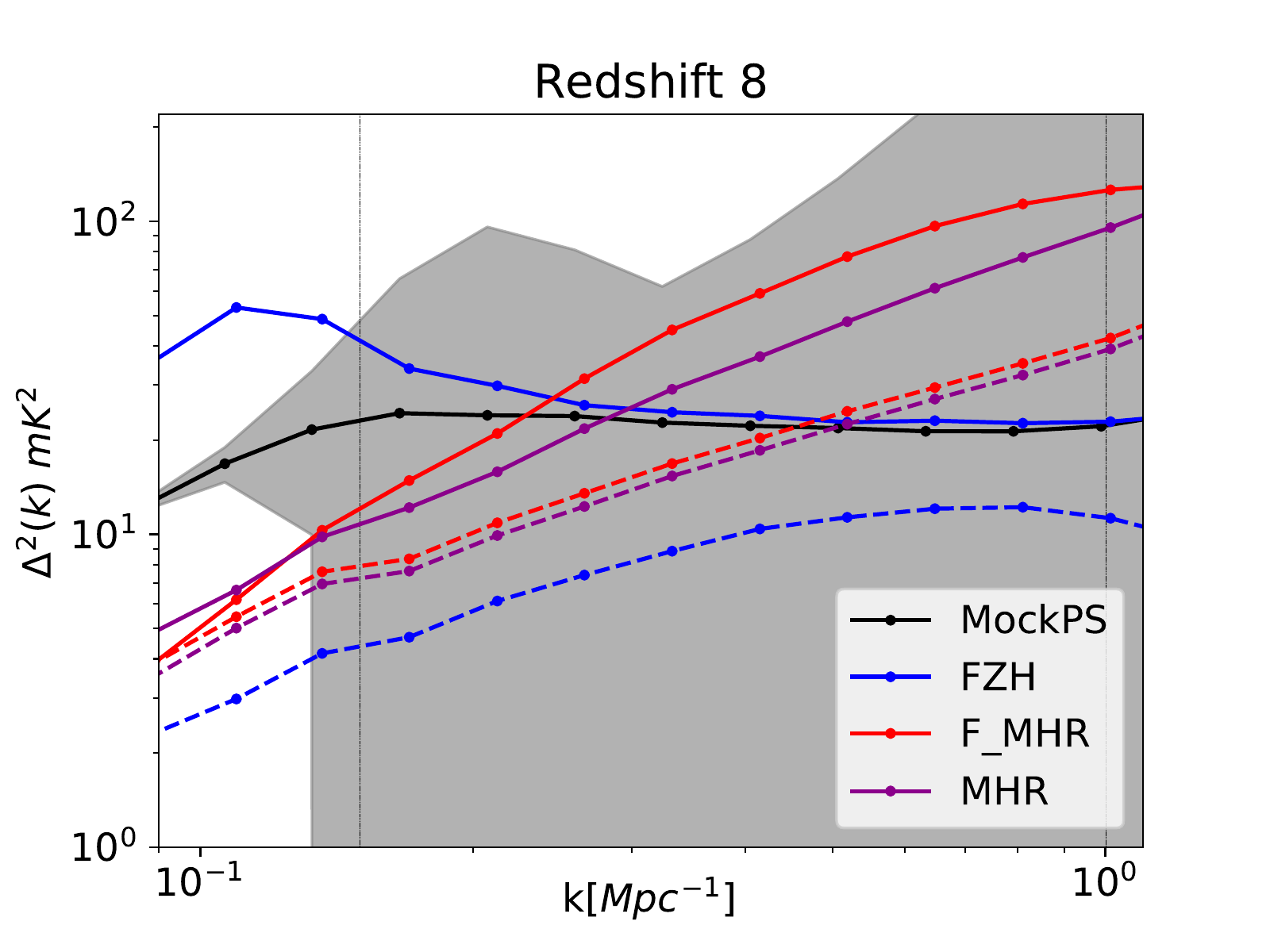}}
    \subfigure[\label{fig: lofarbfactor}]{\includegraphics[scale=0.44]{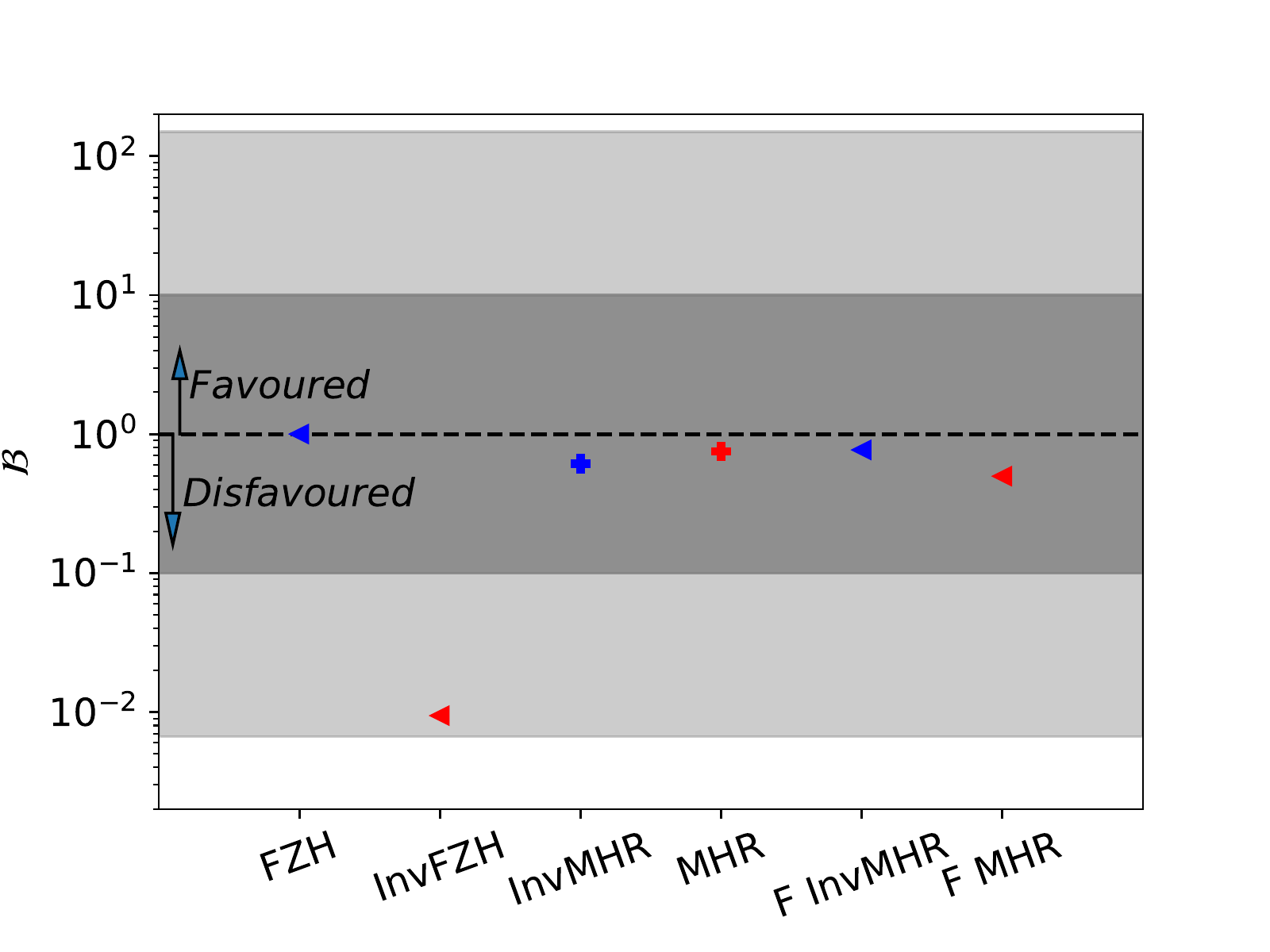}}
    \caption{This Figure refers to LOFAR-48 with 1080 hours of observing time (see Table \ref{Tscopetable} for a summary of the telescope specifications). 
    Plot \ref{fig: lofarPSs} shows best fit power spectra at $z=8$ with dotted lines representing the inverse of the model stated with colour in the legend, the observational error is shown in grey.  
    \ref{fig: lofarbfactor} shows the Bayes factors obtained using an 1080 hour LOFAR observation - white, light and dark grey regions represent \textit{strong}, \textit{moderate}, and \textit{weak} respectively on the Jeffreys' scale.
    Note how much larger the power spectrum error bars are for LOFAR in comparison to that of HERA-331 in Figure \ref{fig: errors}. 
    This is reflected in plot \ref{fig: lofarbfactor} as no models are \textit{strongly} ruled out (all points are within the grey regions). 
    The flexibility in fitting the shape of the best fit power spectra can compensate broadly, except for the case of the Inv FZH model which is disfavoured \textit{moderately} in \ref{fig: lofarbfactor}.}
    \label{fig:my_label}
\end{figure}

Next we ask the question: what length of LOFAR observations is needed to perform Bayesian model selection with the toy EoR models?

\begin{figure}
    \centering      
    \subfigure[\label{1080hrslofar}]{\includegraphics[scale=0.3]{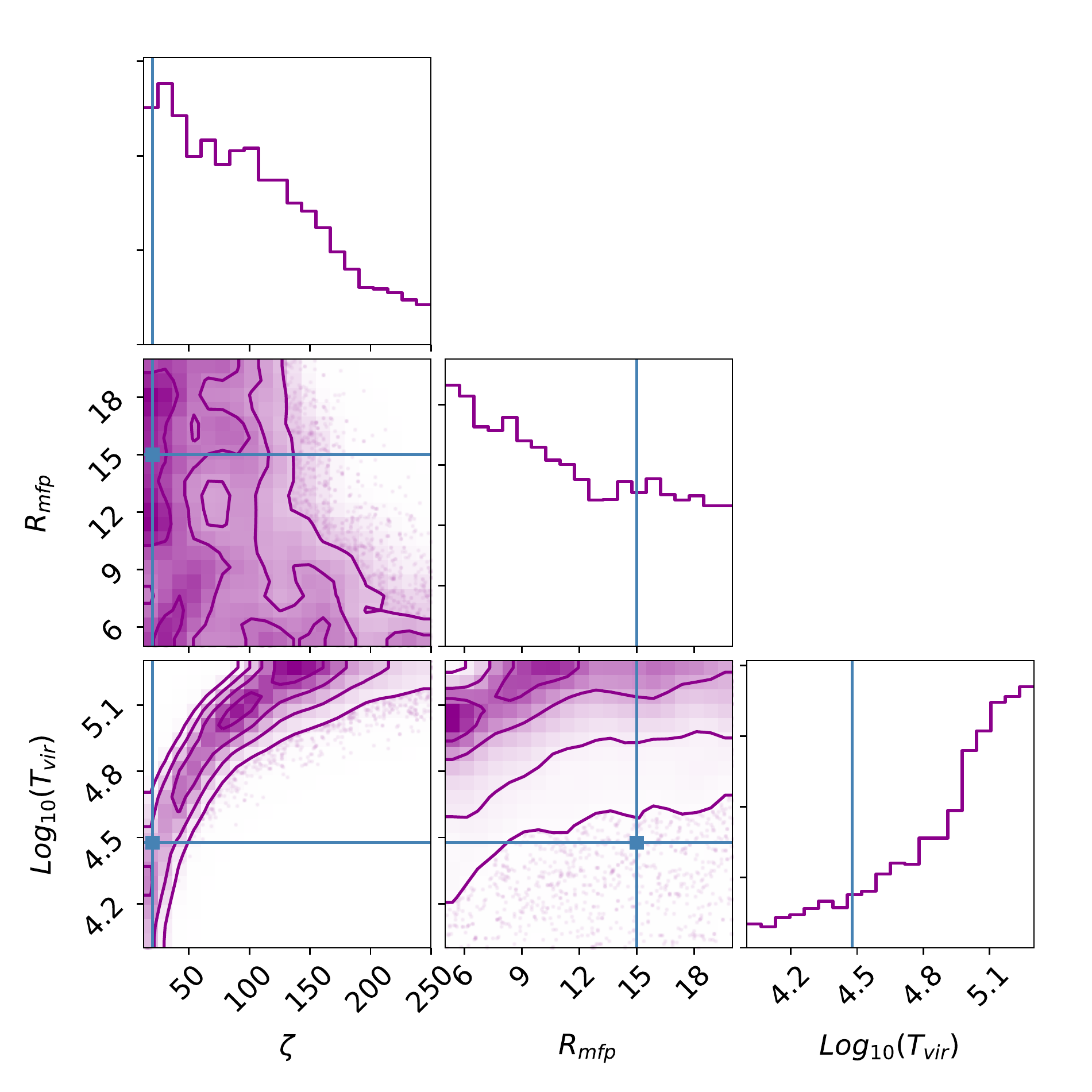}\hfill}
    \subfigure[\label{21600hrslofar}]{\includegraphics[scale=0.3]{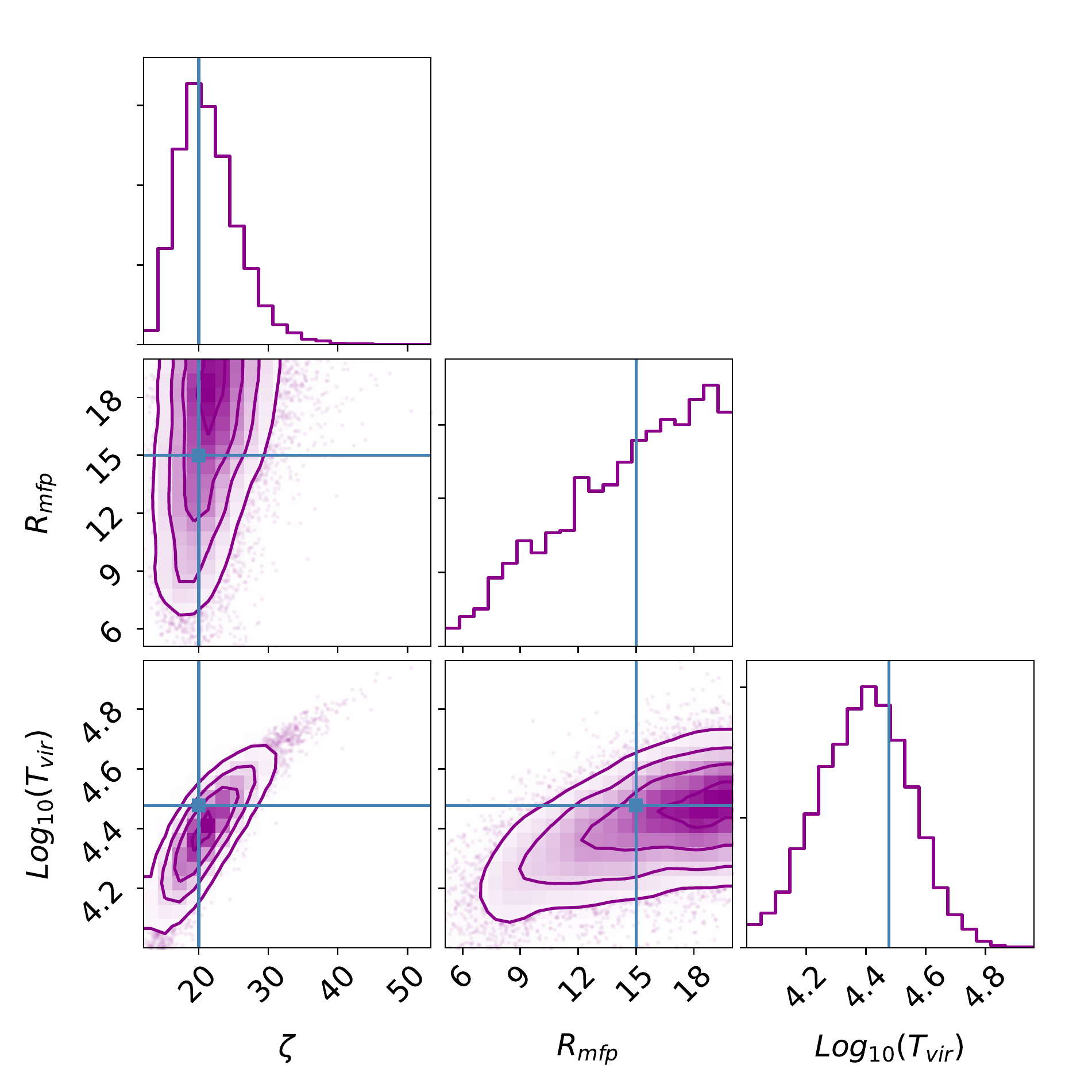}\hfill}
    \subfigure[\label{MONEYlofar}]{\includegraphics[scale=0.5]{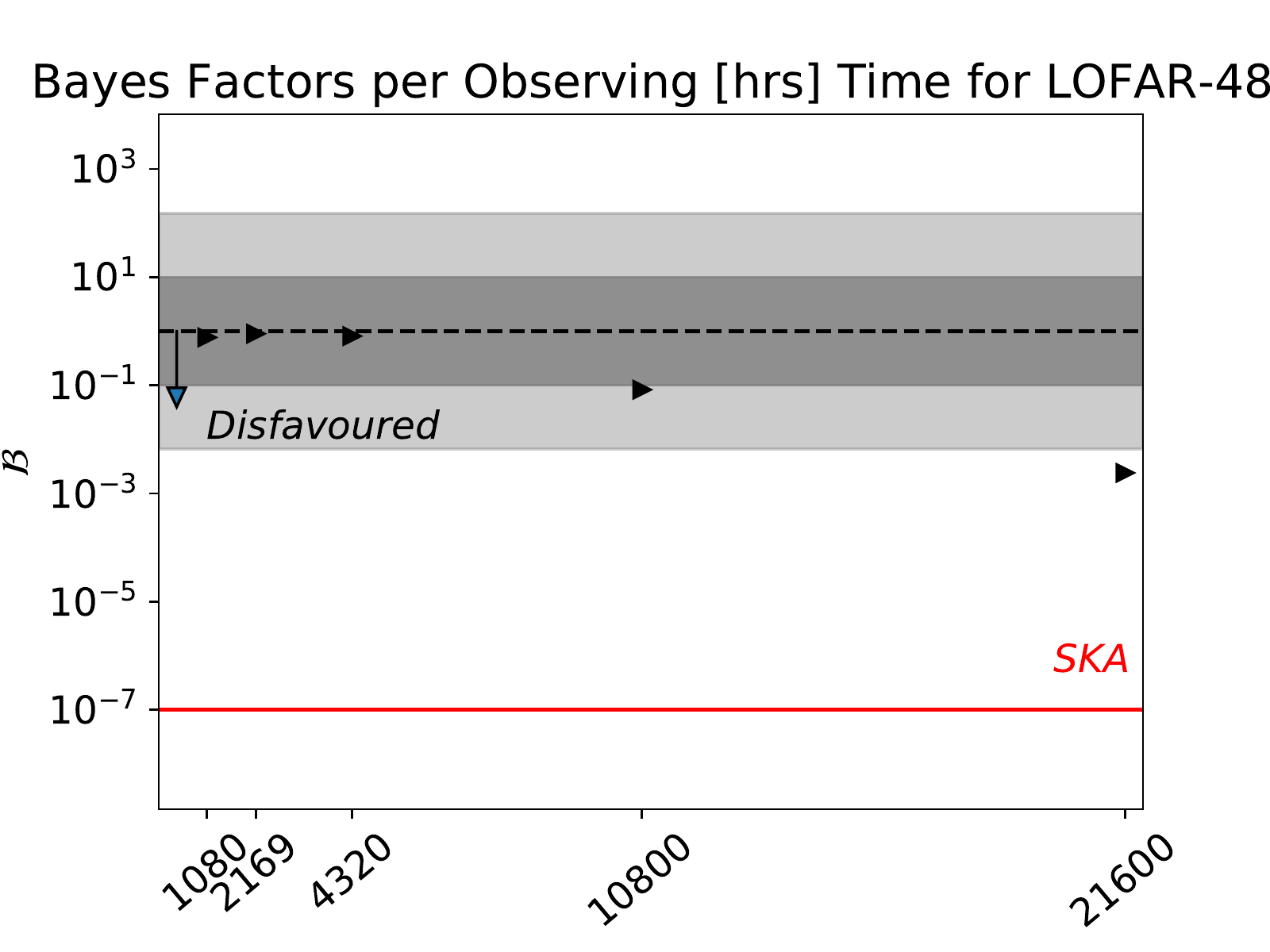}\hfill}
    \caption{\label{fig:lofars} \ref{1080hrslofar} and \ref{21600hrslofar} are the posterior distributions for the FZH parameters used to recover the f1 power spectrum with LOFAR-48 for the observing times 1080 and 21600 hours respectively. 
    \ref{MONEYlofar} shows the Bayes factors of 3pFZHf1 against F Inv MHRf1. 
    \change{For comparison the red line shows the Bayes factor using the SKA-512 with 1080 observing hours.}
    Note that in order to obtain a \textit{strong} distinction between these toy models LOFAR must observe for 21600 hours (indicated by the marker passing into the white disfavoured region). 
    }
\end{figure}

Up until now we have used 1080 hours observing time with all the telescopes considered, as shown in Table \ref{Tscopetable}. 
Each observation is performed with a 1 hour tracked scan of the f1 power spectrum with 6 hours of observing time per day, corresponds to 6 different  fields. 
The Bayes factors and best fit power spectra ($z=8$) are shown for each toy model in Figure \ref{fig:my_label}. 
Due to the large error bars on each power spectrum \textit{weak} results are obtained for all except the InvFZH model.
For the InvFZH model \textit{moderate} results are obtained because of penalties from the observational priors (see Section \ref{split_likelihoods} for further discussion).
The breadth of power spectrum that are capable of fitting within the grey error region of the LOFAR-48 mock signal is reflected in the width of the posteriors in plot \ref{1080hrslofar} - the posterior distributions for $R_{\rm mfp}$ and $Log_{10}[T_{\rm vir}]$ have not peaked at the fiducial values. 

We now vary the observation times. 
LOFAR achieves\footnote{These are a very ambitious trajectory considering LOFAR has observed a total of $\sim 1300$ hours of the NCP field in its 7 years of EoR activity.} signal to noise of 1, 2, 3, 7, and 11 with 1080, 2160, 4320, 10800, and 21600 hours of integration time respectively at $z=8$ (to the nearest integer). 
A total of 21600 hours of observing time is necessary to obtain \textit{strong} conclusions for the toy models considered in this work - Figure \ref{MONEYlofar}.
The posteriors shown in Figure \ref{21600hrslofar} now show peaks that agree with HERA-331 in Figure \ref{fig:5ai}.
In summary model selection is unlikely to be achievable with LOFAR because of the limited k-range in which it is sufficiently sensitive. 

These calculations use 21cmSense's \textit{moderate} foreground settings (Section \ref{sec:noise}) with LOFAR for ease of comparison with the other telescopes. 
In principle LOFAR aims to observe within the foreground wedge (i.e. in a k range that extends lower than $k=0.15~\rm Mpc^{-1}$), meaning these results are conservative estimates. 
The results published in \citet{2017ApJ...838...65P} are within $k=[0.05,~0.13]~\rm Mpc^{-1}$. 
We repeated our 1080 hour LOFAR analyses within this k range to obtain parameter posteriors that are comparable to the 21600 hour observations in the previous k range ($[0.15,~1.]~\rm Mpc^{-1}$). 
In the context of model selection, the results become \textit{moderate} but not \textit{strong} in ruling out F Inv MHR ($\mathcal{B} = 148$) and so model selection would still be challenging for LOFAR.
\change{Ideally LOFAR will be operational without any foreground contamination (the \textit{optimum} 21cmSense setting) in a total possible k range of $k=[0.05,~1.0]~\rm Mpc^{-1}$.
If these criteria are achievable, LOFAR can obtain $\mathcal{B} \approx 120000$ and is likely able to perform model selection.
We stick with our conservative estimates to have a consistent analysis between the telescopes.}

\subsection{Model selection of reionisation scenarios with HERA}\label{sec: HERAmain} 

\begin{figure*}
\centering
\subfigure[\label{fig:7a} 3pFZH f1]{\includegraphics[width=0.3\textwidth]{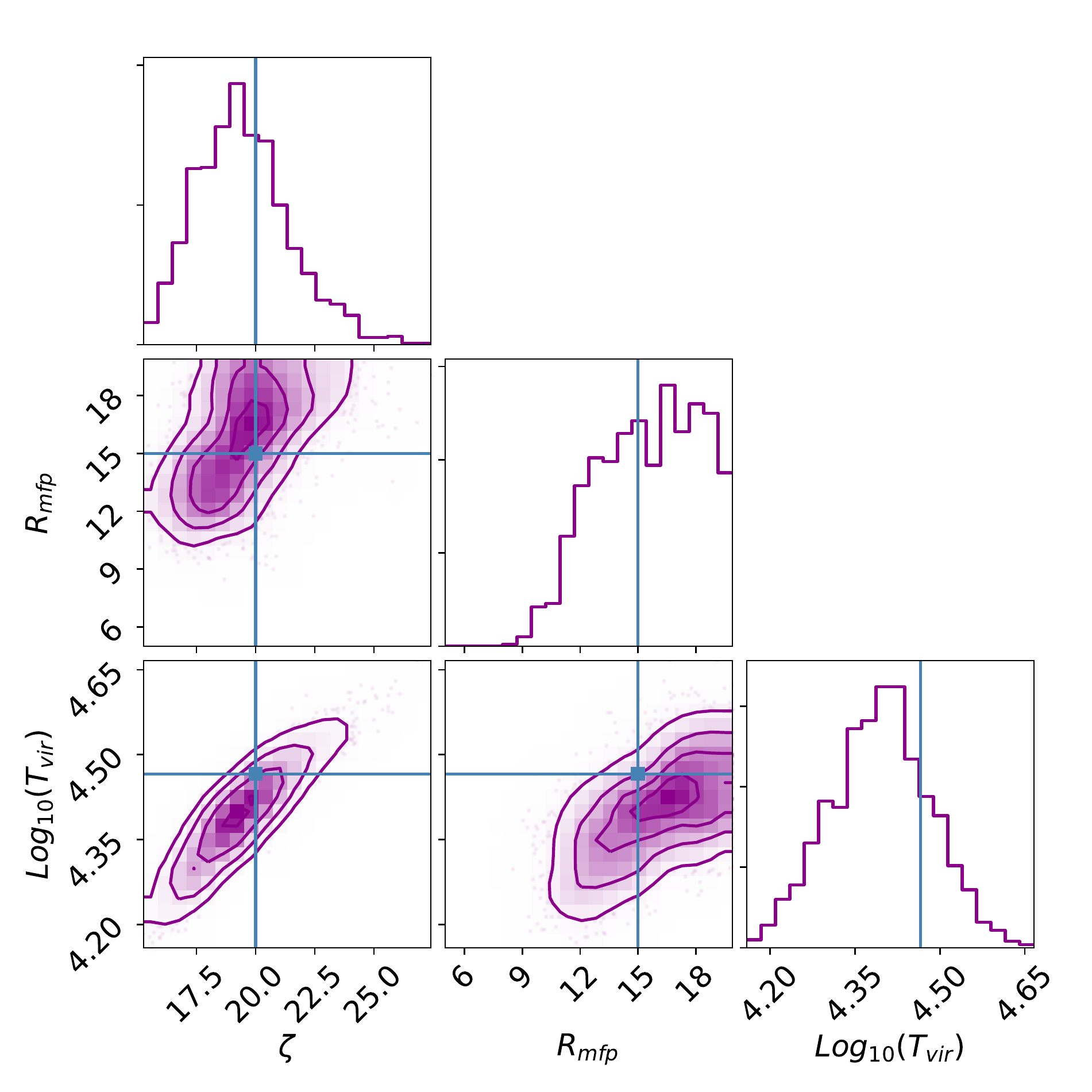}}\hfill
\subfigure[\label{fig:7ai} Inv FZH f1]{\includegraphics[width=0.3\textwidth]{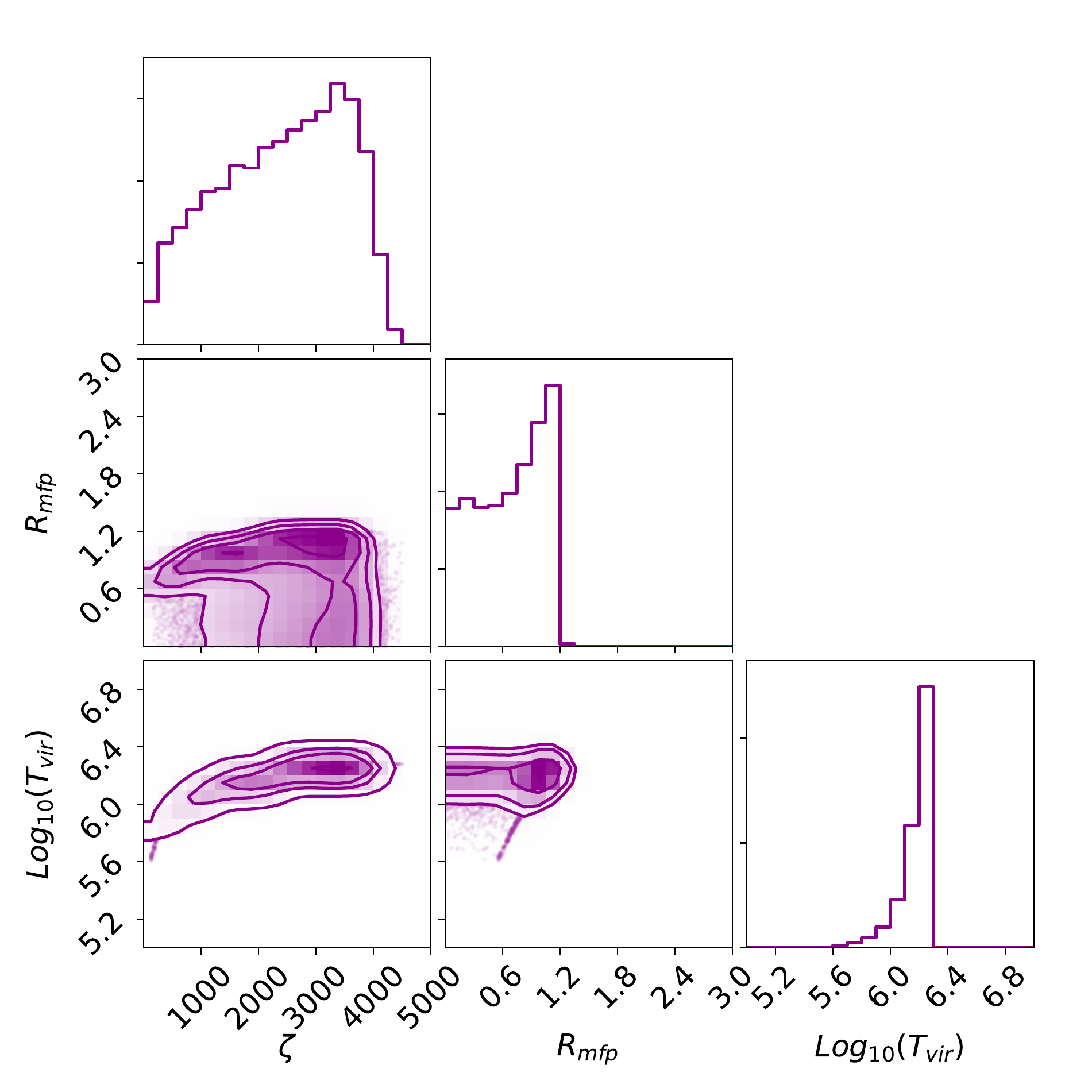}}\hfill 
\subfigure[\label{fig:7b} MHR f1]{\includegraphics[width=0.3\textwidth]{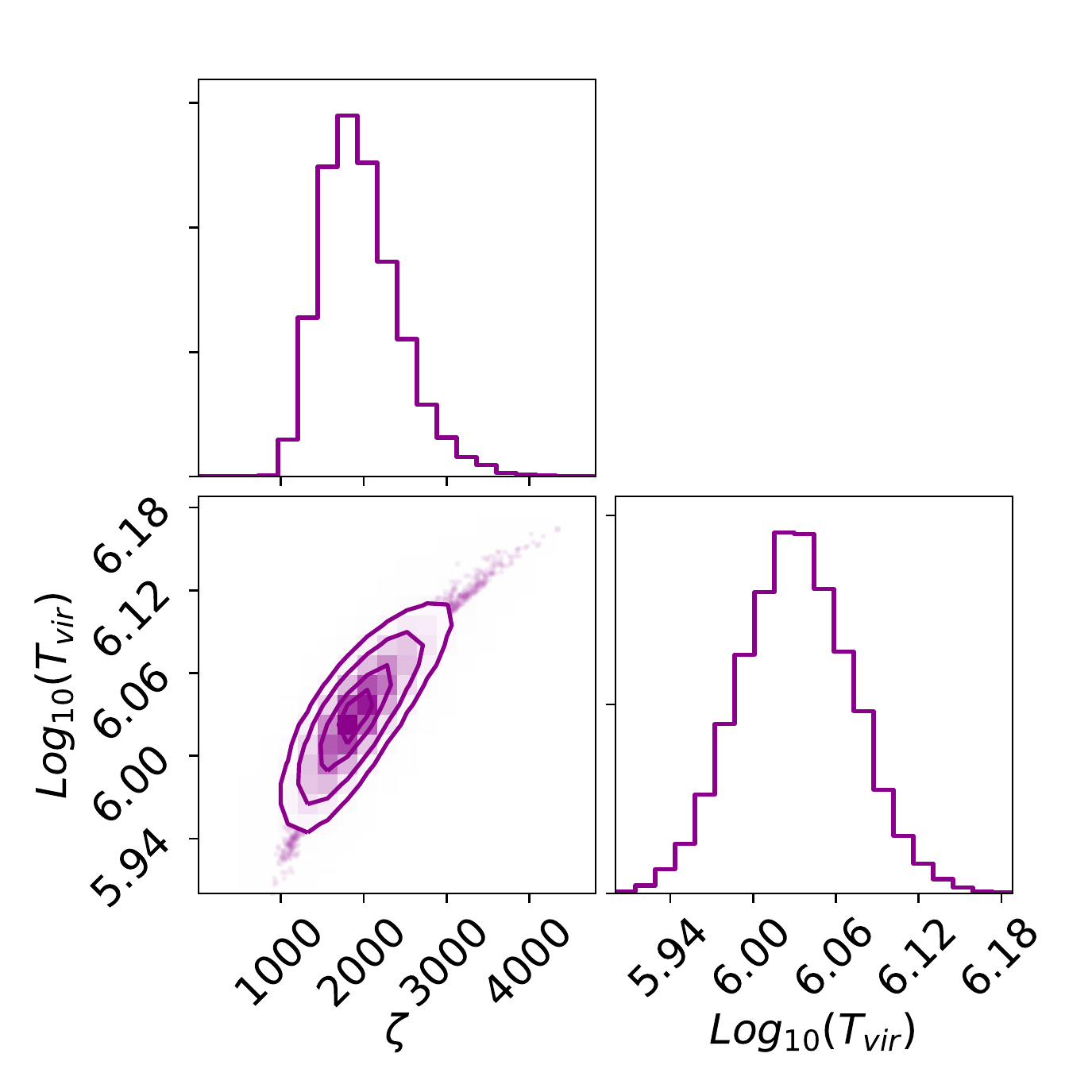}}\hfill
\subfigure[\label{fig:7c} Inv MHR f1]{\includegraphics[width=0.32\textwidth]{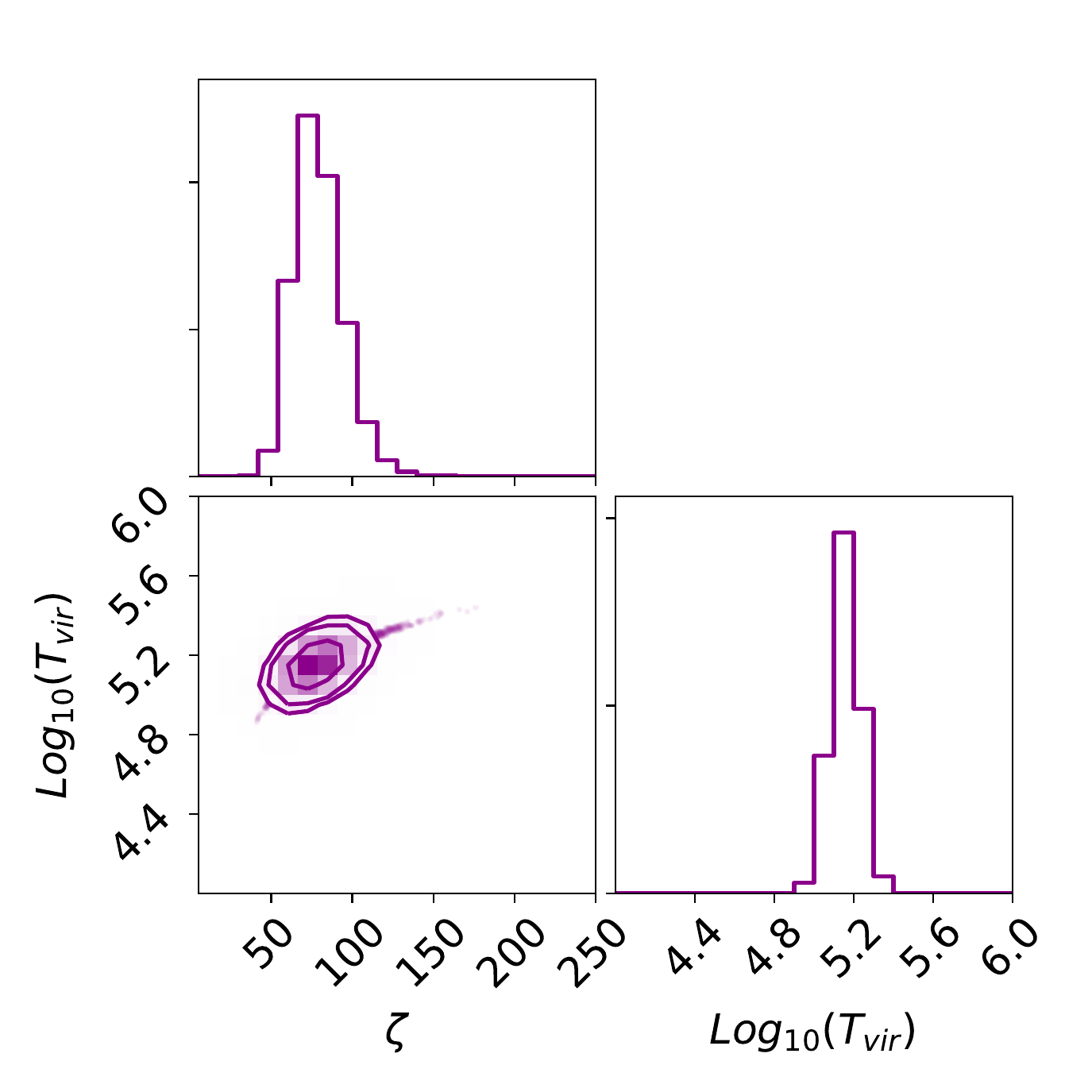}}\hfill
\subfigure[\label{fig:7d} F MHR f1]{\includegraphics[width=0.32\textwidth]{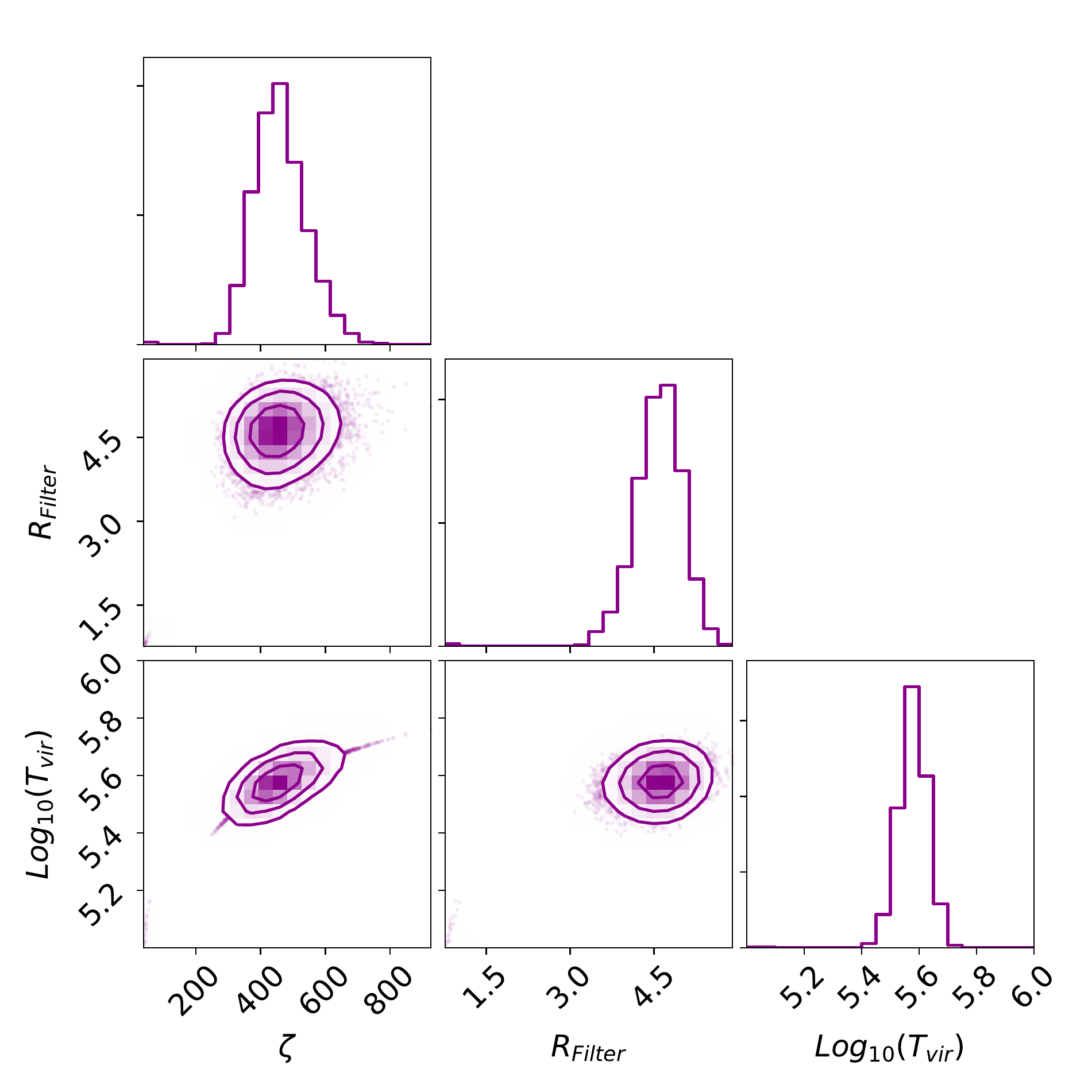}}\hfill
\subfigure[\label{fig:7e} F Inv MHR f1]{\includegraphics[width=0.32\textwidth]{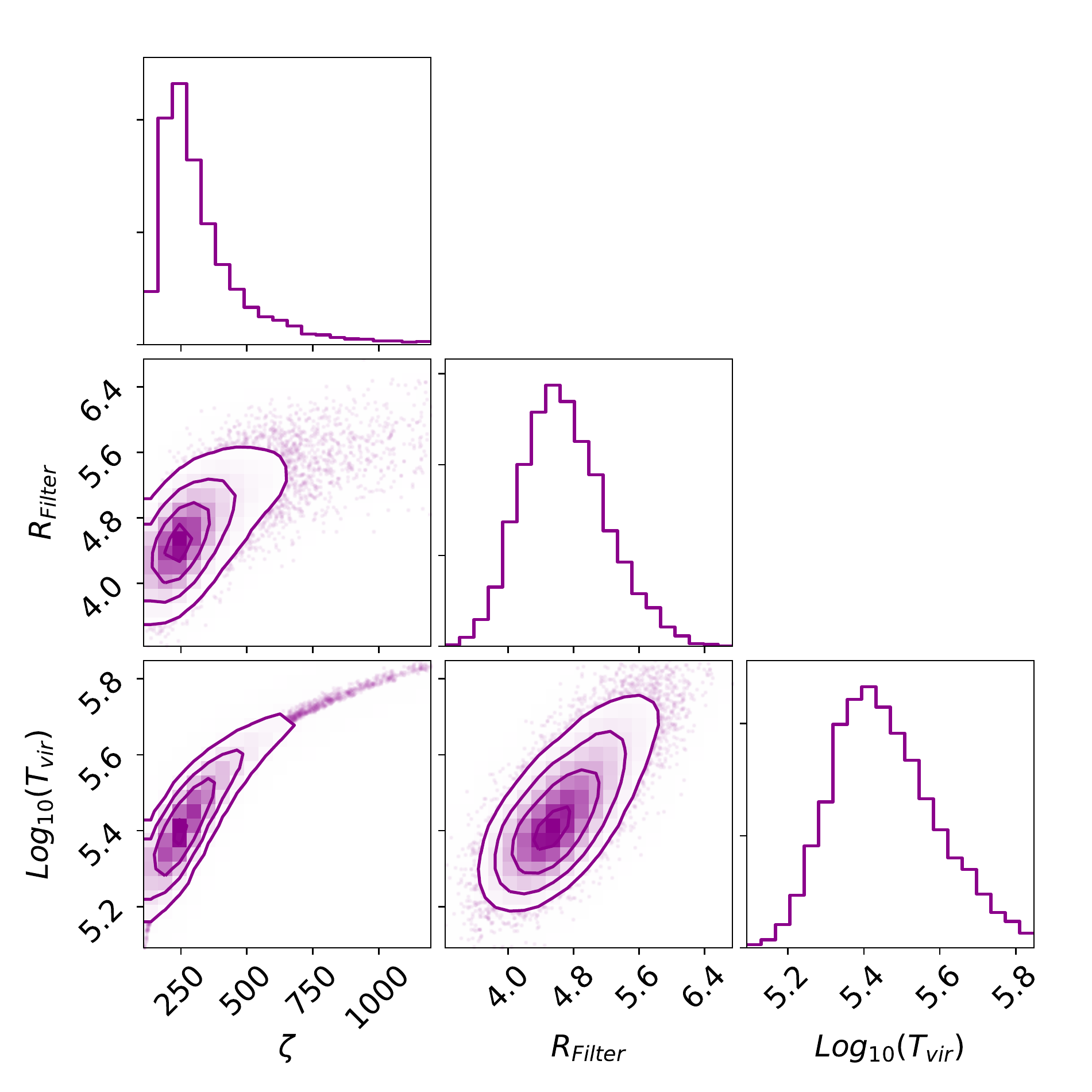}}\hfill
\caption{\label{mydata2} Parameter posteriors for the toy EoR models, as found by 21cmNest using HERA-331 with 1080hrs of observation (averaged over $z=8,9,10$). The MAP values of these distributions are shown in Table \ref{datapoints2}, their power spectra are shown in Figure \ref{fig: PS_comparison}. 
The blue cross hairs on the FZH plot correspond to the values of f1 data set, they are not comparable to the other models.}
\end{figure*}

\begin{figure*}
\centering
\subfigure[]{\includegraphics[scale=0.35]{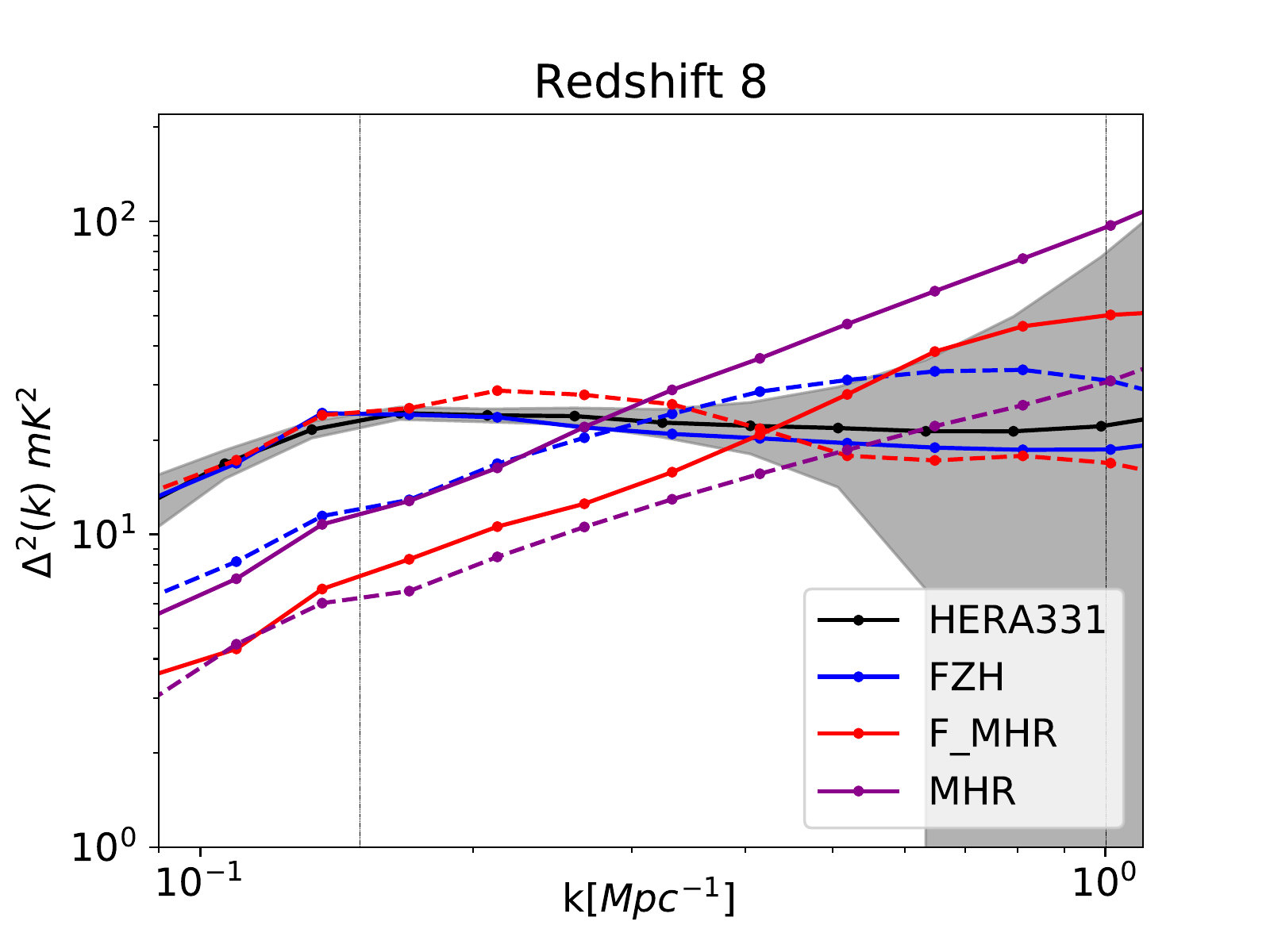}} \hfill
\centering
\subfigure[]{\includegraphics[scale=0.35]{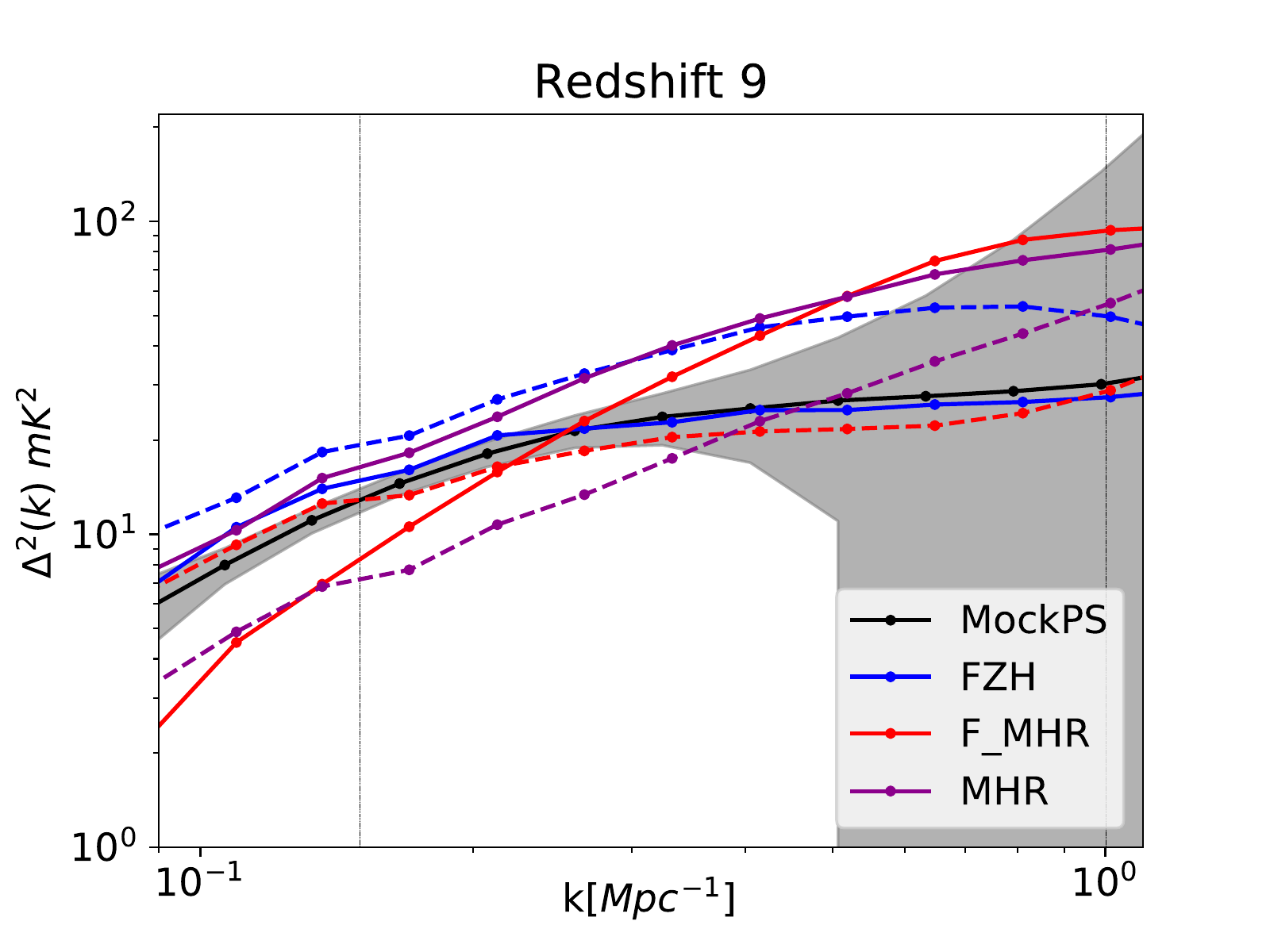}} \hfill
\centering
\subfigure[]{\includegraphics[scale=0.35]{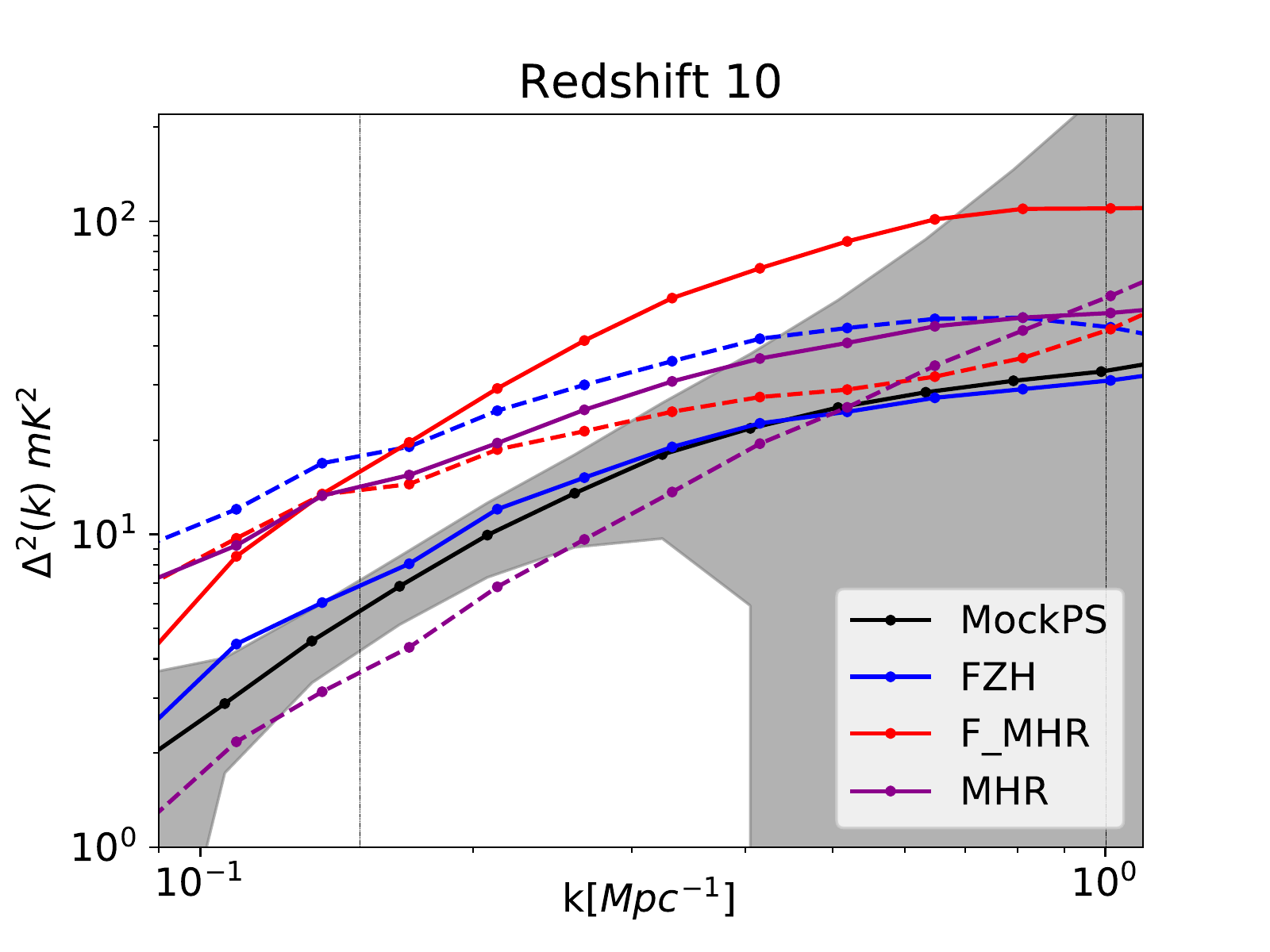}} 
\caption{\label{fig: PS_comparison} Power spectra at redshifts 8, 9 and 10 (a, b and c respectively) for the MAP parameter values of the FZH (blue), Inv FZH (dashed blue), MHR (magenta), Inv MHR (dashed magenta), F MHR (red), and F Inv MHR (dashed red) models (shown in Figure \ref{mydata2}). Only the power spectrum between $k = [0.15,~1.0]~ \rm Mpc$ (between the vertical lines) is included when calculating the likelihood. 
The f1 mock-observed power spectrum, created by 21cmFAST (3pFZH) with fiducial parameters, are shown in black with errors corresponding to HERA-331 shown in grey. 
Note that these power spectra are very different i.e. only the fiducial FZH model is within the error bars for each individual redshift. This is echoed in the huge disparity in Bayes factors shown in Figure \ref{fig: HERAbfactors}.}
\end{figure*}

Here we attempt to distinguish the toy models against the f1 power spectrum (mock data generated with 3pFZH). 
The obtained parameter posteriors and power spectra are shown in Figures \ref{mydata2} and \ref{fig: PS_comparison} respectively. 
As mentioned in Section \ref{sec: jeff} the prior ranges on these plots have been expanded to show the peaks of the likelihood distributions. 
Note that adding the filter scale to the MHR models smooths the density field, producing over-densities that are fewer in number but larger on all scales. 
The background ionising efficiency $\zeta$ must compensate by being bigger in the F Inv MHR case (plot \ref{fig:7c} has a larger $\zeta$ than \ref{fig:7e}) so that these over-densities can be ionised before recombining again. 
The opposite is true for the MHR and its filtered counterpart (plot \ref{fig:7d} has a smaller $\zeta$ than \ref{fig:7b}); the background ionising radiation is less efficient when ionising a smoother density field. 
A qualitative reflection can be interpreted from the sizes of the bubbles (signal-less regions), the smaller and more numerous bubbles require lots of faint small galaxies.
By construction, the MHR models contain a degeneracy between the two main parameters ($\zeta$ and $\rm Log_{10}[T_{\rm vir}]$).
This can be observed by the length of the arc in the corresponding 2d posterior plots. 

\begin{figure}
\includegraphics[scale=0.5]{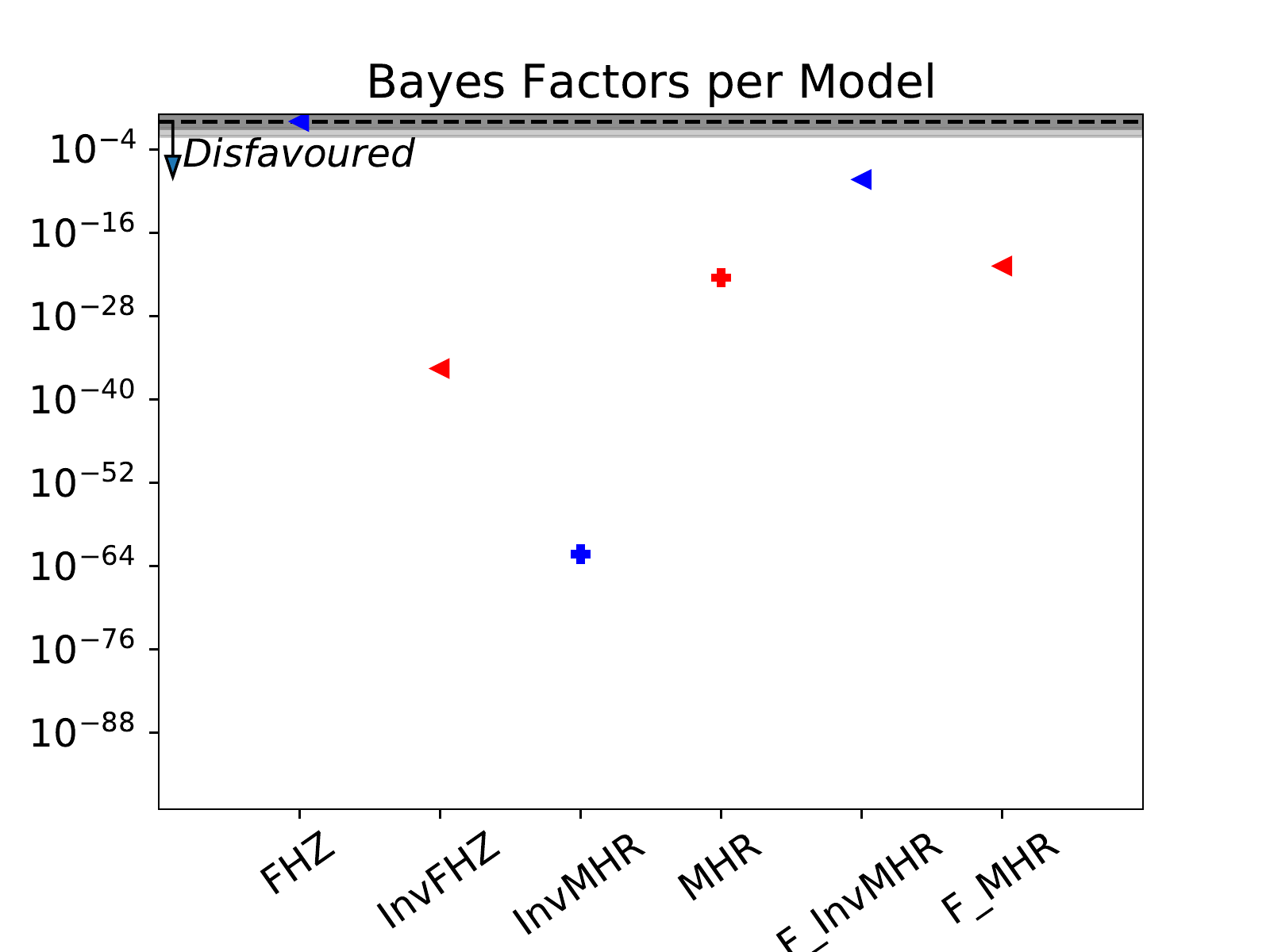}
\caption{ \label{fig: HERAbfactors} The Bayes factors for the toy models as seen by HERA-331 with 1080 hours of observation.  
Blue and red colours indicate an \textit{inside-out} or \textit{outside-in} morphology respectively. 
The arrow heads (\textit{global}) and crosses (\textit{local}) represent EoR scales.
All of the toy models can be distinguished \textit{strongly} (all non-fiducial models are in the white region). 
The corresponding power spectra and parameter posteriors are plotted in Figures \ref{mydata2} and \ref{fig: PS_comparison} respectively. 
}
\end{figure}

The toy models predict vastly different reionisation morphologies and can all be distinguished \textit{strongly} on the Jeffrey's scale, as shown by the Bayes factors in Figure \ref{fig: HERAbfactors}.
\change{The} F Inv MHR proves to be the hardest to distinguish. 
FZH and F Inv MHR share both morphology and scale of reionisation (refer to Figure \ref{slices} for slices of the brightness temperature). 
In contrast, Inv MHR has the \textit{inside-out} morphology, but is \textit{local} rather than \textit{global} in scale - this explains why it is rejected more heavily than its filtered counterpart in Figure \ref{fig: HERAbfactors}.
The filter scale added to the MHR models can clearly increase the flexibility of our \textit{local} based models, resolving this issue.
Our \textit{local inside out} model proves a challenging test only when the \textit{local} constraint is relaxed by smoothing the density field. 
The relative similarity of Inv FZH, MHR and F MHR (red points) suggest that if the morphology is different, there is little compensation achievable by the scale of reionisation.


\begin{table}
    \centering
    \begin{tabular}{| l | c | c | c | r |}
    \hline 
    \small \textbf{Model} & \small$\zeta$ &  \small R  & \small $\rm Log [T_{\rm vir}]$  \\ \hline \hline
    FZH & 18.8 $\pm$ 2.0 & 14.4 & 4.38 $\pm$ 0.09 \\ \hline
    Inv FZH & 1680 $\pm$ 1080 &  1.05 & 6.16 $\pm$ 0.1 \\ \hline
    MHR & 1690 $\pm$ 470 &  & 6.02 $\pm$ 0.04 \\ \hline
    Inv MHR  & 71. $\pm$ 15 &  & 5.12 $\pm$ 0.07 \\ \hline
    F MHR & 432 $\pm$ 80 & 4.65 & 5.56 $\pm$ 0.06 \\ \hline
    F Inv MHR & 203 $\pm$ 170 & 4.35 & 5.3 $\pm$ 0.1 \\ \hline
    \end{tabular}
    \vspace{0.2cm}
    \caption{The MAP values ($\pm 1 \sigma$) for the toy model simulations shown in Figures \ref{mydata2}, \ref{fig: PS_comparison}, and \ref{fig: HERAbfactors} when observing for 1080 hours with HERA-331.}
    \label{datapoints2}
\end{table}

\subsection{QSO data verses the 21cm power spectrum with HERA}\label{sec: 21cmvsQSOs}

\begin{figure}
\subfigure[\label{split_likelihoods}]{\includegraphics[scale=0.5]{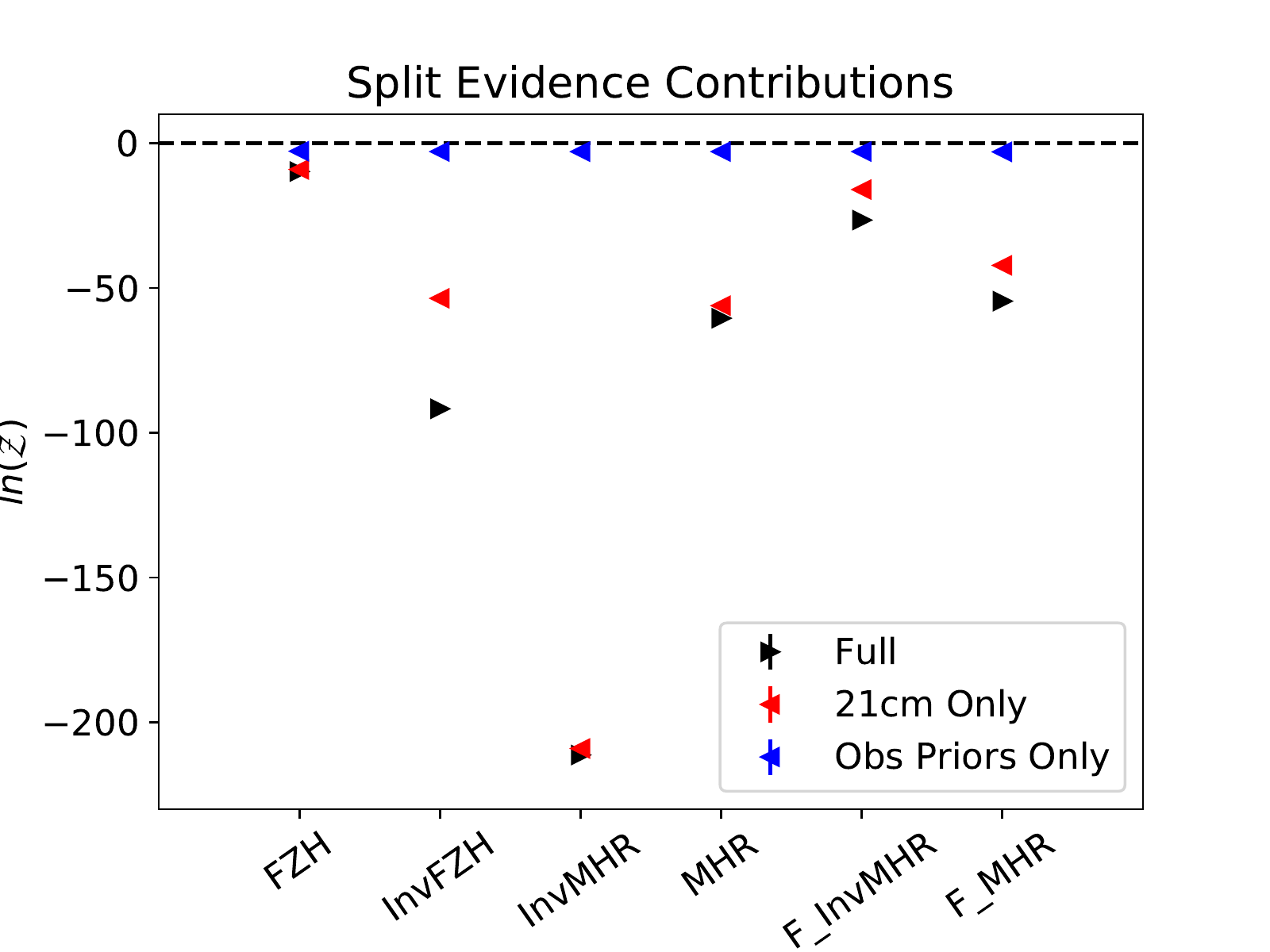}\hfill}
\subfigure[\label{NF_interpolation}]{\includegraphics[scale=0.5]{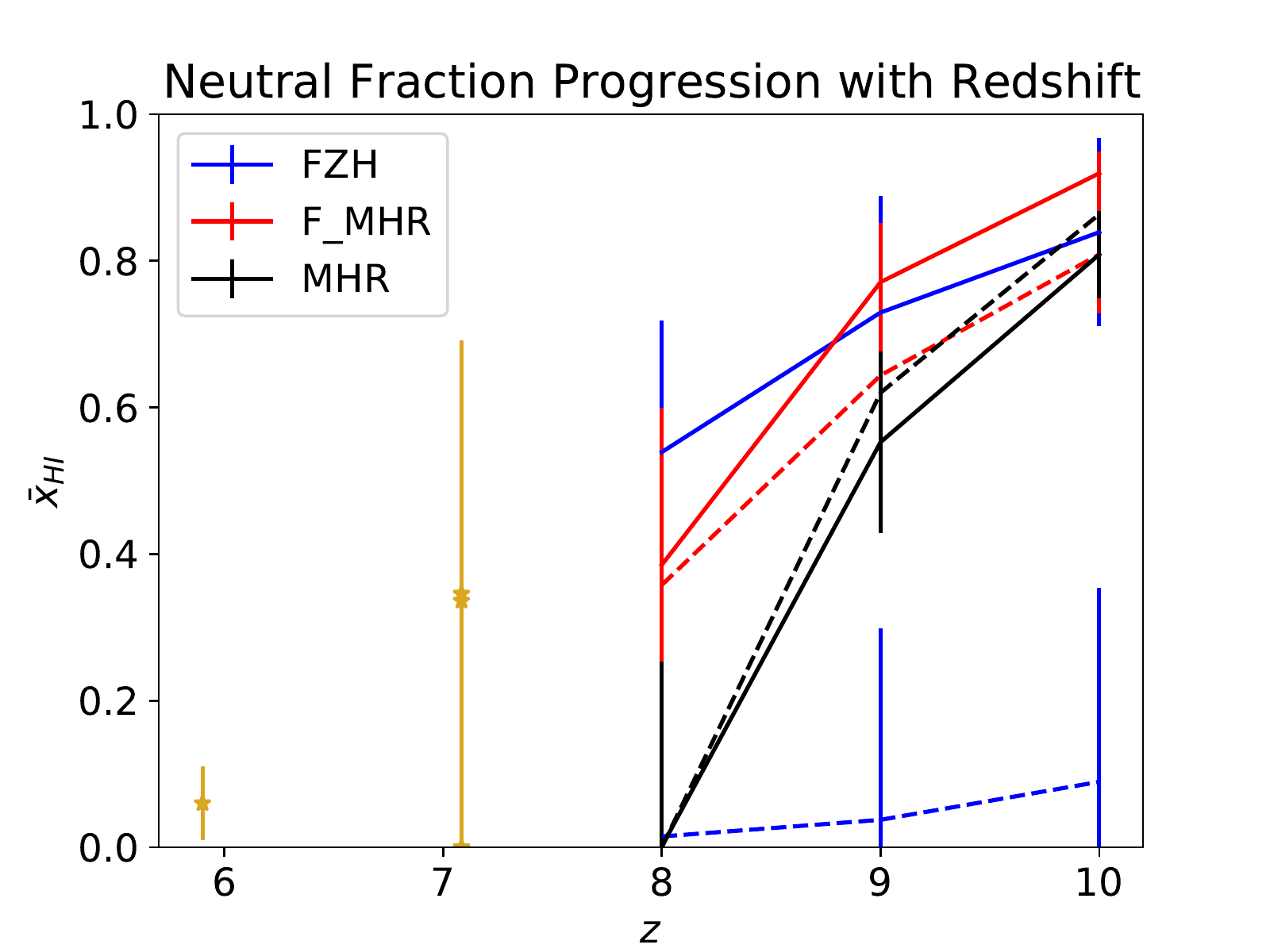}\hfill}
\caption{\ref{split_likelihoods} Due to the modular nature of the combined likelihood contributions (see Figure \ref{lhoodchain}) we can quantify the Bayesian Evidence for the 21cm power spectrum (by itself) compared to that of only the observational priors separately with HERA-331. 
Since the blue points lie close to zero and the red and black points are similar in all cases, this contribution is negligible. 
\ref{NF_interpolation} The progression of neutral fraction (reionisation histories) for all the toy models, due to the width of the error bars on the data points the observational priors penalise no models. 
The observing telescope has a negligible effect on the variance of neutral fractions obtained. 
The 4p FZH model fits within the error bars of the blue solid line (3pFZH) i.e. $\alpha$ has a small impact on the neutral fraction. 
The yellow points at $z=5.9$ and $7.1$ are the McGreer and Greig data points respectively with error bars derived from the respective observation (inverse models are represented by dotted lines of the colours in the legend). }
\end{figure}

In this section we discuss the observational prior contribution towards the total Bayesian Evidence values calculated for the HERA-331. 
Since the observational priors are linearly combined into the likelihood (Figure \ref{lhoodchain}) they are easily separable. 
By separating out the 21cm power spectrum contribution from the QSO data it becomes very apparent how little quantitative impact this observational data has on differentiating between models (the lack of deviation of the blue points from the black dashed line, Figure \ref{split_likelihoods}). 
Note that these blue points have been run separately from the simulation with only the 21cm power spectrum and can be fit easily by all of our toy models (Table \ref{j_ob_params}).
All of the toy models posses a region in their parameter space that agrees with the constraints of the observational priors.
To break this model degeneracy the 21cm power spectrum must be used.
An example of this can be seen from the MAP parameters of the Inv FZH model - 
the values obtained differ significantly (by a few $\sigma$, see Figure \ref{mydata2}) depending on what contributions to the likelihood are included. 
The MAP parameters found by the 21cm power spectrum can be heavily penalised by those found by the observational prior checks, and vice versa. 
The discrepancies of the black (full likelihood,  Figure \ref{split_likelihoods}) Evidence points compared with the red (only the 21cm Likelihood) show that these models are un-physical as they are penalised by these observational prior checks. 
Despite this the F Inv MHR model still proves the hardest to distinguish from FZH (with odds $\sim 10^7:1$), hence a better way of harnessing the non-21cm data is required if a cross check is to be performed.
How the observational prior checks are so poorly constraining can be visualised for the McGreer and Greig neutral fraction checks (Figure \ref{NF_interpolation}). 
The plotted points at the simulated redshifts are interpolated to see if they fit within the error bars on the observed neutral fractions (shown in gold). 
Since the error bars on these neutral fraction values are large it is difficult for them to add any constraining power to our toy models of the EoR. 
Similarly small likelihood contributions come from the Planck prior which utilised the optical depth.
This quantitative approach agrees with the discussion in \citet{2017MNRAS.465.4838G}.
When recovering f1, we indirectly observe this via $\sigma$ values that are of a comparable size to that obtained parameter value (see Table \ref{datapoints2}). 
This should be interpreted such that other data\footnote{This includes more numerous and precise neutral fraction measurements from quasars.}, as well as the 21cm power spectrum, is necessary to break model degeneracies for EoR inference.

A recent publication \citep{Luminosity21cmmc} attempts to harness more information via galaxy luminosity functions, in which the parameterisation of $\zeta_{0}$ in Equation \ref{ioneff} is expanded via power laws in $f_{\rm esc}$ and $f_*$. 
Equation \ref{eq: zetaB} is no longer used due to its inadequacy at reproducing observed UV luminosity functions.
We defer a model selection analysis including this improved framework to future work. 

\subsection{Varying the number of HERA dipoles}\label{sec: heradipoles}

 In this section we illustrate what number of dipoles are required for HERA to attain the model selection capabilities of LOFAR or the SKA. 
 As mentioned previously HERA is organised in a hexagonal structure with redundant baselines aligned for the purpose of noise reduction, for more details on this see  \citet{redundentantenna}. 
 HERA achieves signal to noise of 3, 9, 18, 28, 39, and 50 with 19, 61, 127, 217, 331, and 469 dipoles respectively at $z=8$ on the f1 power spectrum. 

\begin{figure}
    \centering      
    \subfigure[\label{MONEYhera}]{\includegraphics[scale=0.5]{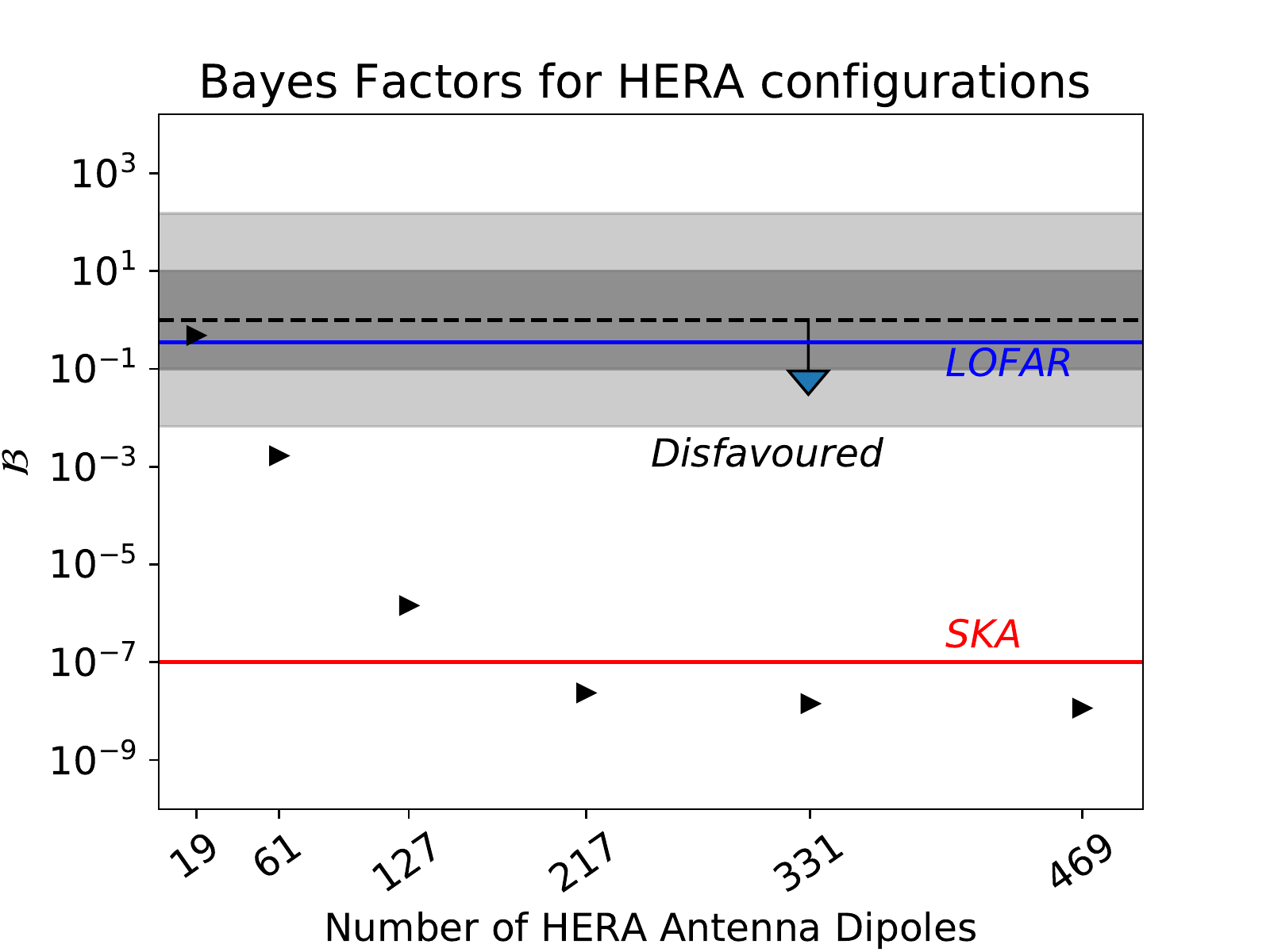}\hfill}
    \caption{This plot is similar to that for LOFAR in Figure \ref{MONEYlofar} but with varying HERA \change{configurations. 
    We} show the Bayes factors for F Inv MHR verses the fiducial 3pFZH mock data power spectrum f1.
    LOFAR (the blue line) and HERA-19 score \textit{weak} on the Jeffreys' scale while all telescopes (including HERA-61) score \textit{strong} results. 
    Above 217 dipoles HERA becomes more sensitive than the SKA (red line). 
   }
    \label{fig:heras}
\end{figure}

As shown in figure \ref{fig:heras} HERA-19 (the lowest dipole configuration) performs similarly to LOFAR. Model selection between the models we consider becomes \textit{strong} with HERA-61. HERA-217 is comparable to SKA (see Section \ref{sec: SKA}).

\subsection{How well will the SKA-512 perform at model selection?}\label{sec: SKA}

The analyses of the SKA-512 are shown in this section for a 1 hour tracked scan of the f1 power spectrum with 6 hours of observing time per day (split across 6 fields). 
We repeat the analyses that have been done for LOFAR-48 in Section \ref{sec: LOFAR}. 
At $z=8$ SKA achieves signal to noise of 4, 7, 10, 14, and 22 with 108, 216, 324, 540, and 1080 hours of integration time.

\begin{figure}
\centering
\includegraphics[scale=0.5]{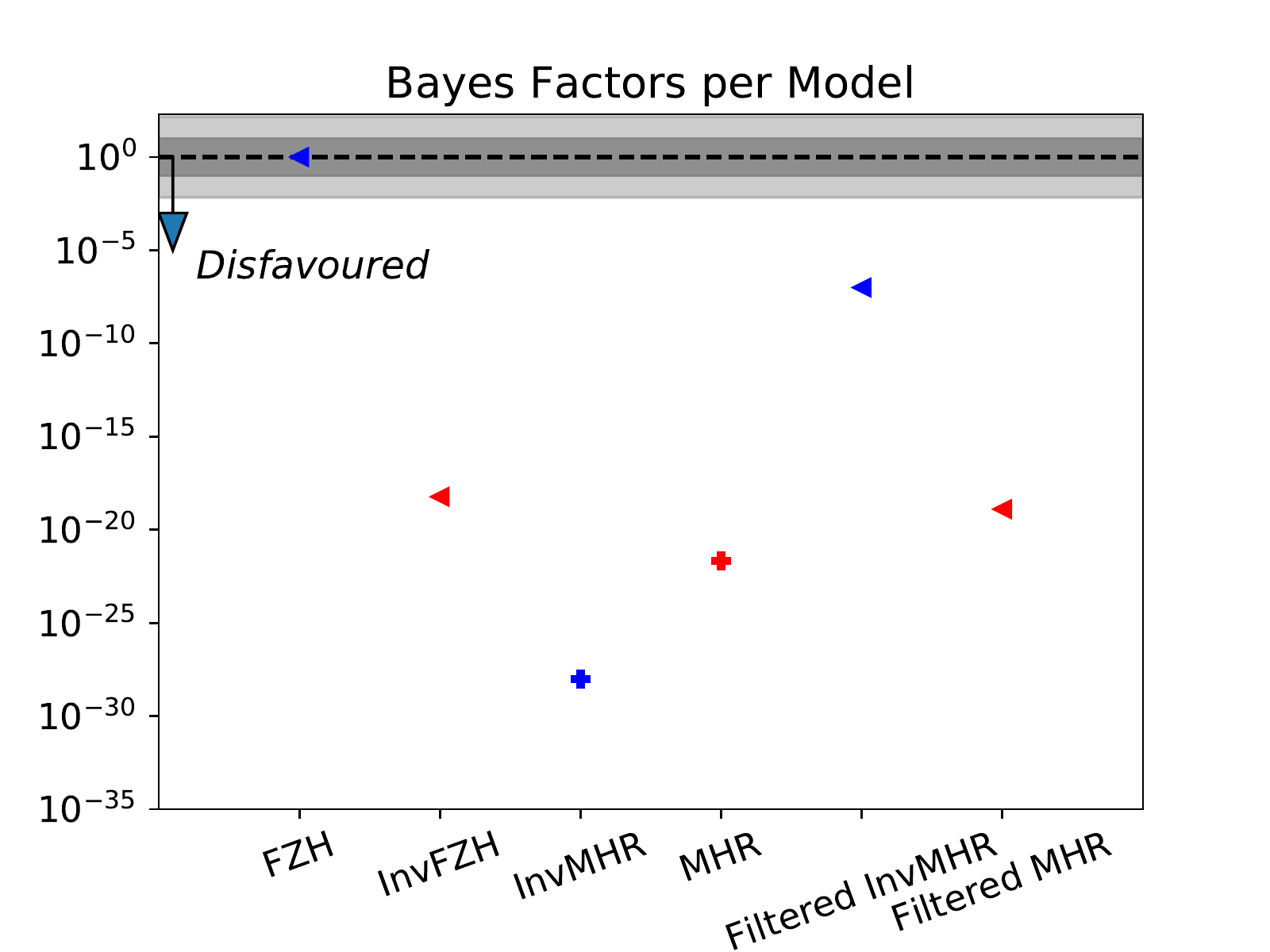}
\caption{\label{SKA_data}
The Bayes factors obtained by SKA with 1080 hours of integration time on the f1 power spectrum. All toy models are \textit{strongly} ruled out by margins similar to those obtained by HERA-331. For details of the plot style please see Figures \ref{fig: lofarbfactor} and \ref{fig: HERAbfactors}. 
}
\end{figure}

Similarly to Section \ref{sec: LOFAR}, we calculate the Bayes factors between 3pFZHf1 and the F Inv MHRf1, with decreasing the observation times. 
we assume the moderate foreground model for the wedge (and a buffer, Section \ref{sec:noise}) within 21cmSense. 
The toy EoR models we consider can be separated out with 324 hours of observing with \change{the} SKA.
Figure \ref{SKA_data} shows the Bayes factors obtained for the toy models considered. 
\change{The} F Inv MHR is used because it is the toy model that has been hardest to distinguish from 3pFZH in Figure \ref{fig: HERAbfactors}. 

In comparison to HERA, \change{SKA} has less repeated baselines and therefore has a marginally lower signal to noise. 
This is compensated for by having a wider range of baselines, meaning a fuller interferometric U-V coverage. 
Despite this SKA does not achieve tighter posterior contours on parameters, when considering the power spectrum - this might not remain the case when considering higher order statistics such as the bispectrum \citep{2019MNRAS.482.2653W}. 
For a direct comparison of HERA-331 and SKA's observational error on the power spectrum see Figure \ref{fig: errors}.

In conclusion, SKA performs comparably at EoR model selection purposes as HERA-217. 



\begin{figure}
    \centering      
    \subfigure[\label{MONEYska}]{\includegraphics[scale=0.5]{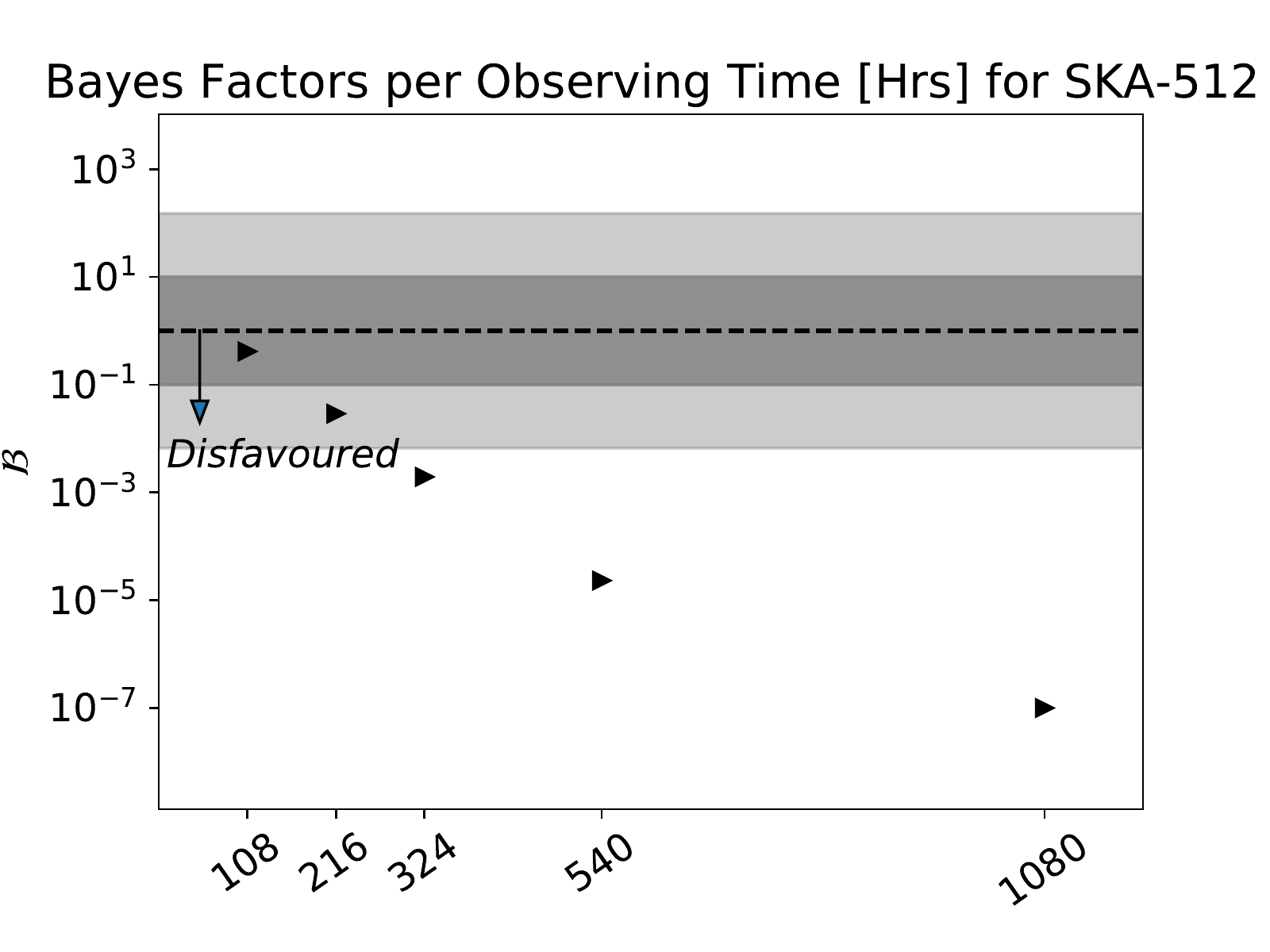}\hfill}
    \caption{\label{fig:skas} This figure follows the same structure as Figures \ref{fig:heras} and \ref{MONEYlofar} but with SKA-512 observing times. 
    The Bayes factors are plotted for F Inv MHR against FZH for the f1 mock data set. 
    Note that in order to obtain a \textit{strong} distinction between these toy models SKA must observe for $\sim 324$ hours. 
    }
\end{figure}

\subsection{Model selection on the number of parameters in 21cmFAST}\label{sec: 21cmfastmodels}

\begin{figure}
\includegraphics[scale=0.5]{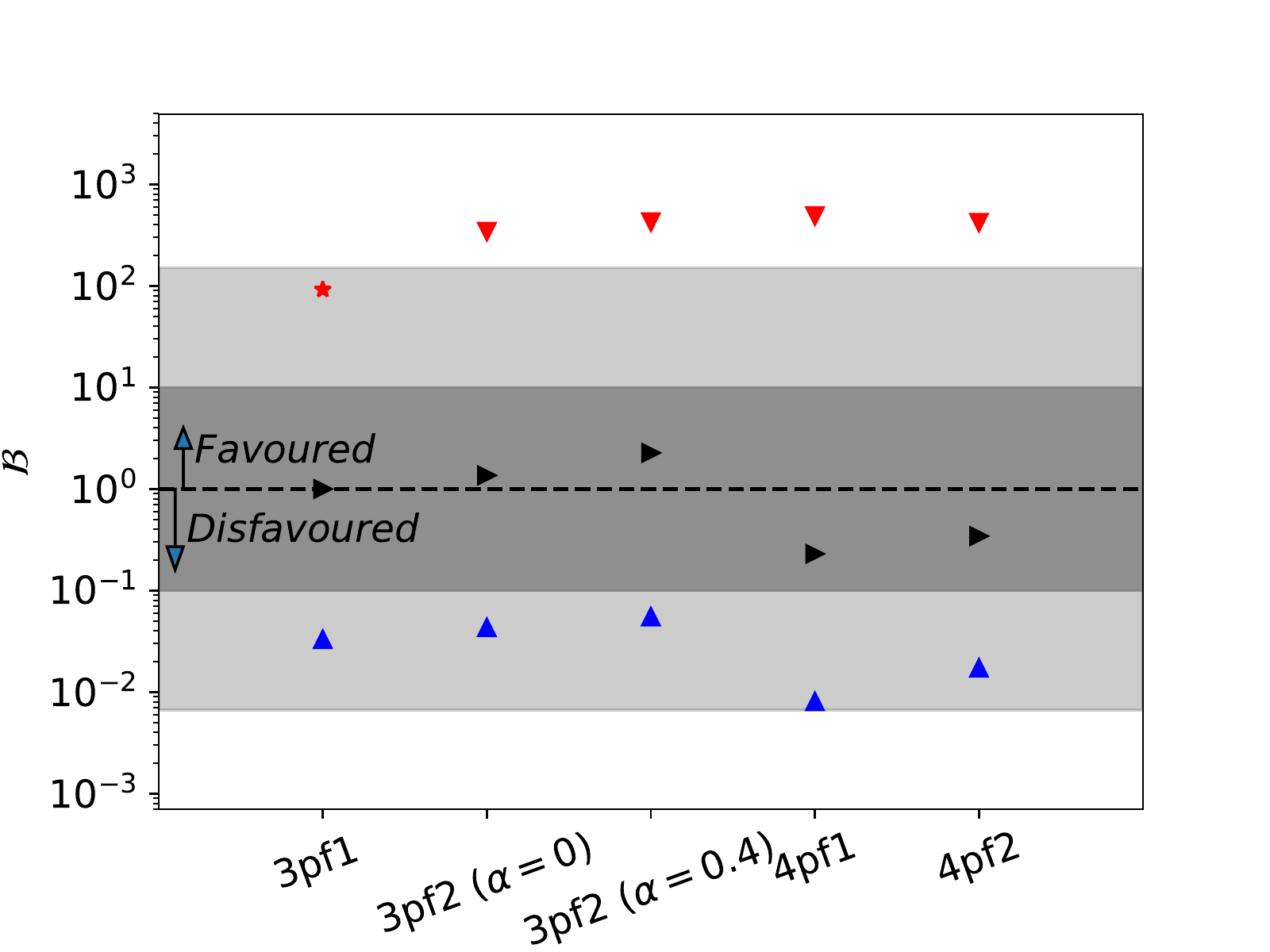}
\caption{ \label{fig: bfactors+bias} The Bayes factors for all FZH based models relative to 3pFZHf1 ($\alpha=0.$). The \textit{weak}, \textit{moderate} and \textit{strong} scores on the Jeffreys' scale are indicated by the dark grey, light grey and white regions respectively. Black points indicate the use of sensible prior volumes - all of 21cmFAST models considered are shown to score \textit{weak} results on the Jeffreys' scale. The secondary aim of this plot is to show the effect of a prior skewing Evidence results. Red indicates the use of a $\delta$-function across either the fiducial values (starred point - 3pf1 only) or the MAP values (red arrows). See Section \ref{sec: 21cmfastmodels} for the discussion as to why the MAP parameters are favoured above the fiducial choice. Blue points have their priors widened to skew against the models (see text for details). Note that it is possible for this skewing of Bayes factor to achieve  \textit{strong} evidence scores - the Jeffreys' scale is intended as a guide only.}
\end{figure}

Having shown that HERA \& SKA can easily determine the most suitable reionisation scenario we now explore refining the number of parameters. Namely testing the ability to determine the use of a 3 or 4 parameter FZH model (detailed in Section \ref{sec: 21cmmc}).
We then proceed to inspect the effect of skewing these results with a naive choice of prior. 

Figure \ref{fig: bfactors+bias} shows the Bayes factors for 3p and 4p FZH models tested with $\delta$ function, used (Table \ref{priortable}) and widened priors in red, black and blue respectively. This figure therefore serves both aforementioned purposes. 

Primarily, the black triangles show that the different parameterisations of 21cmFast considered score \textit{weakly} on the Jeffrey's scale with HERA-331. 
In other words the power spectrum from this 1080 hour observation is not sensitive to changes of the fiducial parameters - nor the change of a constant $\zeta$ to a power law in halo mass (Equations \ref{eq: zetaA} and \ref{eq: zetaB}). 
Even when increasing the SKA's observation times (not shown) we score \textit{weak} results recovering the f1 and f2 power spectra with the 3p and 4p FZH models. 
Applying larger values of $\alpha$ makes these models more distinguishable however this is because the observational priors become dominant in the Evidence.
Increasing $\alpha$ increases the ionising efficiency of the galaxies (Equation \ref{ioneff}) and therefore the EoR will finish too early if $\alpha$ is largely positive. 


Secondly, Figure \ref{fig: bfactors+bias} contains the information obtained when attempting to skew our own results with prior choice (see Section \ref{sec: Pconsiderations} for the full discussion) - this is the only way one can be sure the results are prior independent. 
Namely the blue triangles have uniform priors expanded to reduce the posterior density (and therefore the Evidence).
The red shapes have the prior volume reduced to a $\delta$ function for the opposite effect. 
These blue results use unphysical prior ranges and still fall short of the \textit{strong} Bayes factor score; we consider this a success of the machinery. 
In other words if one chose wide uniform prior distributions out of naivety (a sensible approach for a first attempt), the result could be \textit{moderate}. 
The user should look closer at the physics at hand and try to reduce the wide priors (chosen for the blue points) to those resembling Table \ref{priortable} - which were chosen via physical reasoning and familiarisation with the problem (used for the black points).
It is worth noting that the interpolated shape of the black points is translated down for the blue points - in the uniform case the parameter prior volume is factored out of the Evidence integral as a constant (Equation \ref{Z}).

For the red shapes (using $\delta$ function priors), the maximum possible Evidence is obtained by choosing the MAP parameters, $\mathcal{Z} = \mathcal{L}(\theta_{\rm MAP})$, and is observed by the alignment of the red triangles in Figure \ref{fig: bfactors+bias}. 
The starred red point (only above 3pf1) is calculated with the fiducial parameters. 
It is worse than this maximum because no consideration of the observational priors was made when choosing these values. 
Therefore there is a disagreement between the parameters that produce the fiducial power spectrum and those satisfying the observational priors. 
This was also observed when cross checking the methodology against the Inv MHR fiducial data set in Section \ref{sec: f3} (and is discussed in Section \ref{sec: 21cmvsQSOs}).
It is worth noting here that if a prior is chosen to be too narrow, the posterior plots will show close to uniform distributions and suspicion should be applied - either the parameters dynamic range is not expressed by the chosen prior (the case here), or the parameter is redundant (Section \ref{sec: nested parameters}).

\subsubsection{The Savage-Dickey density ratio: how useful are a model's parameters?}\label{sec: nested parameters}

\begin{figure*}
    \centering
    \subfigure[\label{R_PSs}]{\includegraphics[width=0.4\textwidth]{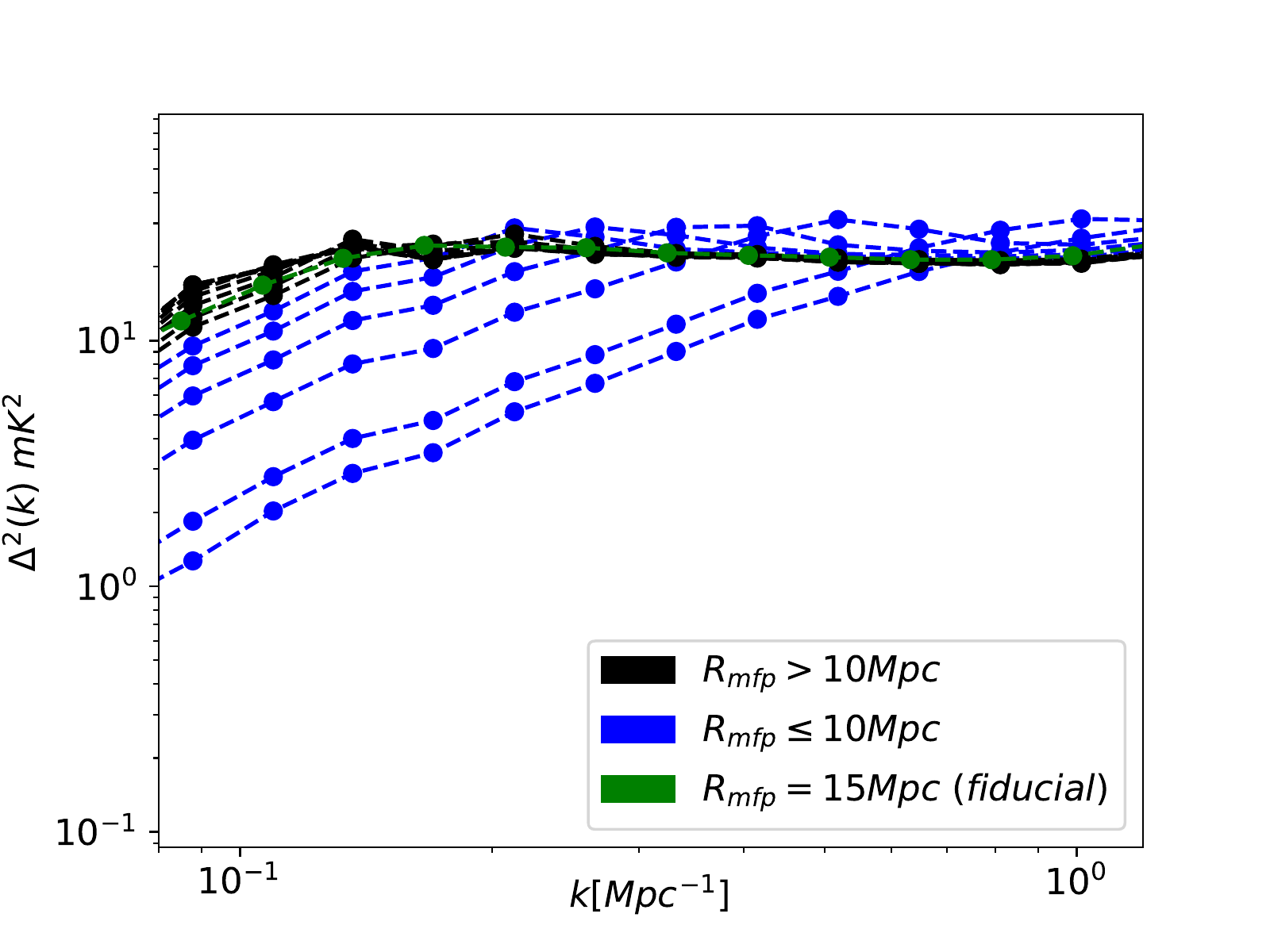}}
    \subfigure[\label{R_Bfactors}]{\includegraphics[width=0.4\textwidth]{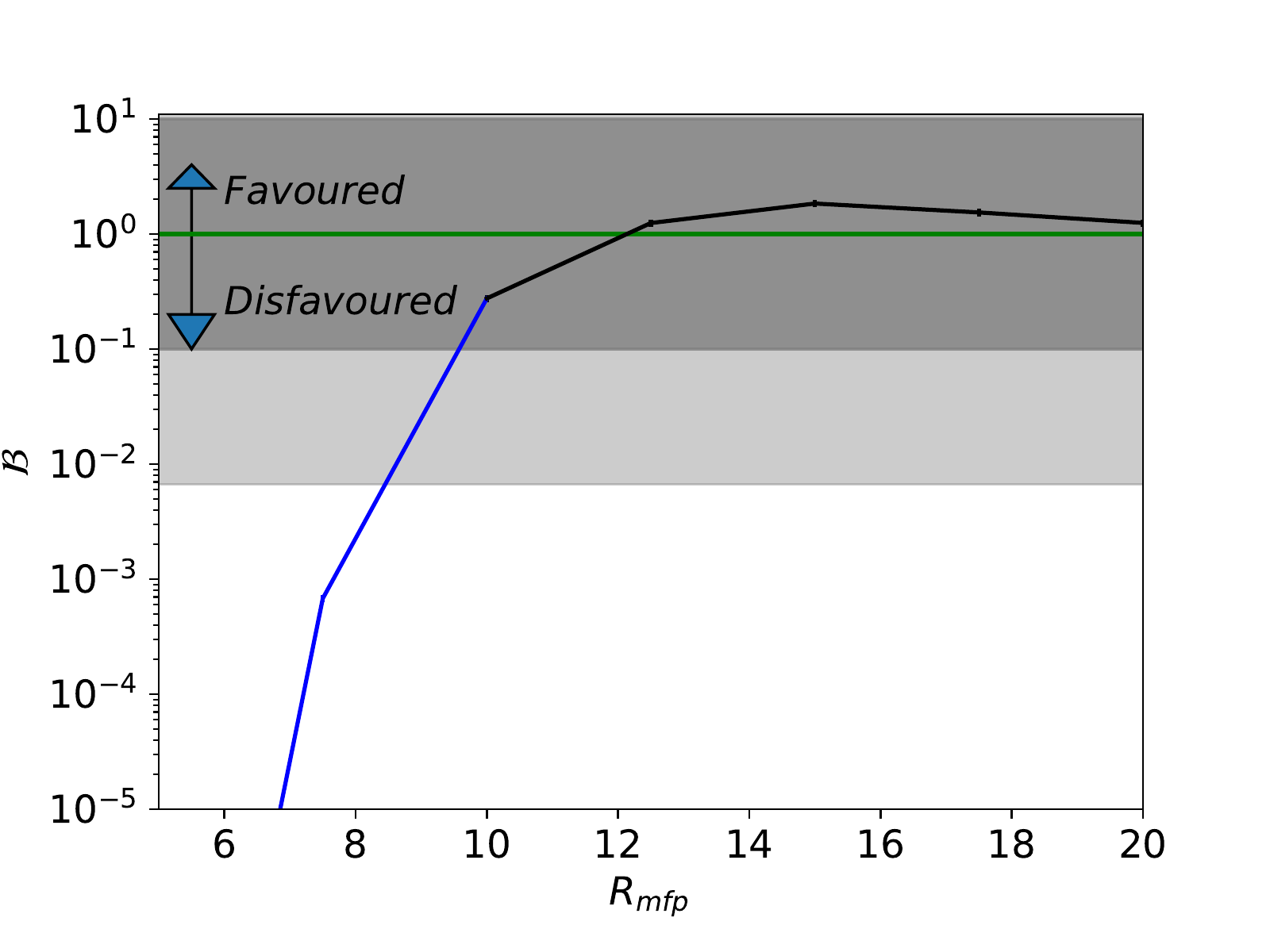}}
    \subfigure[\label{Z_PSs}]{\includegraphics[width=0.4\textwidth]{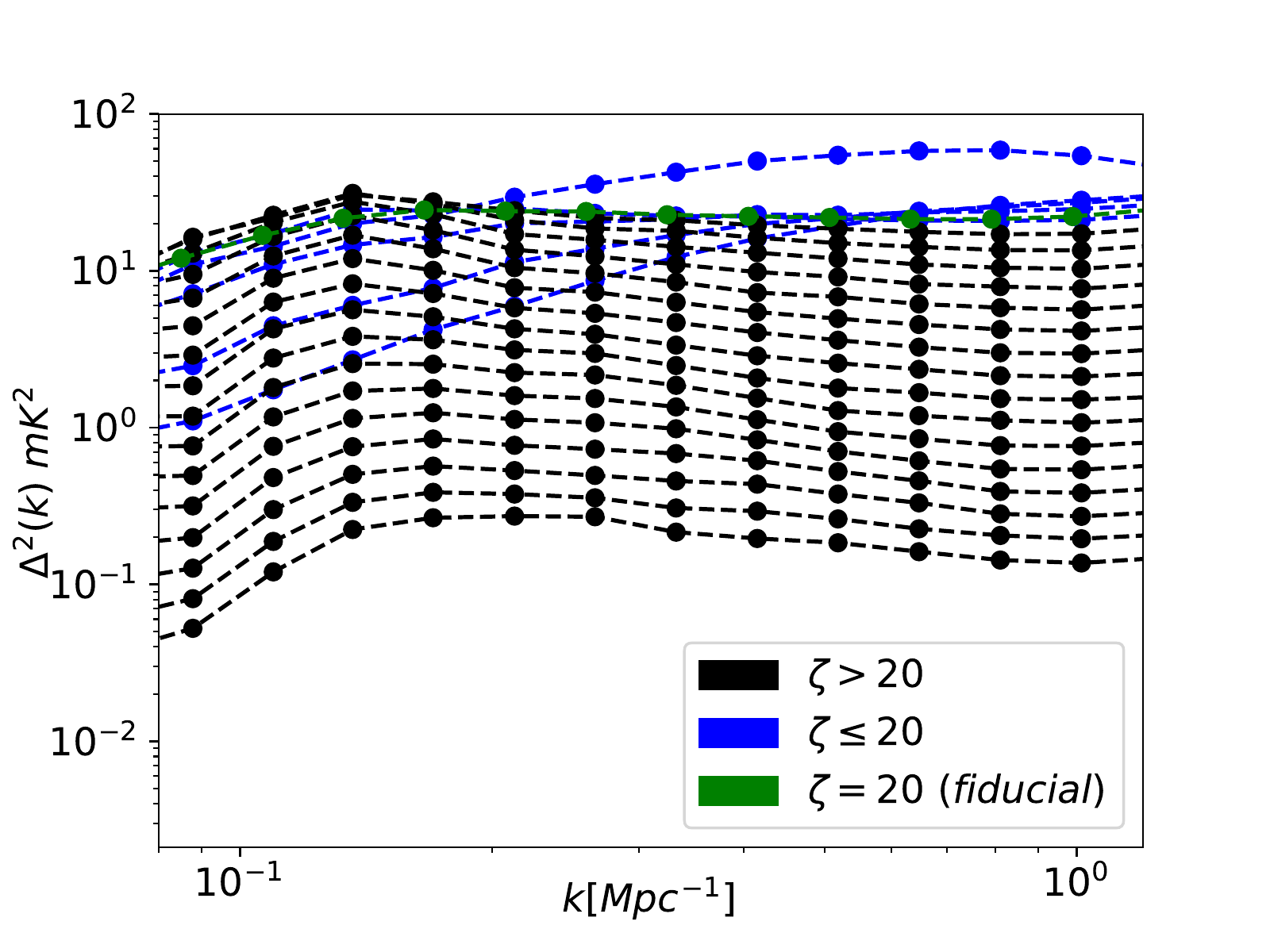}}
    \subfigure[\label{Z_Bfactors}]{\includegraphics[width=0.4\textwidth]{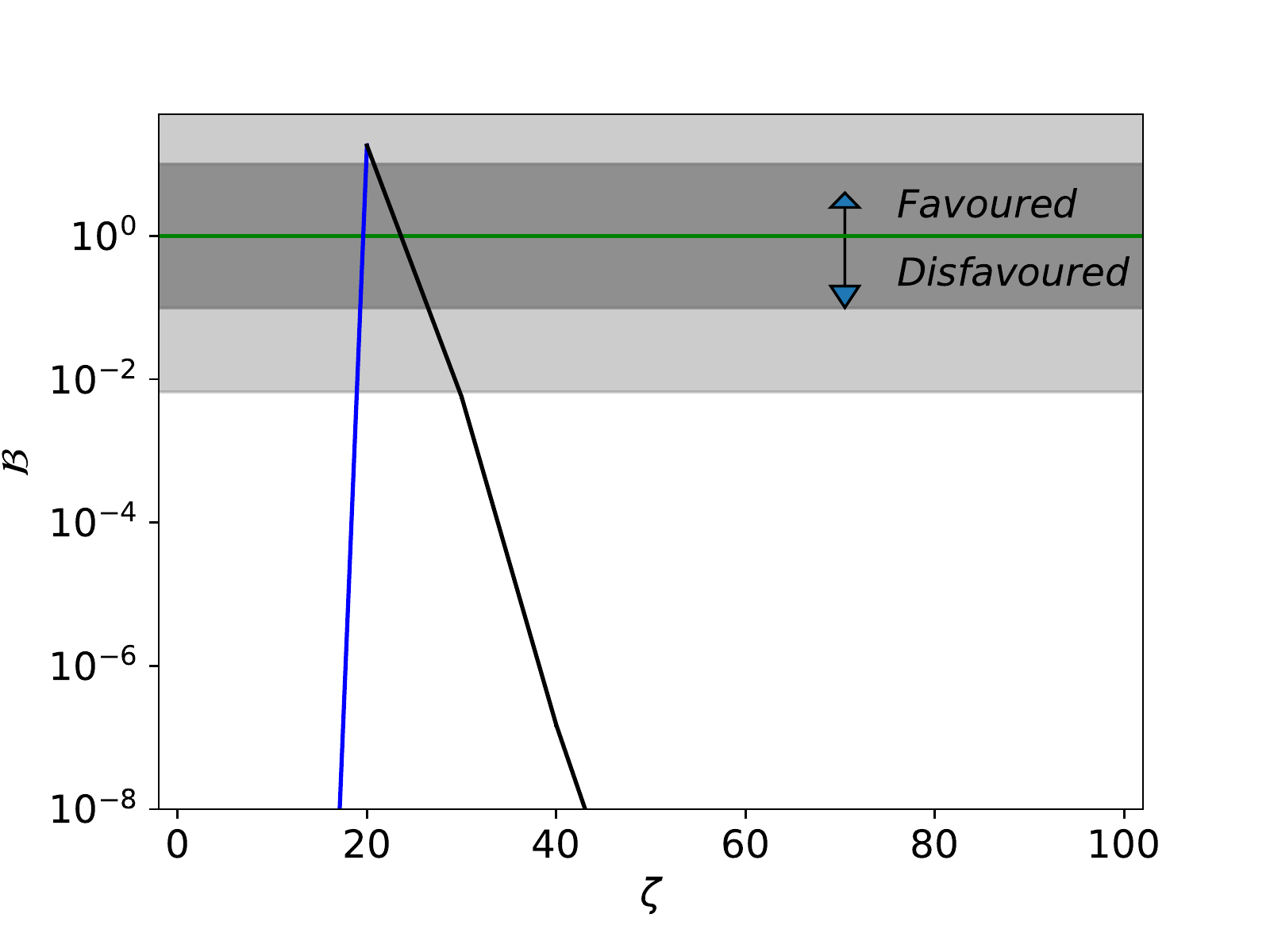}}
    \caption{ Using the Savage-Dickey Density ratio - The impact of varying one fixed parameter as a set of models nested from 3pf1 on the Bayes factor. 
Plots \ref{R_PSs} and \ref{R_Bfactors} both have $R_{\rm mfp}$ fixed, while \ref{Z_PSs} and \ref{Z_Bfactors} fix $\zeta$. 
The colour coding is consistent across all plots in this figure: the green indicates the fiducial simulation. 
Plots \ref{R_Bfactors} and \ref{Z_Bfactors} are the Bayes Factor plots, with \textit{strong}, \textit{moderate} and \textit{weak} scores indicated by the white, light and dark grey regions respectively.
\ref{R_PSs} and \ref{Z_PSs} show how the power spectra vary for $z=8$ given the different nested model (The plots for $z=9$ and $10$ are similar). 
Notice that for $R_{\rm mfp} > 10 \rm Mpc$, the Bayes factor provided \protect\textit{weak} inference, but \protect$\zeta$ shows \protect\textit{strong} results against all but the fiducial values.}
\label{2p_RZ}
\end{figure*}

Here we explore redundancy of parameters within models via the Bayesian Evidence (see Section \ref{S-D} for the methodology). 
We consider each parameter of the 3pf1 simulation in turn ($R_{\rm mfp}$ and  $\zeta$ are shown in Figure \ref{2p_RZ}). 
Since the $R_{\rm mfp}$ parameter is actually the maximum smoothing scale (a top-hat in k-space via the excursion-set formalism to be precise, see Section \ref{sec: 21cmmc}) it is an easy target for this evaluation. 
Above a certain (model dependent) threshold the maximum smoothing scale will capture all of the structural detail within the model. 
Below this scale we expect the model to perform poorly in model selection (against a larger $R_{\rm mfp}$) due to a loss of large-scale structure. 
As is shown (Figure \ref{R_PSs}, the power spectrum at different $R_{\rm mfp}$) by the descending left hand side of the blue power spectra while the right hand side remains fixed. 
This corresponds to the Bayes factor results $\mathcal{B}(R_{\rm mfp})$ from Equation \ref{eq: sddr} is plotted in Figure \ref{R_Bfactors}. 
All values of $R_{\rm mfp}>10$ agree with the fiducial results in green - showing negligible inference is provided by varying $R_{\rm mfp}$ in this range.
For a comparison $\zeta$ produces no degenerate power spectra and is therefore a useful parameter (Figure \ref{Z_PSs}, the power spectrum at different $\zeta$). 
In the Bayes factor results, $\mathcal{B}(\zeta)$ from Equation \ref{eq: sddr} is plotted in Figure \ref{Z_Bfactors}, this leads to the sharp peaked value that agrees with the fiducial result shown in green. 
When $\zeta$ is small, the power spectra show that the simulation attempts to compensate by increasing the number of small structures (the blue power spectrum is the largest at the small scale end of Figure \ref{Z_PSs}). 
On the other hand if galaxies have a high ionising efficiency the simulation responds by decreasing the size of structures on all scales. 
This can be identified from the shape of each black power spectrum being maintained as the amplitude decreases. 
Within all of $\zeta$'s prior range, there are dynamic changes to the simulation that are observable with the power spectrum.

The lines in Plots \ref{R_Bfactors} and \ref{Z_Bfactors} rise above the green fiducial result due to the reduction in prior volume of the parameter being nested. 
In \ref{Z_Bfactors} the evidence rises above the \textit{moderate} Jeffreys' scale threshold - an even wider choice of uniform prior distribution for this parameter could have provided \textit{strong} evidence however this should not be taken seriously. 
This draws light on the implicit nature of Occam's razor within the Bayesian Evidence, as mentioned in Section \ref{sec: jeff}. 
Similarly with $R_{\rm mfp}$, we are able to see the increase in Bayes factor as the number of parameters (and therefore prior volume) is reduced. 
Here we obtain an \textit{weak} scoring on the Jeffrey's scale due to a smaller prior distribution ($[5,20]\rm Mpc$ compared with $\zeta$'s $[5,100]$).
If in doubt, comfort is found in the variability of the power spectra per parameter as this quantitatively reveals the dynamic range of that parameter - \ref{R_PSs} and \ref{Z_PSs}.
Figures \ref{R_Bfactors} and \ref{Z_Bfactors} reflect the posterior distributions of each parameter, as derived in Section \ref{S-D}.

\section{Conclusions}
\label{sec: conclusion}

Initially we show that Multinest can produce matching posterior distributions to those produced by Cosmohammer in 21CMMC (Section \ref{sec: agreement}).

The literature warns of two potential heffalump traps within Bayesian model selection: these are the prior dependency of results and the overcompensation of redundant parameters via Occam's Razor. 
We address carefully these issues in Sections \ref{sec: jeff} and \ref{sec: Pconsiderations} to show the validity of the Bayesian model selection we apply. For a rough interpretation of our results we adopt the Jeffreys scale to distinguish \textit{weak}, \textit{moderate} and \textit{strong} conclusions.

This work considers toy EoR models as well as those in the original 21CMMC model. Namely: an inverted version of the original 3 parameter FZH (\textit{global inside-out}) model used in 21cmFAST (within which the excursion set formalism solves for reionised bubbles); a simpler model MHR (\textit{local outside-in}), in which ionisation is defined via a density criteria; both of these models have had their ionisation criterion mathematical inverted (Inv FZH, \textit{global outside-in}; and Inv MHR, \textit{local inside-out}); and finally a 3rd parameter is added to both the MHR and Inv MHR models in the form of a Gaussian density field filter - F MHR, \textit{global outside-in}; and F Inv MHR (\textit{global inside-out}). By using the specifications of HERA-331 (see Table \ref{Tscopetable}) for 1080 hour observations we can distinguish these toy models \textit{strongly} on the Jeffreys' scale (Section \ref{sec: further}).

By calculating Evidences for the separate likelihood contributions of 21CMMC - we quantitatively show that more QSO data is required for effective non-21cm observational checks than only that of neutral fractions (surrounging high $z$ quasars) and the optical depth (Section \ref{sec: 21cmvsQSOs}).
In future work, we will quantitatively evaluate the inference capabilities of including luminosity function data \change{- following the methodology of \citet{Luminosity21cmmc}}.

On a subtler level the considered version of 21CMMC contains the addition of a 4th parameter, enabling a mass dependent ionising efficiency, in other words relaxing the assumption of a constant mass to light ratio for the simulated galaxies. We show in Section \ref{sec: 21cmfastmodels} that these cannot be distinguished with 1080 hour observations as they score \textit{weak} results on the Jeffreys' scale, as does changing the fiducial parameters within these models.

Using the Savage-Dickey density ratio we proceeded to show the redundancy of the $R_{\rm mfp} > 10 h^{-1}~\rm Mpc$ in 21CMMC (Section \ref{sec: nested parameters}) - hence justifying its omission from the newest parameterisations of 21CMMC and 21cmFAST \change{in favour of inhomogeneous recombinations \citep{sobacchimesinger2014}}.

The most difficult to distinguish of the toy models is F Inv MHR because it shares scale and morphology with FZH. We therefore use it as a bench mark for our primary questions on feasibility:\footnote{Assuming the `moderate' foreground wedge model within 21cmSense in all cases (Section \ref{sec:noise}).} 
\begin{itemize}
    \item  LOFAR-48 would struggle to perform model selection. We  require 21600 hours of observation to \textit{strongly} disfavour the toy models (with odds $\sim 400:1$).
    \item  Using 1080 observing hours, HERA would require configurations of at least 61 dipoles in order to rule out any of the toy models with \textit{strong} evidence on the Jeffreys' scale (scoring odds of $\sim 600:1$).
    \item  Finally, with the SKA-512: The \textit{strong} rejection of the considered EoR toy models can be achieved quickly with 324 observing hours at $\sim 500:1$.
\end{itemize}

At observing the 21cm EoR power spectrum HERA-217 becomes comparable to the SKA. 
As can be seen in Figure \ref{fig: errors}, SKA is better at observing small scales while HERA dominates the sensitivity at large scale. 
Assuming the \textit{moderate} noise settings within 21cmSense can be acheived, the redundant baseline instrumental method is very promising at performing model selection with the 21cm EoR power spectrum \citep{2018arXiv181101378B, 2019AAS...23341301D}.

We have now set the scene for ruling out toy models of the EoR with Bayesian Model Selection.
This work shows that modest 21cm experiments, such as HERA-61, are likely to observationally pin down the correct morphology and scale of reionisation.
In future work we intend to look more closely at the level of precision that can be obtained from more involved models of the EoR using the newer versions of 21CMMC. 
An updated version of 21cmNest\footnote{\textcolor{blue}{https://binnietom@bitbucket.org/binnietom/21cmnest\_1.0.git}} will be released soon.

Once the desired reionisation scenario can be quantitatively chosen, parameter inference will be performed rigorously to infer the involved astrophysics. 


\section*{Acknowledgments}
\change{The authors would like to especially thank Catherine Watkinson for her contributions to an earlier draft of this work.} 
We would also like to thank Daniel Mortlock, Conor O'Riordan, Brad Greig, Andrei Mesinger, Emma Chapman, Roberto Trotta, Alan Heavens, Suman Majumdar and the anonymous referee for their helpful conversations and suggestions. 
We acknowledge use of Linux OS, C, C++, Fortran90, gnu, gsl, Overleaf, GitHub, Python, and the Python packages numpy, scipy, cornerplot, matplotlib. 
T Binnie thanks STFC for their studentship funding. 
JRP is pleased to acknowledge support from the European Research Council under ERC grant number 638743-FIRSTDAWN. 

\bibliography{21CMN}

\begin{thebibliography}{}
\makeatletter
\relax
\def\mn@urlcharsother{\let\do\@makeother \do\$\do\&\do\#\do\^\do\_\do\%\do\~}
\def\mn@doi{\begingroup\mn@urlcharsother \@ifnextchar [ {\mn@doi@}
  {\mn@doi@[]}}
\def\mn@doi@[#1]#2{\def\@tempa{#1}\ifx\@tempa\@empty \href
  {http://dx.doi.org/#2} {doi:#2}\else \href {http://dx.doi.org/#2} {#1}\fi
  \endgroup}
\def\mn@eprint#1#2{\mn@eprint@#1:#2::\@nil}
\def\mn@eprint@arXiv#1{\href {http://arxiv.org/abs/#1} {{\tt arXiv:#1}}}
\def\mn@eprint@dblp#1{\href {http://dblp.uni-trier.de/rec/bibtex/#1.xml}
  {dblp:#1}}
\def\mn@eprint@#1:#2:#3:#4\@nil{\def\@tempa {#1}\def\@tempb {#2}\def\@tempc
  {#3}\ifx \@tempc \@empty \let \@tempc \@tempb \let \@tempb \@tempa \fi \ifx
  \@tempb \@empty \def\@tempb {arXiv}\fi \@ifundefined
  {mn@eprint@\@tempb}{\@tempb:\@tempc}{\expandafter \expandafter \csname
  mn@eprint@\@tempb\endcsname \expandafter{\@tempc}}}

\bibitem[\protect\citeauthoryear{Akaike}{Akaike}{1974}]{1100705}
Akaike H.,  1974, \mn@doi [IEEE Transactions on Automatic Control]
  {10.1109/TAC.1974.1100705}, 19, 716

\bibitem[\protect\citeauthoryear{{Akeret}, {Seehars}, {Amara}, {Refregier}  \&
  {Csillaghy}}{{Akeret} et~al.}{2013}]{2013A&C.....2...27A}
{Akeret} J.,  {Seehars} S.,  {Amara} A.,  {Refregier} A.,   {Csillaghy} A.,
  2013, \mn@doi [Astronomy and Computing] {10.1016/j.ascom.2013.06.003}, \href
  {http://adsabs.harvard.edu/abs/2013A%26C.....2...27A} {2, 27}

\bibitem[\protect\citeauthoryear{{Ali} et~al.,}{{Ali}
  et~al.}{2015}]{2015ApJ...809...61A}
{Ali} Z.~S.,  et~al., 2015, \mn@doi [\apj] {10.1088/0004-637X/809/1/61}, \href
  {http://adsabs.harvard.edu/abs/2015ApJ...809...61A} {809, 61}

\bibitem[\protect\citeauthoryear{{Baek, S.}, {Semelin, B.}, {Di Matteo, P.},
  {Revaz, Y.}  \& {Combes, F.}}{{Baek, S.} et~al.}{2010}]{baek}
{Baek, S.} {Semelin, B.} {Di Matteo, P.} {Revaz, Y.}  {Combes, F.} 2010,
  \mn@doi [A\&A] {10.1051/0004-6361/201014347}, 523, A4

\bibitem[\protect\citeauthoryear{{Barkana} \& {Loeb}}{{Barkana} \&
  {Loeb}}{2001}]{barkanaloeb2001}
{Barkana} R.,  {Loeb} A.,  2001, \mn@doi [\physrep]
  {10.1016/S0370-1573(01)00019-9}, \href
  {http://adsabs.harvard.edu/abs/2001PhR...349..125B} {349, 125}

\bibitem[\protect\citeauthoryear{{Barkana} \& {Loeb}}{{Barkana} \&
  {Loeb}}{2005}]{barkanaloeb2005}
{Barkana} R.,  {Loeb} A.,  2005, \mn@doi [\apj] {10.1086/429954}, \href
  {http://adsabs.harvard.edu/abs/2005ApJ...626....1B} {626, 1}

\bibitem[\protect\citeauthoryear{{Beardsley}, {Morales}, {Lidz}, {Malloy}  \&
  {Sutter}}{{Beardsley} et~al.}{2015}]{2015ApJ...800..128B}
{Beardsley} A.~P.,  {Morales} M.~F.,  {Lidz} A.,  {Malloy} M.,   {Sutter}
  P.~M.,  2015, \mn@doi [\apj] {10.1088/0004-637X/800/2/128}, \href
  {http://adsabs.harvard.edu/abs/2015ApJ...800..128B} {800, 128}

\bibitem[\protect\citeauthoryear{{Bond}, {Cole}, {Efstathiou}  \&
  {Kaiser}}{{Bond} et~al.}{1991}]{excursionset}
{Bond} J.~R.,  {Cole} S.,  {Efstathiou} G.,   {Kaiser} N.,  1991, \mn@doi
  [\apj] {10.1086/170520}, \href
  {http://adsabs.harvard.edu/abs/1991ApJ...379..440B} {379, 440}

\bibitem[\protect\citeauthoryear{Bowman, Rogers, Monsalve, Mozdzen  \&
  Mahesh}{Bowman et~al.}{2018}]{2bc8363417fb44159194d3775d6e51bd}
Bowman J.,  Rogers A.,  Monsalve R.,  Mozdzen T.,   Mahesh N.,  2018, \mn@doi
  [Nature] {10.1038/nature25792}, 555, 67

\bibitem[\protect\citeauthoryear{{Byrne} et~al.,}{{Byrne}
  et~al.}{2018}]{2018arXiv181101378B}
{Byrne} R.,  et~al., 2018, arXiv e-prints, \href
  {http://adsabs.harvard.edu/abs/2018arXiv181101378B} {}

\bibitem[\protect\citeauthoryear{{Chapman} et~al.,}{{Chapman}
  et~al.}{2012}]{2012MNRAS.423.2518C}
{Chapman} E.,  et~al., 2012, \mn@doi [\mnras]
  {10.1111/j.1365-2966.2012.21065.x}, \href
  {http://adsabs.harvard.edu/abs/2012MNRAS.423.2518C} {423, 2518}

\bibitem[\protect\citeauthoryear{{Chen} \& {Miralda-Escud{\'e}}}{{Chen} \&
  {Miralda-Escud{\'e}}}{2004}]{ChenMiralde-Ecude}
{Chen} X.,  {Miralda-Escud{\'e}} J.,  2004, \mn@doi [\apj] {10.1086/380829},
  \href {http://adsabs.harvard.edu/abs/2004ApJ...602....1C} {602, 1}

\bibitem[\protect\citeauthoryear{{Dayal} \& {Ferrara}}{{Dayal} \&
  {Ferrara}}{2018}]{2018PhR...780....1D}
{Dayal} P.,  {Ferrara} A.,  2018, \mn@doi [\physrep]
  {10.1016/j.physrep.2018.10.002}, \href
  {http://adsabs.harvard.edu/abs/2018PhR...780....1D} {780, 1}

\bibitem[\protect\citeauthoryear{{Dayal}, {Ferrara}, {Dunlop}  \&
  {Pacucci}}{{Dayal} et~al.}{2014}]{2014MNRAS.445.2545D}
{Dayal} P.,  {Ferrara} A.,  {Dunlop} J.~S.,   {Pacucci} F.,  2014, \mn@doi
  [\mnras] {10.1093/mnras/stu1848}, \href
  {http://ukads.nottingham.ac.uk/abs/2014MNRAS.445.2545D} {445, 2545}

\bibitem[\protect\citeauthoryear{{DeBoer} et~al.,}{{DeBoer}
  et~al.}{2017}]{2017PASP..129d5001D}
{DeBoer} D.~R.,  et~al., 2017, \mn@doi [\pasp]
  {10.1088/1538-3873/129/974/045001}, \href
  {http://adsabs.harvard.edu/abs/2017PASP..129d5001D} {129, 045001}

\bibitem[\protect\citeauthoryear{Dickey}{Dickey}{1971}]{10.2307/2958475}
Dickey J.~M.,  1971, The Annals of Mathematical Statistics, 42, 204

\bibitem[\protect\citeauthoryear{Dillon \& Parsons}{Dillon \&
  Parsons}{2016}]{redundentantenna}
Dillon J.~S.,  Parsons A.~R.,  2016, The Astrophysical Journal, 826, 181

\bibitem[\protect\citeauthoryear{{Dillon} et~al.,}{{Dillon}
  et~al.}{2015}]{2015PhRvD..91l3011D}
{Dillon} J.~S.,  et~al., 2015, \mn@doi [\prd] {10.1103/PhysRevD.91.123011},
  \href {http://adsabs.harvard.edu/abs/2015PhRvD..91l3011D} {91, 123011}

\bibitem[\protect\citeauthoryear{{Dillon}, {Orosz}, {Parsons}, {Ewall-Wice}  \&
  {Thyagarajan}}{{Dillon} et~al.}{2019}]{2019AAS...23341301D}
{Dillon} J.~S.,  {Orosz} N.,  {Parsons} A.,  {Ewall-Wice} A.,   {Thyagarajan}
  N.,  2019, in American Astronomical Society Meeting Abstracts 233. p. 413.01

\bibitem[\protect\citeauthoryear{{Fan} et~al.,}{{Fan}
  et~al.}{2006}]{2006AJ....132..117F}
{Fan} X.,  et~al., 2006, \mn@doi [\aj] {10.1086/504836}, \href
  {http://adsabs.harvard.edu/abs/2006AJ....132..117F} {132, 117}

\bibitem[\protect\citeauthoryear{{Feroz}, {Hobson}  \& {Bridges}}{{Feroz}
  et~al.}{2009}]{Multinest}
{Feroz} F.,  {Hobson} M.~P.,   {Bridges} M.,  2009, \mn@doi [\mnras]
  {10.1111/j.1365-2966.2009.14548.x}, \href
  {http://adsabs.harvard.edu/abs/2009MNRAS.398.1601F} {398, 1601}

\bibitem[\protect\citeauthoryear{{Ferrara} \& {Loeb}}{{Ferrara} \&
  {Loeb}}{2013}]{LoebFerrara2013}
{Ferrara} A.,  {Loeb} A.,  2013, \mn@doi [\mnras] {10.1093/mnras/stt381}, \href
  {http://adsabs.harvard.edu/abs/2013MNRAS.431.2826F} {431, 2826}

\bibitem[\protect\citeauthoryear{{Fialkov}, {Barkana}, {Tseliakhovich}  \&
  {Hirata}}{{Fialkov} et~al.}{2012}]{2012MNRAS.424.1335F}
{Fialkov} A.,  {Barkana} R.,  {Tseliakhovich} D.,   {Hirata} C.~M.,  2012,
  \mn@doi [\mnras] {10.1111/j.1365-2966.2012.21318.x}, \href
  {http://adsabs.harvard.edu/abs/2012MNRAS.424.1335F} {424, 1335}

\bibitem[\protect\citeauthoryear{{Fialkov}, {Barkana}, {Pinhas}  \&
  {Visbal}}{{Fialkov} et~al.}{2014}]{complete21cmhistory}
{Fialkov} A.,  {Barkana} R.,  {Pinhas} A.,   {Visbal} E.,  2014, \mn@doi
  [\mnras] {10.1093/mnrasl/slt135}, \href
  {http://adsabs.harvard.edu/abs/2014MNRAS.437L..36F} {437, L36}

\bibitem[\protect\citeauthoryear{{Finlator}, {Keating}, {Oppenheimer},
  {Dav{\'e}}  \& {Zackrisson}}{{Finlator} et~al.}{2018}]{technicolor}
{Finlator} K.,  {Keating} L.,  {Oppenheimer} B.~D.,  {Dav{\'e}} R.,
  {Zackrisson} E.,  2018, \mn@doi [\mnras] {10.1093/mnras/sty1949}, \href
  {http://adsabs.harvard.edu/abs/2018MNRAS.480.2628F} {480, 2628}

\bibitem[\protect\citeauthoryear{{Foreman-Mackey}, {Hogg}, {Lang}  \&
  {Goodman}}{{Foreman-Mackey} et~al.}{2013}]{emcee}
{Foreman-Mackey} D.,  {Hogg} D.~W.,  {Lang} D.,   {Goodman} J.,  2013, \mn@doi
  [\pasp] {10.1086/670067}, \href
  {http://adsabs.harvard.edu/abs/2013PASP..125..306F} {125, 306}

\bibitem[\protect\citeauthoryear{{Furlanetto}}{{Furlanetto}}{2006a}]{2006NewAR..50..157F}
{Furlanetto} S.~R.,  2006a, \mn@doi [\nar] {10.1016/j.newar.2005.11.022}, \href
  {http://adsabs.harvard.edu/abs/2006NewAR..50..157F} {50, 157}

\bibitem[\protect\citeauthoryear{{Furlanetto}}{{Furlanetto}}{2006b}]{furlanettoTb}
{Furlanetto} S.~R.,  2006b, \mn@doi [\mnras]
  {10.1111/j.1365-2966.2006.10725.x}, \href
  {http://adsabs.harvard.edu/abs/2006MNRAS.371..867F} {371, 867}

\bibitem[\protect\citeauthoryear{{Furlanetto}, {Zaldarriaga}  \&
  {Hernquist}}{{Furlanetto} et~al.}{2004}]{FHZ}
{Furlanetto} S.~R.,  {Zaldarriaga} M.,   {Hernquist} L.,  2004, \mn@doi [\apj]
  {10.1086/423025}, \href {http://adsabs.harvard.edu/abs/2004ApJ...613....1F}
  {613, 1}

\bibitem[\protect\citeauthoryear{{Furlanetto}, {Oh}  \& {Briggs}}{{Furlanetto}
  et~al.}{2006}]{furlanetto06b}
{Furlanetto} S.~R.,  {Oh} S.~P.,   {Briggs} F.~H.,  2006, \mn@doi [\physrep]
  {10.1016/j.physrep.2006.08.002}, \href
  {http://adsabs.harvard.edu/abs/2006PhR...433..181F} {433, 181}

\bibitem[\protect\citeauthoryear{{GAMBIT Scanner Workgroup} et~al.,}{{GAMBIT
  Scanner Workgroup} et~al.}{2017}]{gambit}
{GAMBIT Scanner Workgroup} T.,  et~al., 2017, preprint, \href
  {http://adsabs.harvard.edu/abs/2017arXiv170507959G} {} (\mn@eprint {arXiv}
  {1705.07959})

\bibitem[\protect\citeauthoryear{{Garaldi}, {Compostella}  \&
  {Porciani}}{{Garaldi} et~al.}{2019}]{2019MNRAS.483.5301G}
{Garaldi} E.,  {Compostella} M.,   {Porciani} C.,  2019, \mn@doi [\mnras]
  {10.1093/mnras/sty3414}, \href
  {http://adsabs.harvard.edu/abs/2019MNRAS.483.5301G} {483, 5301}

\bibitem[\protect\citeauthoryear{{Gelman} \& {Rubin}}{{Gelman} \&
  {Rubin}}{1992}]{1992StaSc...7..457G}
{Gelman} A.,  {Rubin} D.~B.,  1992, \mn@doi [Statistical Science]
  {10.1214/ss/1177011136}, \href
  {http://adsabs.harvard.edu/abs/1992StaSc...7..457G} {7, 457}

\bibitem[\protect\citeauthoryear{{Gillet}, {Mesinger}, {Greig}, {Liu}  \&
  {Ucci}}{{Gillet} et~al.}{2018}]{2018arXiv180502699G}
{Gillet} N.,  {Mesinger} A.,  {Greig} B.,  {Liu} A.,   {Ucci} G.,  2018,
  preprint, \href {http://adsabs.harvard.edu/abs/2018arXiv180502699G} {}
  (\mn@eprint {arXiv} {1805.02699})

\bibitem[\protect\citeauthoryear{{Gnedin}, {Kravtsov}  \& {Chen}}{{Gnedin}
  et~al.}{2008}]{Gnedin}
{Gnedin} N.~Y.,  {Kravtsov} A.~V.,   {Chen} H.-W.,  2008, \mn@doi [\apj]
  {10.1086/524007}, \href {http://adsabs.harvard.edu/abs/2008ApJ...672..765G}
  {672, 765}

\bibitem[\protect\citeauthoryear{{Gorce}, {Douspis}, {Aghanim}  \&
  {Langer}}{{Gorce} et~al.}{2017}]{2017arXiv171004152G}
{Gorce} A.,  {Douspis} M.,  {Aghanim} N.,   {Langer} M.,  2017, preprint, \href
  {http://adsabs.harvard.edu/abs/2017arXiv171004152G} {} (\mn@eprint {arXiv}
  {1710.04152})

\bibitem[\protect\citeauthoryear{{Greig} \& {Mesinger}}{{Greig} \&
  {Mesinger}}{2015}]{21CMMC}
{Greig} B.,  {Mesinger} A.,  2015, \mn@doi [\mnras] {10.1093/mnras/stv571},
  \href {http://adsabs.harvard.edu/abs/2015MNRAS.449.4246G} {449, 4246}

\bibitem[\protect\citeauthoryear{{Greig} \& {Mesinger}}{{Greig} \&
  {Mesinger}}{2017a}]{21CMMC_EoH}
{Greig} B.,  {Mesinger} A.,  2017a, preprint, \href
  {http://adsabs.harvard.edu/abs/2017arXiv170503471G} {} (\mn@eprint {arXiv}
  {1705.03471})

\bibitem[\protect\citeauthoryear{{Greig} \& {Mesinger}}{{Greig} \&
  {Mesinger}}{2017b}]{2017MNRAS.465.4838G}
{Greig} B.,  {Mesinger} A.,  2017b, \mn@doi [\mnras] {10.1093/mnras/stw3026},
  \href {http://adsabs.harvard.edu/abs/2017MNRAS.465.4838G} {465, 4838}

\bibitem[\protect\citeauthoryear{{Greig}, {Mesinger}, {Haiman}  \&
  {Simcoe}}{{Greig} et~al.}{2017}]{Grieg}
{Greig} B.,  {Mesinger} A.,  {Haiman} Z.,   {Simcoe} R.~A.,  2017, \mn@doi
  [\mnras] {10.1093/mnras/stw3351}, \href
  {http://adsabs.harvard.edu/abs/2017MNRAS.466.4239G} {466, 4239}

\bibitem[\protect\citeauthoryear{{Gunn} \& {Peterson}}{{Gunn} \&
  {Peterson}}{1965}]{1965ApJ...142.1633G}
{Gunn} J.~E.,  {Peterson} B.~A.,  1965, \mn@doi [\apj] {10.1086/148444}, \href
  {http://adsabs.harvard.edu/abs/1965ApJ...142.1633G} {142, 1633}

\bibitem[\protect\citeauthoryear{{Handley}, {Hobson}  \& {Lasenby}}{{Handley}
  et~al.}{2015}]{polychord}
{Handley} W.~J.,  {Hobson} M.~P.,   {Lasenby} A.~N.,  2015, \mn@doi [\mnras]
  {10.1093/mnras/stv1911}, \href
  {http://adsabs.harvard.edu/abs/2015MNRAS.453.4384H} {453, 4384}

\bibitem[\protect\citeauthoryear{{Hassan}, {Dav{\'e}}, {Finlator}  \&
  {Santos}}{{Hassan} et~al.}{2017}]{2017MNRAS.468..122H}
{Hassan} S.,  {Dav{\'e}} R.,  {Finlator} K.,   {Santos} M.~G.,  2017, \mn@doi
  [\mnras] {10.1093/mnras/stx420}, \href
  {http://adsabs.harvard.edu/abs/2017MNRAS.468..122H} {468, 122}

\bibitem[\protect\citeauthoryear{{Hassan}, {Liu}, {Kohn}  \& {La
  Plante}}{{Hassan} et~al.}{2019}]{2019MNRAS4832524H}
{Hassan} S.,  {Liu} A.,  {Kohn} S.,   {La Plante} P.,  2019, \mn@doi [\mnras]
  {10.1093/mnras/sty3282}, \href
  {http://adsabs.harvard.edu/abs/2019MNRAS.483.2524H} {483, 2524}

\bibitem[\protect\citeauthoryear{{Heavens}, {Fantaye}, {Mootoovaloo}, {Eggers},
  {Hosenie}, {Kroon}  \& {Sellentin}}{{Heavens} et~al.}{2017}]{ALAN}
{Heavens} A.,  {Fantaye} Y.,  {Mootoovaloo} A.,  {Eggers} H.,  {Hosenie} Z.,
  {Kroon} S.,   {Sellentin} E.,  2017, preprint, \href
  {http://adsabs.harvard.edu/abs/2017arXiv170403472H} {} (\mn@eprint {arXiv}
  {1704.03472})

\bibitem[\protect\citeauthoryear{{Higson}, {Handley}, {Hobson}  \&
  {Lasenby}}{{Higson} et~al.}{2017}]{Dynamic_NS}
{Higson} E.,  {Handley} W.,  {Hobson} M.,   {Lasenby} A.,  2017, preprint,
  \href {http://adsabs.harvard.edu/abs/2017arXiv170403459H} {} (\mn@eprint
  {arXiv} {1704.03459})

\bibitem[\protect\citeauthoryear{{Higson}, {Handley}, {Hobson}  \&
  {Lasenby}}{{Higson} et~al.}{2018}]{nestcheck}
{Higson} E.,  {Handley} W.,  {Hobson} M.,   {Lasenby} A.,  2018, preprint,
  \href {http://adsabs.harvard.edu/abs/2018arXiv180406406H} {} (\mn@eprint
  {arXiv} {1804.06406})

\bibitem[\protect\citeauthoryear{{Hobson}, {Jaffe}, {Liddle}, {Mukherjee}  \&
  D.}{{Hobson} et~al.}{2009}]{al2009bayesian}
{Hobson} H.,  {Jaffe} A.,  {Liddle} A.,  {Mukherjee} P.,   D. P.,  2009,
  Bayesian Methods in Cosmology.
Cambridge University Press, \url
  {https://books.google.co.uk/books?id=SMvBQwAACAAJ}

\bibitem[\protect\citeauthoryear{Jaynes}{Jaynes}{2003}]{jaynes03}
Jaynes E.~T.,  2003, Probability theory: The logic of science.
Cambridge University Press, Cambridge

\bibitem[\protect\citeauthoryear{{Kern}, {Liu}, {Parsons}, {Mesinger}  \&
  {Greig}}{{Kern} et~al.}{2017}]{2017ApJ...848...23K}
{Kern} N.~S.,  {Liu} A.,  {Parsons} A.~R.,  {Mesinger} A.,   {Greig} B.,  2017,
  \mn@doi [\apj] {10.3847/1538-4357/aa8bb4}, \href
  {http://adsabs.harvard.edu/abs/2017ApJ...848...23K} {848, 23}

\bibitem[\protect\citeauthoryear{{Koopmans} et~al.,}{{Koopmans}
  et~al.}{2015}]{2015aska.confE...1K}
{Koopmans} L.,  et~al., 2015, Advancing Astrophysics with the Square Kilometre
  Array (AASKA14), \href {http://adsabs.harvard.edu/abs/2015aska.confE...1K}
  {p.~1}

\bibitem[\protect\citeauthoryear{{La Plante} \& {Ntampaka}}{{La Plante} \&
  {Ntampaka}}{2018}]{2018arXiv181008211L}
{La Plante} P.,  {Ntampaka} M.,  2018, arXiv e-prints, \href
  {http://adsabs.harvard.edu/abs/2018arXiv181008211L} {}

\bibitem[\protect\citeauthoryear{{Liddle}}{{Liddle}}{2007}]{2007MNRAS.377L..74L}
{Liddle} A.~R.,  2007, \mn@doi [\mnras] {10.1111/j.1745-3933.2007.00306.x},
  \href {http://adsabs.harvard.edu/abs/2007MNRAS.377L..74L} {377, L74}

\bibitem[\protect\citeauthoryear{Loeb \& Furlanetto}{Loeb \&
  Furlanetto}{2013}]{FGU}
Loeb A.,  Furlanetto S.~R.,  2013, The First Galaxies in the Universe, stu -
  student edition edn.
Princeton University Press, \url {http://www.jstor.org/stable/j.ctt24hrpv}

\bibitem[\protect\citeauthoryear{{Magueijo} \& {Sorkin}}{{Magueijo} \&
  {Sorkin}}{2007}]{2007MNRAS.377L..39M}
{Magueijo} J.,  {Sorkin} R.~D.,  2007, \mn@doi [\mnras]
  {10.1111/j.1745-3933.2007.00299.x}, \href
  {http://adsabs.harvard.edu/abs/2007MNRAS.377L..39M} {377, L39}

\bibitem[\protect\citeauthoryear{{McGreer}, {Mesinger}  \&
  {D'Odorico}}{{McGreer} et~al.}{2015}]{McGreer}
{McGreer} I.~D.,  {Mesinger} A.,   {D'Odorico} V.,  2015, \mn@doi [\mnras]
  {10.1093/mnras/stu2449}, \href
  {http://adsabs.harvard.edu/abs/2015MNRAS.447..499M} {447, 499}

\bibitem[\protect\citeauthoryear{{Mellema} et~al.,}{{Mellema}
  et~al.}{2013}]{2013ExA....36..235M}
{Mellema} G.,  et~al., 2013, \mn@doi [Experimental Astronomy]
  {10.1007/s10686-013-9334-5}, \href
  {http://adsabs.harvard.edu/abs/2013ExA....36..235M} {36, 235}

\bibitem[\protect\citeauthoryear{{Mesinger} \& {Furlanetto}}{{Mesinger} \&
  {Furlanetto}}{2007}]{DEXM}
{Mesinger} A.,  {Furlanetto} S.,  2007, \mn@doi [\apj] {10.1086/521806}, \href
  {http://adsabs.harvard.edu/abs/2007ApJ...669..663M} {669, 663}

\bibitem[\protect\citeauthoryear{{Mesinger}, {Furlanetto}  \& {Cen}}{{Mesinger}
  et~al.}{2011}]{21cmFAST}
{Mesinger} A.,  {Furlanetto} S.,   {Cen} R.,  2011, \mn@doi [\mnras]
  {10.1111/j.1365-2966.2010.17731.x}, \href
  {http://adsabs.harvard.edu/abs/2011MNRAS.411..955M} {411, 955}

\bibitem[\protect\citeauthoryear{{Miralda-Escud{\'e}}, {Haehnelt}  \&
  {Rees}}{{Miralda-Escud{\'e}} et~al.}{2000}]{MHR}
{Miralda-Escud{\'e}} J.,  {Haehnelt} M.,   {Rees} M.~J.,  2000, \mn@doi [\apj]
  {10.1086/308330}, \href {http://adsabs.harvard.edu/abs/2000ApJ...530....1M}
  {530, 1}

\bibitem[\protect\citeauthoryear{{Nesseris} \&
  {Garc{\'{\i}}a-Bellido}}{{Nesseris} \&
  {Garc{\'{\i}}a-Bellido}}{2013}]{2013JCAP...08..036N}
{Nesseris} S.,  {Garc{\'{\i}}a-Bellido} J.,  2013, \mn@doi [\jcap]
  {10.1088/1475-7516/2013/08/036}, \href
  {http://adsabs.harvard.edu/abs/2013JCAP...08..036N} {8, 036}

\bibitem[\protect\citeauthoryear{{Park}, {Mesinger}, {Greig}  \&
  {Gillet}}{{Park} et~al.}{2018}]{Luminosity21cmmc}
{Park} J.,  {Mesinger} A.,  {Greig} B.,   {Gillet} N.,  2018, preprint, \href
  {http://adsabs.harvard.edu/abs/2018arXiv180908995P} {} (\mn@eprint {arXiv}
  {1809.08995})

\bibitem[\protect\citeauthoryear{{Parsons} et~al.,}{{Parsons}
  et~al.}{2014}]{2014ApJ...788..106P}
{Parsons} A.~R.,  et~al., 2014, \mn@doi [\apj] {10.1088/0004-637X/788/2/106},
  \href {http://adsabs.harvard.edu/abs/2014ApJ...788..106P} {788, 106}

\bibitem[\protect\citeauthoryear{{Patil} et~al.,}{{Patil}
  et~al.}{2017}]{2017ApJ...838...65P}
{Patil} A.~H.,  et~al., 2017, \mn@doi [\apj] {10.3847/1538-4357/aa63e7}, \href
  {http://adsabs.harvard.edu/abs/2017ApJ...838...65P} {838, 65}

\bibitem[\protect\citeauthoryear{{Peebles}}{{Peebles}}{1968}]{1968ApJ...153....1P}
{Peebles} P.~J.~E.,  1968, \mn@doi [\apj] {10.1086/149628}, \href
  {http://adsabs.harvard.edu/abs/1968ApJ...153....1P} {153, 1}

\bibitem[\protect\citeauthoryear{{Planck Collaboration} et~al.,}{{Planck
  Collaboration} et~al.}{2016a}]{2016A&A...594A..13P}
{Planck Collaboration} et~al., 2016a, \mn@doi [\aap]
  {10.1051/0004-6361/201525830}, \href
  {http://adsabs.harvard.edu/abs/2016A%26A...594A..13P} {594, A13}

\bibitem[\protect\citeauthoryear{{Planck Collaboration} et~al.,}{{Planck
  Collaboration} et~al.}{2016b}]{Planck}
{Planck Collaboration} et~al., 2016b, \mn@doi [\aap]
  {10.1051/0004-6361/201628897}, \href
  {http://adsabs.harvard.edu/abs/2016A%26A...596A.108P} {596, A108}

\bibitem[\protect\citeauthoryear{{Pober}}{{Pober}}{2015}]{2015MNRAS.447.1705P}
{Pober} J.~C.,  2015, \mn@doi [\mnras] {10.1093/mnras/stu2575}, \href
  {http://adsabs.harvard.edu/abs/2015MNRAS.447.1705P} {447, 1705}

\bibitem[\protect\citeauthoryear{{Pober}}{{Pober}}{2016}]{21cmsense}
{Pober} J.,  2016, {21cmSense: Calculating the sensitivity of 21cm experiments
  to the EoR power spectrum}, Astrophysics Source Code Library (\mn@eprint
  {ascl} {1609.013})

\bibitem[\protect\citeauthoryear{{Pober} et~al.,}{{Pober}
  et~al.}{2013}]{2013ApJ...768L..36P}
{Pober} J.~C.,  et~al., 2013, \mn@doi [\apjl] {10.1088/2041-8205/768/2/L36},
  \href {http://adsabs.harvard.edu/abs/2013ApJ...768L..36P} {768, L36}

\bibitem[\protect\citeauthoryear{{Pober} et~al.,}{{Pober}
  et~al.}{2014a}]{2014ApJ...782...66P}
{Pober} J.~C.,  et~al., 2014a, \mn@doi [\apj] {10.1088/0004-637X/782/2/66},
  \href {http://adsabs.harvard.edu/abs/2014ApJ...782...66P} {782, 66}

\bibitem[\protect\citeauthoryear{{Pober} et~al.,}{{Pober}
  et~al.}{2014b}]{2014ApJ...788...96P}
{Pober} J.~C.,  et~al., 2014b, \mn@doi [\apj] {10.1088/0004-637X/788/1/96},
  \href {http://adsabs.harvard.edu/abs/2014ApJ...788...96P} {788, 96}

\bibitem[\protect\citeauthoryear{{Press} \& {Schechter}}{{Press} \&
  {Schechter}}{1974}]{P-S}
{Press} W.~H.,  {Schechter} P.,  1974, \mn@doi [\apj] {10.1086/152650}, \href
  {http://adsabs.harvard.edu/abs/1974ApJ...187..425P} {187, 425}

\bibitem[\protect\citeauthoryear{{Pritchard} \& {Loeb}}{{Pritchard} \&
  {Loeb}}{2012}]{2012RPPh...75h6901P}
{Pritchard} J.~R.,  {Loeb} A.,  2012, \mn@doi [Reports on Progress in Physics]
  {10.1088/0034-4885/75/8/086901}, \href
  {http://adsabs.harvard.edu/abs/2012RPPh...75h6901P} {75, 086901}

\bibitem[\protect\citeauthoryear{{Santos}, {Ferramacho}, {Silva}, {Amblard}  \&
  {Cooray}}{{Santos} et~al.}{2010}]{2010MNRAS.406.2421S}
{Santos} M.~G.,  {Ferramacho} L.,  {Silva} M.~B.,  {Amblard} A.,   {Cooray} A.,
   2010, \mn@doi [\mnras] {10.1111/j.1365-2966.2010.16898.x}, \href
  {http://adsabs.harvard.edu/abs/2010MNRAS.406.2421S} {406, 2421}

\bibitem[\protect\citeauthoryear{Schwarz}{Schwarz}{1978}]{schwarz1978}
Schwarz G.,  1978, \mn@doi [Ann. Statist.] {10.1214/aos/1176344136}, 6, 461

\bibitem[\protect\citeauthoryear{{Sheth} \& {Tormen}}{{Sheth} \&
  {Tormen}}{1999}]{S-T}
{Sheth} R.~K.,  {Tormen} G.,  1999, \mn@doi [\mnras]
  {10.1046/j.1365-8711.1999.02692.x}, \href
  {http://adsabs.harvard.edu/abs/1999MNRAS.308..119S} {308, 119}

\bibitem[\protect\citeauthoryear{{Shimabukuro} \& {Semelin}}{{Shimabukuro} \&
  {Semelin}}{2017}]{2017MNRAS.468.3869S}
{Shimabukuro} H.,  {Semelin} B.,  2017, \mn@doi [\mnras]
  {10.1093/mnras/stx734}, \href
  {http://adsabs.harvard.edu/abs/2017MNRAS.468.3869S} {468, 3869}

\bibitem[\protect\citeauthoryear{Sivia \& Skilling}{Sivia \&
  Skilling}{2006}]{sivia2006data}
Sivia D.,  Skilling J.,  2006, Data Analysis: A Bayesian Tutorial.
Oxford science publications, OUP Oxford, \url
  {https://books.google.co.uk/books?id=lYMSDAAAQBAJ}

\bibitem[\protect\citeauthoryear{{Skilling}}{{Skilling}}{2004}]{skilling}
{Skilling} J.,  2004, in {Fischer} R.,  {Preuss} R.,   {Toussaint} U.~V.,  eds,
   American Institute of Physics Conference Series Vol. 735, American Institute
  of Physics Conference Series. pp 395--405, \mn@doi{10.1063/1.1835238}

\bibitem[\protect\citeauthoryear{{Sobacchi} \& {Mesinger}}{{Sobacchi} \&
  {Mesinger}}{2014}]{sobacchimesinger2014}
{Sobacchi} E.,  {Mesinger} A.,  2014, \mn@doi [\mnras] {10.1093/mnras/stu377},
  \href {http://adsabs.harvard.edu/abs/2014MNRAS.440.1662S} {440, 1662}

\bibitem[\protect\citeauthoryear{{Thompson}, {Moran}  \& {Swenson}}{{Thompson}
  et~al.}{2017}]{2017isra.book.....T}
{Thompson} A.~R.,  {Moran} J.~M.,   {Swenson} Jr. G.~W.,  2017, {Interferometry
  and Synthesis in Radio Astronomy, 3rd Edition},
  \mn@doi{10.1007/978-3-319-44431-4.
}

\bibitem[\protect\citeauthoryear{Verdinelli \& Wasserman}{Verdinelli \&
  Wasserman}{1995}]{doi:10.1080/01621459.1995.10476554}
Verdinelli I.,  Wasserman L.,  1995, \mn@doi [Journal of the American
  Statistical Association] {10.1080/01621459.1995.10476554}, 90, 614

\bibitem[\protect\citeauthoryear{{Watkinson} \& {Pritchard}}{{Watkinson} \&
  {Pritchard}}{2014}]{Catherine}
{Watkinson} C.~A.,  {Pritchard} J.~R.,  2014, \mn@doi [\mnras]
  {10.1093/mnras/stu1384}, \href
  {http://adsabs.harvard.edu/abs/2014MNRAS.443.3090W} {443, 3090}

\bibitem[\protect\citeauthoryear{{Watkinson}, {Giri}, {Ross}, {Dixon}, {Iliev},
  {Mellema}  \& {Pritchard}}{{Watkinson} et~al.}{2019}]{2019MNRAS.482.2653W}
{Watkinson} C.~A.,  {Giri} S.~K.,  {Ross} H.~E.,  {Dixon} K.~L.,  {Iliev}
  I.~T.,  {Mellema} G.,   {Pritchard} J.~R.,  2019, \mn@doi [\mnras]
  {10.1093/mnras/sty2740}, \href
  {http://adsabs.harvard.edu/abs/2019MNRAS.482.2653W} {482, 2653}

\bibitem[\protect\citeauthoryear{{Wise} \& {Cen}}{{Wise} \&
  {Cen}}{2009}]{Wise_Cen}
{Wise} J.~H.,  {Cen} R.,  2009, \mn@doi [\apj] {10.1088/0004-637X/693/1/984},
  \href {http://adsabs.harvard.edu/abs/2009ApJ...693..984W} {693, 984}

\bibitem[\protect\citeauthoryear{{Zahn}, {Mesinger}, {McQuinn}, {Trac}, {Cen}
  \& {Hernquist}}{{Zahn} et~al.}{2011}]{zahn}
{Zahn} O.,  {Mesinger} A.,  {McQuinn} M.,  {Trac} H.,  {Cen} R.,   {Hernquist}
  L.~E.,  2011, \mn@doi [\mnras] {10.1111/j.1365-2966.2011.18439.x}, \href
  {http://adsabs.harvard.edu/abs/2011MNRAS.414..727Z} {414, 727}

\bibitem[\protect\citeauthoryear{{Zeldovich} \& {Sunyaev}}{{Zeldovich} \&
  {Sunyaev}}{1969}]{1969Ap&SS...4..301Z}
{Zeldovich} Y.~B.,  {Sunyaev} R.~A.,  1969, \mn@doi [\apss]
  {10.1007/BF00661821}, \href
  {http://adsabs.harvard.edu/abs/1969Ap%26SS...4..301Z} {4, 301}

\bibitem[\protect\citeauthoryear{{van Haarlem} et~al.,}{{van Haarlem}
  et~al.}{2013}]{2013A&A...556A...2V}
{van Haarlem} M.~P.,  et~al., 2013, \mn@doi [\aap]
  {10.1051/0004-6361/201220873}, \href
  {http://adsabs.harvard.edu/abs/2013A\%26A...556A...2V} {556, A2}

\makeatother
\end{thebibliography}


\label{lastpage}
\end{document}